\documentclass[fleqn]{article}
\usepackage[utf8]{inputenc}
\usepackage{fleqn}
\usepackage{graphics}
\usepackage{epsfig}
\usepackage{amsmath}
\usepackage{amssymb}
\usepackage{calc}
\usepackage{mathtools}
\usepackage{caption}
\usepackage{subcaption}
\usepackage{comment}
\usepackage[english]{babel}
\newtheorem{theorem}{Theorem}

\newcommand{\beq}{\begin{equation}}
\newcommand{\eeq}{\end{equation}}
\newcommand{\bea}{\begin{eqnarray}}
\newcommand{\eea}{\end{eqnarray}}
\begin{document}
\Large
\begin{center}
{\bf Segmented strings and holography}
\end{center}
\large
\vspace*{-.1cm}
\begin{center}
Bercel Boldis$^{1,2}$ and P\'eter L\'evay$^{2}$

 \end{center}
 \vspace*{-.4cm} \normalsize
 \begin{center}
$^{1}$ HUN-REN Wigner Research Centre for Physics\\Konkoly-Thege Miklós u. 29-33, 1121 Budapest, Hungary

\vspace{10pt}
$^{2}$ MTA-BME Quantum Dynamics and Correlations Research Group\\
Budapest University of Technology and Economics
\\M\H uegyetem rkp. 3., H-1111 Budapest, Hungary

\vspace*{.0cm}
\vspace*{.2cm} (22 November 2023)
\end{center}
\vspace*{-.3cm} \noindent \hrulefill

\vspace*{.1cm} \noindent {\bf Abstract}
In this paper we establish a connection between segmented strings propagating in $AdS_{d+1}$ and $CFT_d$ subsystems  in
Minkowski spacetime characterized by quantum information theoretic quantities calculated for the vacuum state. We show that the area of the world sheet of a string segment on the AdS side can be connected to fidelity susceptibility
(the real part of the quantum geometric tensor)
on the CFT side.
This quantity has another interpretation as the computational complexity for infinitesimally separated states corresponding to causal diamonds
that are displaced in a spacelike manner according to the metric of kinematic space.
These displaced causal diamonds encode information for a unique reconstruction of the string world sheet segments in a holographic manner.  Dually the bulk segments are representing causally ordered sets of consecutive boundary events 
in boosted inertial frames or in noninertial ones proceeding with constant acceleration. 
For the special case of $AdS_3$ one can also see the segmented stringy area in units of $4GL$ ($G$ is Newton's constant and $L$ is the AdS length) as the conditional mutual information $I(A,C\vert B)$ calculated for a trapezoid configuration arising from boosted spacelike intervals $A$,$B$ and $C$.
In this special case the variation of the discretized Nambu-Goto action leads to an equation for entanglement entropies in the boundary theory of the form of a Toda equation.
For arbitrary $d$ the string world sheet patches are living in the modular slices of the entanglement wedge. They seem to provide some sort of tomography of the entanglement wedge where the patches are linked together by the interpolation ansatz, i.e. the discretized version of the equations of motion for the Nambu-Goto action.

\vspace*{.3cm}

\section{Introduction}

The AdS/CFT correspondence\cite{Maldacena1} makes it possible to understand quantum theories of gravity in terms of dual non-gravitational theories. The interesting feature of this correspondence is that in this framework
spacetime and gravitational physics emerge from ordinary non-gravitational quantum systems. 
The correspondence employs the holographic principle\cite{holo1,holo2}.
This states that the quantum gravitational theories are defined in terms of
ordinary non-gravitational quantum theories (typically quantum field theories) on a
fixed lower-dimensional spacetime.
Recent elaborations on this correspondence have also uncovered connections between quantum information theory and quantum gravity. Such ideas culminated in the discovery \cite{RT,Raamsdonk1,Raamsdonk2} that even classical spacetime geometry is an emergent entity encoded into the entanglement structure of the quantum states of the underlying non-gravitational quantum mechanical degrees of freedom.
An especially nice manifestation of this idea is the Ryu-Takayanagi formula\cite{RT} and its covariant generalizations\cite{HRT,Wall} which are telling us
that spacetime could be a geometrical representation of the entanglement structure of certain CFT
states.

The basic question underlying all of these considerations is that 
precisely how a classical spacetime geometry $M$ is
encoded into a particular CFT state?
Of course in order to study such a question one should restrict to states which have sensible dual classical descriptions.
Then according to the covariant  holographic entanglement entropy proposal\cite{HRT} in principle one should be able to recover the dual spacetime geometry by calculating entanglement entropies for many different regions and then
searching for a geometry with extremal surface areas matching with the entropies. This
appears to be a highly overconstrained problem, due to the fact that entanglement entropies give us
some function on the space of causal diamonds\cite{Myers} associated to subsets. On the other hand the dual geometries 
are specified by  the much smaller space of a small number of functions of a few coordinates \cite{Raamsdonk2}.

In order to remedy this situation in this paper we propose that for a consistent recovery for an $M$ one should  take it together with its propagating strings.
As is well-known, unlike point-like objects, the quantization of extended ones like strings
puts nontrivial constraints on $M$.
Moreover, taking strings into account there are much more
degrees of freedom to be considered than in the point particle limit. These degrees of freedom are naively expected to be the ones that should somehow be  connected to subsystems and their causal diamonds in a holographic manner.
Finally we note that entertaining this idea seems natural since after all it was string theory  as a consistent quantum gravity theory which produced the first successful implementation for the idea of holography. Then it is time to put strings back into the mixture of ideas  on quantum entanglement and emergent spacetime.

Looking at the space of subsystems or more precisely on the moduli space of causal diamonds for ball-shaped regions\cite{Myers} (or kinematic space) regarded as the space of extremal surfaces\cite{Czech},
amounts to discretization of the boundary into causal sets. If we would like to relate them to stringy information on the bulk side, then one should consider a discretized version of string theory. 
Luckily a discretization of that kind already exists under the name as the theory of segmented strings\cite{Callebaut, DV1,Gubser,DV2}.

In this paper we are establishing a connection between segmented strings of the bulk and causal diamonds of the boundary by studying the simplest example of a quantum state which has a classical holographic dual namely the $CFT_d$ vacuum in $d$-dimensional Minkowski space ${\mathbb R}^{d-1,1}$.
As is well-known this state is dual to pure $AdS_{d+1}$.

We consider the propagation of string segments in $AdS_{d+1}$. The rectangular world sheet of such segments is bounded by four lightlike vectors. This means that the neighbouring four vertices of the world sheet are lightlike separated, or in other words the edges are labeled 
by the lightlike vectors $p_i$, $i=1,2,3,4$ satisfying the conservation law $p_1+p_2=p_3+p_4$. This equation shows that the dynamics of string segments can also be regarded as the scattering problem of light like objects: kinks\cite{Callebaut,DV1,Gubser}. The string segments defined in this manner have  constant normal vectors.
Given a vertex $V_1$ and two lightlike vectors $p_1$ and $p_4$ starting from it 
the remaining vertices can be calculated from the so called interpolation ansatz\cite{Gubser}. This ansatz satisfies the discretized version of the equation of motion for the string derived from the Nambu-Goto action. One can then argue  that we can patch together the individual world sheets of a segmented string consistently\cite{Callebaut,DV1,Gubser}. 

The aim of this paper is to relate data characterizing segmented strings in the bulk to quantum information theoretic data associated to causal diamonds in the boundary patch ${\mathbb R}^{d-1,1}$.
For this purpose we establish a connection between extremal surfaces of $AdS_{d+1}$ anchored to boundary subregions 
and the vertices of the world sheets of segmented strings in a holographic manner.
Our main result for $d$ even is a formula explicitly relating the areas of rectangular world sheets of bulk string segments and a combination of entanglement entropies reminiscent of conditional mutual information for boundary causal diamonds.
For $d=2$ we indeed manage to show that 
the segmented stringy area in units of 4GL (G is Newton’s constant and
L is the AdS length) is just the conditional mutual information $I(A, C\vert B)$ calculated for
a trapezoid configuration arising from suitable boosted spacelike intervals A,B and C. 
For $d>2$ and even $d$ we did not manage to arrive at a similar interpretation.
However, one can arrive at another quantum information theoretic understanding of this combination as fidelity susceptibility. More precisely we prove that for $d$ even the area of an infinitesimal string world sheet segment multiplied by the volume of a $d-2$ dimensional Euclidean ball with radius L, is  dual to the fidelity susceptibility calculated from the real part of the quantum geometric tensor\cite{Provost}.
For $d$ odd this fidelity susceptibility interpretation still holds, though the area of the string world sheet in this case is not related to the area of extremal surfaces.
It turns out that one can arrive at yet another interpretation for the world sheet area as the computational complexity for infinitesimally separated states corresponding to causal diamonds
that are displaced in a spacelike manner according to the metric of kinematic space.
This interpretation based on quantum complexity
holds for arbitrary $d$, however the one based on a special combination of entanglement entropies
is valid only for $d$ even.
We then point out that this result is reminiscent in form to the 
"complexity equals volume" proposal\cite{Suss,Stanford} where unlike complexity entanglement is not enough to account for all the properties of spacetime structures in a holographic manner. This observation is deserving further elaboration. 

The organization of this paper is as follows.
In order to present our ideas in the simplest way
the first half of the paper is a detailed case study of the $d=2$ i.e. $AdS_3$ case. 
The second half is devoted to generalizations for $AdS_{d+1}$.
In Section 2. we recall basic information on $AdS_3$ and its extremal surfaces (geodesics). In Section 3.1 we introduce segmented strings in $AdS_3$ and consider an illustrative example.
In 3.2 we show how lightlike geodesics emanating from the vertices of the world sheet of the string segment give rise to a causally ordered set of boundary points. Subsection 3.3. is devoted to a very detailed investigation on the issue of how string segments can be reconstructed in a unique manner from  causally ordered sets of boundary points. The conclusion is that a string world sheet segment is emerging as a holographic image from an ordered set of boundary data provided by the geometry of future and past tips of causal diamonds. 
The boundary points are causally ordered
in the sense that they are representing consecutive events in boosted inertial frames or in noninertial ones proceeding with constant acceleration, i.e. exhibiting hyperbolic motion in ${\mathbb R}^{1,1}$.
The acceleration of such frames is related to the normal vector of the world sheet of the corresponding string segment.
It turns out that amusingly the reconstruction of the world sheets in the bulk from boundary data is communicated, precisely as in holography in optics, via the use of lightlike geodesics i.e. light rays.

In 3.4. we calculate the area of the world sheet of a segment, and then in 3.5 for a special arrangement we prove that 
it is related in a holographic manner to the well-known combination of entanglement entropies showing up in strong subadditivity. It then turns out that the nonnegativity of area measured in units of $4GL$ ($G$ is the 3d Newton constand and $L$ is the AdS length) is directly related to the nonnegativity of conditional mutual information for suitable combinations of boundary regions. In order to prove this claim we invoke the trapezoid configuration of boosted boundary regions used in\cite{CasiniHuerta,He} for checking the strong subadditivity for the covariant holographic entanglement entropy proposal\cite{HRT}. 
In order to generalize the results of the previous sections for the most general segmented string arrangements in 3.6. and 3.7. we use the helicity formalism.
Here we make use of the fact that by using this formalism and the $SO(2,2)$ symmetry as decomposed into left mover and right mover $SL(2,{\mathbb R})$ parts our special case  can be transformed to the general one.
This means that our main result of displaying the connection between the area and the conditional  mutual information still holds.

In 3.8. 
the physical meaning of the trapezoid configurations is clarified. They are needed to 
make it possible to relate the special entanglement entropy combinations
to conditional mutual informations.
The key result here is the observation that
displaced causal diamonds  by timelike vectors and spacelike ones are dual ones both of them can be related to areas of world sheets of string segments. 
One of the situations is related  to flow lines in modular time, and the other (related to trapezoids) to flow lines in varying acceleration.
In 3.9 it is shown that the variation of the discretized Nambu-Goto action leads to an equation for entanglement entropies in the boundary theory in the form of a Toda equation.

In Section 4. we turn to the higher dimensional cases and show that a similar correspondence holds in the $AdS_{d+1}/CFT_d$ scenario when $d$ is even.
In order to do this in 4.1 and 4.2 we summarize some basics on $AdS_{d+1}$, and its extremal surfaces.
Using a convenient parametrization in Section 4.3 we calculate the regularized area of such surfaces.
Here
we derive for $AdS_{d+1}$ with $d$ even a formula connecting the area of the world sheet of a string segment with a combination of entanglement entropies. It turns out that this combination from the boundary side, 
equals the area of a string world sheet segment multiplied by the volume of a $d-2$ dimensional
Euclidean ball with radius L, calculated in units $4G^{(d+1)}L$ on the bulk side.
The boundary side of this formula (just like in the $AdS_3$ case)  is again reminiscent of conditional mutual information for suitable boundary regions. 
Subsection 4.4. is devoted to a detailed study on the issue of how to interpret this boundary combination in terms of some quantity of quantum information theoretic meaning.
In the first half of this 
subsection we show that the analogue of considering trapezoids as in the $AdS_3/CFT_2$ case is giving some valuable new insight for regarding this combination as conditional mutual information, but unfortunately this interpretation runs out of steam due to a lack of explicit results in the literature.

In the second half of this subsection however, armed with this insight, we manage to show that there is yet another striking possibility for understanding this combination.
Indeed, for $d\geq 2$  and $d$ even we demonstrate that the area of an
infinitesimal string world sheet segment multiplied by the volume of a $d-2$ dimensional
Euclidean ball with radius L, is dual to the fidelity susceptibility calculated from the real
part of the quantum geometric tensor.
This quantity has another interpretation as the computational complexity for
infinitesimally separated states corresponding to causal diamonds that are displaced in a
spacelike manner according to the metric of kinematic space.
For $d$ odd this fidelity susceptibility interpretation of the area of world sheet segments for strings still holds. However, in this case 
the area is not related to extremal surfaces, hence to any simple universal combination of entanglement entropies.
However, for $d$ arbitrary we observe that the area  of string world-sheet segments is naturally related to quantum complexity.
This result gives rise to a surprising connection with the complexity equals
volume proposal\cite{Suss,Stanford}.

Concluding Section 4.
we point out that
the string world sheet patches are living in the modular slices of the entanglement wedge. They seem to provide some sort of tomography of the entanglement wedge where the patches are linked together by the interpolation ansatz, i.e. the discretized version of the equations of motion for the Nambu-Goto action.
This interpretation is valid for $d$ arbitrary. Hence for the special case explored in this paper segmented strings in the bulk seem to be holographically related not to quantum entanglement but rather to quantum complexity properties of boundary subsystems in a natural manner. 
The conclusions and comments are left for Section 5.
Some calculational details can be found in Appendix A, B and C.

\section{$AdS_3$ and its extremal surfaces}

\subsection{The $AdS_3$ space and its Poincaré patch}

The three dimensional anti de Sitter space $AdS_3$ is the locus of points $X\in\mathbb{R}^{2,2}$ whose coordinates $X^a, a=-1,0,1,2$ 
satisfy the constraint
\begin{equation}
    X\cdot X:=\eta_{ab}X^aX^b:=-(X^{-1})^2-(X^0)^2+(X^1)^2+(X^2)^2=-X\overline{X}+X^+X^-=-L^2
\end{equation}
where 
\begin{equation}
X^{\pm}=X^1\pm X^{-1}\qquad
X=X^0+X^2,\quad \overline{X}=X^0-X^2
\label{pm2}
\end{equation}
and
$L$ is the AdS radius. The two dimensional asymptotic boundary of the $AdS_{3}$ space is defined by the set
\begin{equation}
    {\partial}_{\infty} AdS_{3}\coloneqq {\mathbb P}\{U\in{\mathbb R}^{2,2}\vert U\cdot U=0\}
\end{equation}
where ${\mathbb P}$ means projectivization.

Following the convention of \cite{DV1} one can define the Poincaré patch representation of the $AdS_3$ space
\begin{equation}
X=\left(\frac{t^{2}-z^{2}-x^{2}-L^2}{2 z}, L\frac{t}{z}, \frac{t^{2}-z^{2}-x^{2}+L^2}{2 z}, L\frac{x}{z}\right)
\label{poin}
\end{equation}
The line element in these coordinates $x^{\hat{\mu}}=(x^0,x^1,x^2)=(t,x,z)$, $x^{\mu}=(x^0,x^1)$ is
\begin{equation}
ds^2=g_{\hat{\mu}\hat{\nu}}dx^{\hat{\mu}}dx^{\hat{\nu}}=L^2\frac{dz^2-dt^2+dx^2}{z^2}
\label{pmetric}
\end{equation}
The Poincaré patch coordinates $x$ of an $AdS_3$ point can be expressed by the global coordinates $X$ in the following way:
\begin{equation}
     x^0:=t=L\frac{X^0}{X^-},\quad x^1:=x=L\frac{X^2}{X^-}\quad x^2:=z=\frac{L^2}{X^-} 
\label{patch}
\end{equation}
The boundary of the $AdS$ space in the Poincaré patch is obtained by taking the $z\to 0$ limit. Notice that in this limit the metric is conformally equivalent to the $d=2$ dimensional Minkowki space.
The $\mathbb{R}^{2,2}$ null vectors $U$ representing boundary points have coordinates:
\begin{equation}
    x_u^\mu=(x_u^0,x_u^1):=(t_u,x_u)=\frac{L}{U^-}(U^0,U^2)
\label{boundcord}
\end{equation}

Since the conformal boundary is $d=2$ dimensional Minkowski space it is useful to introduce the product of two vectors $x,y\in{\mathbb R}^{1,1}$ with components $x^{\mu}, y^{\mu}$ 
\begin{equation}
x\bullet y
\label{bullet}=\eta_{\mu\nu}x^{\mu}y^{\nu}=-x^0y^0+x^1y^1
\end{equation}
In particular for two null vectors $U$ and $V$ representing 
boundary points $x_u$ and $x_v$ we have
\begin{equation}
(x_u-x_v)^2:=(x_u-x_v)\bullet (x_u-x_v)=-(t_u-t_v)^2+(x_u-x_v)^2
\end{equation}

Notice that since for an arbitrary vector $K\in {\mathbb R}^{2,2}$ one has
$K\cdot K=K\bullet K+K^+K^-$, then for a special vector with the property $K^-=0$ one has $K\cdot K=K\bullet K$.
As an example for a vector of that kind we take $K\equiv UV^--VU^-$, with $U$ and $V$ null.
Then one has
\begin{equation}
(UV^--VU^-)\bullet (UV^--VU^-)=(UV^--VU^-)\cdot (UV^--VU^-)=-2(U\cdot V)(U^-V^-)
\nonumber
\end{equation} 
Dividing this equation by $(U^-V^-)^2$ and using (\ref{boundcord}) one arrives at the important formula
\begin{equation}
(x_u-x_v)\bullet(x_u-x_v)=-2L^2\frac{U\cdot V}{U^-V^-}
\label{fontos}
\end{equation}

\subsection{Extremal surfaces of $AdS_3$}

In section we wish to work with a special set of codimension two spacelike extremal surfaces of the $AdS_{3}$ space. These surfaces (curves) will be chosen to be homologous to boundary regions with end points spacelike separated.
The surfaces are extremal, meaning that they give rise to stationary points of the area (length) functional.
The elements of this set will be surfaces (geodesics) homologous to spacelike regions not necessarily lying on the same time slices (hyperplanes).
These geodesics are on totally geodesic hyperplanes then according to the covariant holographic entropy proposal\cite{HRT}
the constructions based on the extremal surfaces and light sheets yield the same result, and boils down to the usual calculation of spacelike geodesics on boosted time slices.

A particular surface from this set
is defined by null vectors $U$ and $V$ with components $(U^{-1},U^0,U^1,U^2)$ and $V=(V^{-1},V^0,V^1,V^2)$ such that
\begin{equation}
U\cdot U=V\cdot V=0
\label{nullcond}
\end{equation}
also satisfying the extra constraints
\begin{equation}
    U\cdot V<0,\qquad U^-V^-<0
\label{justlike}
\end{equation}
where $U^\pm=U^1\pm U^{-1}$ and $V^\pm=V^1\pm V^{-1}$. Then the surface in question is the intersection of the following two hypersurfaces 
\begin{equation}
    U\cdot X=0,\qquad V\cdot X=0
\end{equation}
where $X\in AdS_3$ so that $X\cdot X=-L^2$.

Such points $X\in AdS_3$ can alternatively be represented in the Poincar\'e patch coordinates given by Eq.(\ref{patch}).
The null vectors $U$ and $V$ then will represent boundary points, with their corresponding coordinates given by
\begin{equation}
    x_u^\mu=(x_u^0,x_u^1):=(t_u,x_u)=\frac{L}{U^-}(U^0,U^2),\qquad x_v^\mu=(x_v^0,x_v^1):=(t_v,x_v)=\frac{L}{V^-}(V^0,V^2)
\end{equation}

By virtue of the (\ref{fontos}) identity the constraints 
$U\cdot V<0$ and $U^-V^-<0$ show that $x_u$ and $x_v$ are timelike separated. 
These two points will serve as the past and future tips of a causal diamond of a boundary subregion (linear segment) ${\mathcal R}$ with end points spacelike separated. See Figure 1.

In the Poincar\'e patch the equations of the cones $U\cdot X=0$ and $V\cdot X=0$ can be written in the form
\begin{align}\label{eq:cones}
z^2-\left(t-t_u\right)^2+\left(x-x_u\right)^2&=0\\
z^2-\left(t-t_v\right)^2+\left(x-x_v\right)^2&=0
\end{align}
These cones are having centers $(z,t,x)=\left(0,t_u,x_u\right)$ and $(z,t,x)=\left(0,t_v,x_v\right)$. 
Moreover, after introducing the notation
\begin{equation}
\Delta^{\mu}=(\Delta^0,\Delta^1)=(\Delta_t,\Delta_x)=(t_u-t_v,x_u-x_v)
\label{delta}
\end{equation}
\begin{equation}
x_0^{\mu}=\frac{1}{2}(t_u+t_v,x_u+x_v)
\label{iksznull}
\end{equation}
one observes that the intersection of the cones, which is our extremal surface $X_R$, is given by the two equations
\begin{equation}\label{eq:min_line}
\Delta_t(t-t_0)=\Delta_x(x-x_0),\qquad -(t-t_0)^2+(x-x_0)^2+z^2=r^2
\end{equation}
where 
\begin{equation}
r^2=\frac{1}{4}\left(\Delta_t^2-\Delta_x^2\right)>0
\label{er2}
\end{equation} 
due to Eq.(\ref{fontos}).
The first of the equations (\ref{eq:min_line}) having the form $\Delta\bullet (x-x_0)=0$ defines a spacelike hyperplane with timelike normal vector having components $(\Delta_t,\Delta_x,0)$, and the second is a hyperboloid.
Their intersection gives rise to half of an ellipse situated in the hyperplane.

\begin{figure}[!h]
    \centering
    \includegraphics[width=0.6\textwidth]{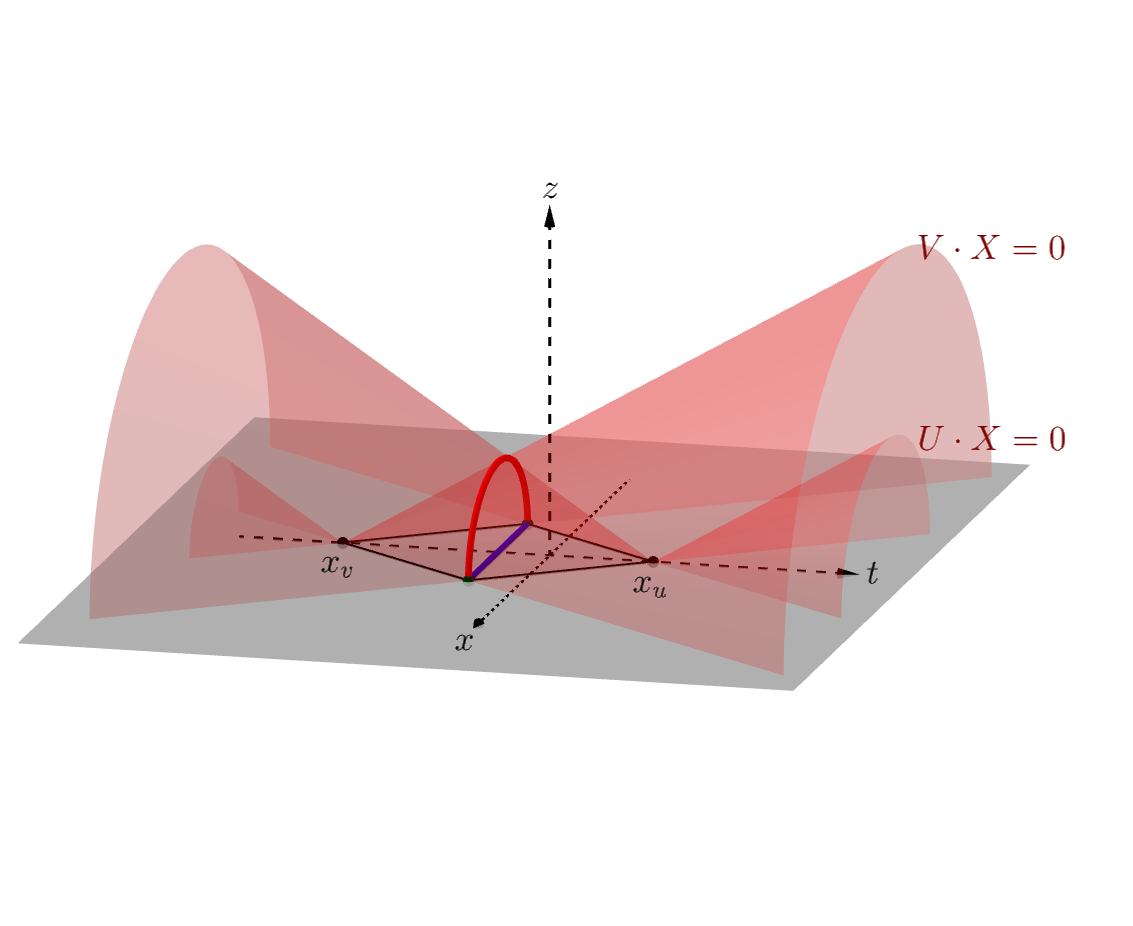}
    \caption{The spacelike minimal surface $X_{\mathcal R}$, a spacelike geodesic (red), defined by the intersection of the cones $U\cdot X=0, \quad V\cdot X=0$
in the Poincar\'e patch. In the figure the $x_u=x_v=x_0=0$ , $t=t_0=\frac{1}{2}(t_u+t_v)$ case is shown.
On the boundary (the gray plane with $z=0$) the cones give rise to a causal diamond (black) of the blue region ${\mathcal R}$.
Note that both the blue region and the red minimal surface is on a $t=t_0={\rm const}$ hypersurface (Cauchy slice) given by the first of Eq.(\ref{eq:min_line}).
 }
\end{figure}

Indeed, let us notice that the intersection of the cones with the boundary gives rise to a causal diamond  of the region ${\mathcal R}$.
The past, future, right and left tips of this diamond are having the light cone coordinates 
($x^{\pm}=x\pm t$) as follows
\begin{equation}
x_u=(x_u^+,x_u^-),\quad x_v=(x_v^+,x_v^-),\quad x_2=(x_v^+,x_u^-),
\quad x_1=(x_u^+,x_v^-)
\label{tips}
\end{equation}
Now writing
\begin{equation}
x^{\mu}(s)=\frac{1}{2}w^{\mu}s+x_0^{\mu},\qquad s\in[-1,1]
\end{equation}
one can get  $x^{\mu}(-1)=x_1^{\mu}$ and $x^{\mu}(1)=x_2^{\mu}$ for the left and right tips.
Then one can verify that the vector $w^{\mu}$ is 
having the explicit form $w^{\mu}=(\Delta_x,\Delta_t)$
hence it is
Minkowski orthogonal to $\Delta^{\mu}$ and having the property $w^2=-\Delta^2=4R^2$.
One then obtains from the second of (\ref{eq:min_line}) the constraint
\begin{equation}
\frac{4\xi^2}{\vert\vert w\vert\vert^2}+\frac{4z^2}{w^2}=1
\end{equation}
where $\xi=\vert\vert w\vert\vert s/2$.
Hence our extremal surface is just half of an ellipse lying in the hyperplane with coordinate axes $(\xi,z)$
and eccentricity
\begin{equation}
\epsilon =\sqrt{1-\frac{w^2}{\vert\vert w\vert\vert^2}}=
\sqrt{1-\frac{(t_u-t_v)^2-(x_u-x_v)^2}{(t_u-t_v)^2+(x_u-x_v)^2}}
\label{eccentricity}
\end{equation}

As an illustration let us consider the familiar $t=\text{const}$ case. From the first of Eq.(\ref{eq:min_line}) one can see that for the existence of such a surface the constraint $x_u=x_v$ is required.
Then we should have $t=t_0$, the eccentricity of (\ref{eccentricity}) is zero
and the equation for the extremal surface is
\begin{equation}
z^2+(x-x_0)^2=r^2,\qquad r=\frac{1}{2}\vert t_u-t_v\vert
\label{err}
\end{equation}
Therefore the bulk surface is a circular arc with origin $x_0$ and radius $r$ and the corresponding boundary region ${\mathcal R}$ is a line segment. Due to the translational invariance in the spatial direction as a further specification one can consider $x_0=0$. Of course our extremal surface $X_{\mathcal R}$ is now a minimal one which is just a spacelike geodesic of $AdS_3$. This is illustrated in Figure 1.

Let us reproduce the well known area formula (geodesic length) of our special surface.  
One can parametrize our surface by the parameter $y=z/r$. Hence the induced metric on the surface is:
\begin{equation}
    h=L^2\frac{\partial_y z\partial_y z+\partial_y x\partial_y x}{z^2}=L^2\frac{1}{y^2(1-y^2)}
\end{equation}
The area (length) of the surface $X_R$ can be calculated by evaluating the following integral:
\begin{equation}
{\mathcal A}(X_R)=\int dS \sqrt{h}= 2L\int_{\delta/r}^1 dy\frac{(1-y^2)^{-1/2}}{y}
\end{equation}
Due to the infinite area (length) a UV cutoff $\delta\ll r$ was introduced. This leads to the familiar result that the area (length) of the $AdS_{3}$ static minimal surface (geodesic) is determined by the data provided by the null vectors $U$ and $V$ in the following manner
\begin{equation}
{\mathcal A}(X_R)=2L\log\frac{2r}{\delta}=2L\log\frac{L}{\delta}\vert\frac{U^0}{U^-}-\frac{V^0}{V^-}\vert=2L\log\frac{\vert t_u-t_v\vert}{\delta}.
\end{equation}

Notice the slightly unusual parametrization of this well-known formula.
Indeed, for $x_0=0$ it is parametrized by the time coordinates of the future and past tips of the corresponding causal diamonds in the boundary.
Explicitly we have the corresponding  past, future, right and left tips of this diamond as follows:
$x_v=(t_v,0), x_u=(t_u,0),  x_2=\frac{1}{2}(t_v+t_u,t_u-t_v)),  x_1=\frac{1}{2}(t_v+t_u,t_v-t_u))$. Since the spacelike separated end points of the subregion (interval) ${\mathcal R}$ are $x_2$ and $x_1$ the length of the geodesic $X_{\mathcal R}$ homologous to ${\mathcal R}$ is proportional to $\log\left(\sqrt{(x_2-x_1)^2}/\delta\right)=\log\left(\vert t_u-t_v\vert/\delta\right)$. This is of course the well-known result.
For the corresponding causal diamond of ${\mathcal R}$ see Figure 2.

\begin{figure}[!h]
    \centering
    \includegraphics[width=0.6\textwidth]{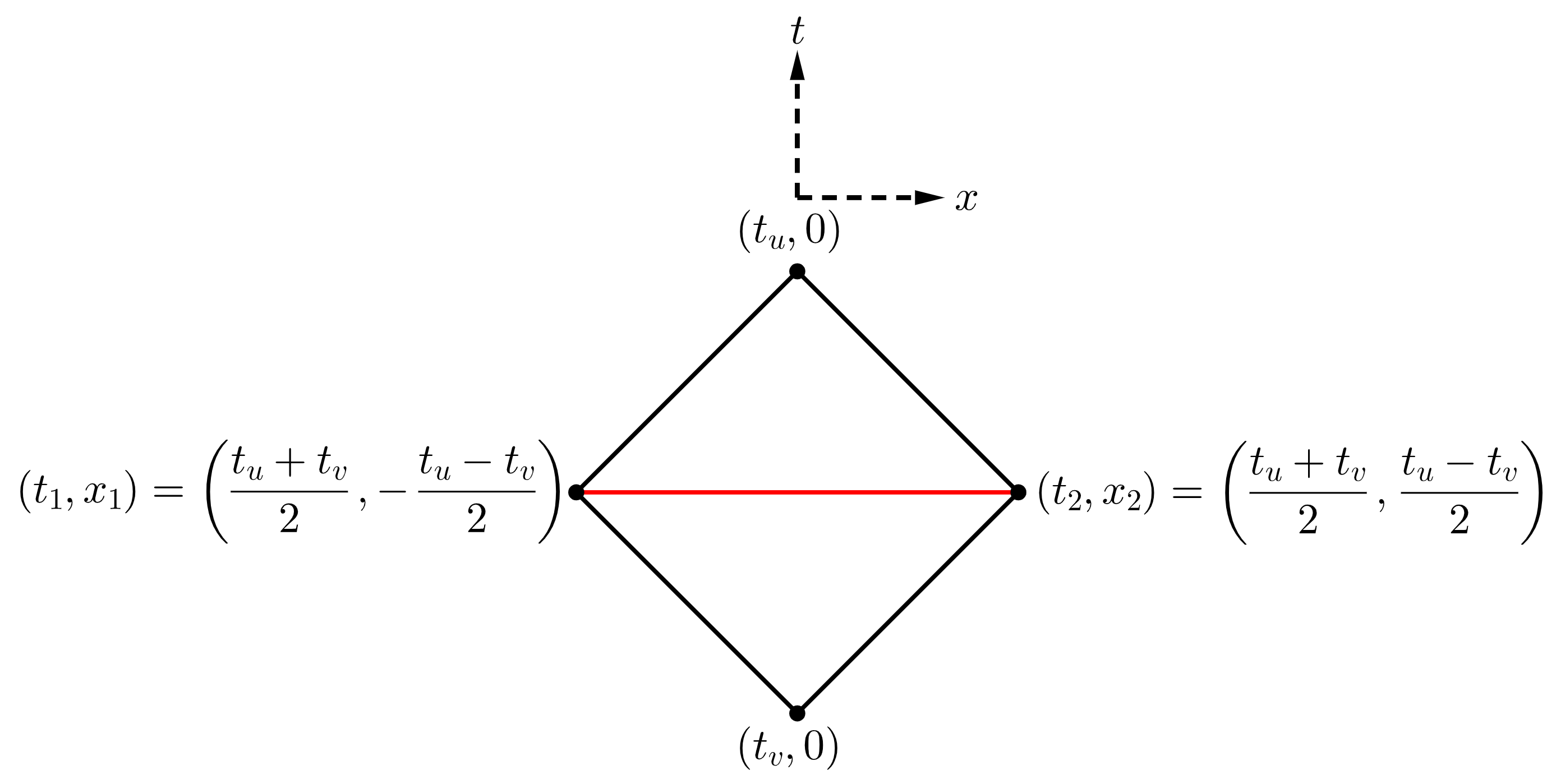}
    \caption{The static subregion $R$ (blue) with its causal diamond. For the extremal surface homologous to it see Figure 1.}
\end{figure}

\section{$AdS_3$ string segments}
\subsection{Definition of the string segments}

Two dimensional strings embedded into $AdS_{3}$ are determined 
by the following equation of motion \cite{Alday}
\begin{equation}\label{eq:eom}
    \partial_+\partial_- X-\frac{1}{L^2}(\partial_-X\cdot\partial_+X)X=0
\end{equation}
Which can be derived via the variation of the Nambu-Goto action. We have parametrized the string by the parameters $\sigma^-$ and $\sigma^+$. The Virasoro constraints for the string are
\begin{equation}
\partial_- X\cdot \partial_- X=\partial_+ X\cdot\partial_+ X=0
\end{equation}
A normal vector of the string can be defined by
\begin{equation}
N_{a}=\frac{\epsilon_{a b c d} X^{b} \partial_- X^{c} \partial_+ X^{d}}{\partial_- X\cdot \partial_+ X}
\label{Ncont}
\end{equation}The most basic solution for the equation of motion is a one that has a constant normal vector. We call a string segmented, if its
world sheet is build up from segments with constant normal vectors. In the following we examine such segments.

Let us define the rectangular world sheet of a string segment in $AdS_3$ by the four 
vectors $V_i$ with $i=1,2,3,4$ with the property that its neighbouring vertices are lightlike separated. This means that we have
\begin{equation}
V_i^2=-L^2,\qquad
 V_i\cdot V_j=-L^2, \qquad (ij)=(12),(23),(34),(14)
\label{propisegment}
\end{equation}
The lightlike vectors of the edges of the world sheet are labelled in the following way \cite{DV1}
\begin{align}
\label{cincin}
p_1=V_2-V_1&&p_2=V_3-V_2\\
p_3=V_3-V_4&&p_4=V_4-V_1
\label{cincin1}
\end{align}
Where $(p_i)^2=0$ and $p_1+p_2=p_3+p_4$. 
See Figure 3. for an illustration.

Notice that the constraint $V_i\cdot V_j=-L^2$ for neighbouring 
vertices can also be written to the form
\begin{align}
    p_1\cdot V_2=p_1\cdot V_1=0&&p_2\cdot V_3=p_2\cdot V_2=0\\
    p_3\cdot V_3=p_3\cdot V_4=0&&p_4\cdot V_4=p_4\cdot V_1=0
\end{align}
The string segments defined this way have got a constant normal vector $N$ which is given by
\begin{equation}
    N_a=\frac{\epsilon_{a b c d} V_1^{b} p_1^{c} p_4^{d}}{p_1\cdot p_4}
\label{normalis}
\end{equation}
Defining string segments this way is insightful, however a string segment can uniquely defined by the initial data: an $AdS_{3}$ vector $V_1$ and two null vectors $p_1$ and $p_4$, and one can prescribe the following initial condition
\begin{align}
    X(\sigma^-,0)&=V_1+\sigma^- p_1\\
    X(0,\sigma^+)&=V_1+\sigma^+ p_4
\end{align}
Where $\sigma^\pm\in[0,1]$. In the following we also assume that $p_1\cdot p_4<0$.

 According to 
the interpolation ansatz\cite{Gubser} points on the surface are given by the equation
\begin{equation}
    X(\sigma^-,\sigma^+)=\frac{L^2+\sigma^-\sigma^+\frac{1}{2}p_1\cdot p_4}{L^2-\sigma^-\sigma^+\frac{1}{2}p_1\cdot p_4}V_1+L^2\frac{\sigma^-p_1+\sigma^+p_4}{L^2-\sigma^-\sigma^+\frac{1}{2}p_1\cdot p_4}
\label{interpol}
\end{equation}
This expression is satisfying the (\ref{eq:eom}) equation of motion and for 
the vertices $V_2,V_3,V_4$ of the string we obtain
\begin{align}
    V_2&=X(1,0)=V_1+p_1\\
    V_4&=X(0,1)=V_1+p_4\\
    V_3&=X(1,1)=\frac{L^2+\frac{1}{2}p_1\cdot p_4}{L^2-\frac{1}{2}p_1\cdot p_4}V_1+\frac{L^2}{L^2-\frac{1}{2}p_1\cdot p_4}(p_1+p_4)
\end{align}
and for the null vectors $p_2,p_3$ 
\begin{align}
    p_2&=V_3-V_2=\frac{\frac{1}{2}p_1\cdot p_4}{L^2-\frac{1}{2}p_1\cdot p_4}\left(2V_1+p_1+\frac{2L^2}{p_1\cdot p_4}p_4\right)\label{eq:intpol1}\\
    p_3&=V_3-V_4=\frac{\frac{1}{2}p_1\cdot p_4}{L^2-\frac{1}{2}p_1\cdot p_4}\left(2V_1+\frac{2L^2}{p_1\cdot p_4}p_1+p_4\right)\label{eq:intpol2}
\end{align}
Note that the "momentum conservation" formula 
\begin{equation}
p_1+p_2=p_3+p_4
\label{momconv2}
\end{equation}
holds.
Moreover, having calculated $V_2$ and $V_4$ from the initial triple $(V_1,p_1,p_4)$ the expression for $V_3$ is given by
\begin{equation}
    V_3=-V_1-4L^2\frac{V_2+V_4}{(V_2+V_4)^2}
\label{scattt}
\end{equation}

\subsubsection*{Example}
We can start with the following initial data \cite{DV2}:
\begin{equation}
    \begin{aligned}
    V_1&=L(-1,0,0,0)\\
    p_1&=L(0,c,-c,0)\\
    p_4&=L(0,\tilde{c},\tilde{c},0)
    \end{aligned}
\label{illex}
\end{equation}
where 
\begin{equation}
c\in(0,1),\qquad \tilde{c}\in{\mathbb R}^+    
\end{equation}
 All other string segments can be generated from this situation by a global $SO(2,2)$ transformation. It can be shown that the other two null vectors are the following
\begin{align}
    p_2&=-L\frac{c\tilde{c}}{1+c\tilde{c}}\left(-2,-\frac{1}{c}+c,-\frac{1}{c}-c,0\right)\\
    p_3&=-L\frac{c\tilde{c}}{1+c\tilde{c}}\left(-2,-\frac{1}{\tilde{c}}+\tilde{c},\frac{1}{\tilde{c}}+\tilde{c},0\right)
\end{align}
Therefore the four centers of the null cones $p_i\cdot X=0$ are
\begin{equation}
    t_1=-L,\qquad
    t_2=L\frac{1+c}{1-c},\qquad
t_3=L\frac{\tilde{c}-1}{\tilde{c}+1},\qquad
    t_4=L
\label{illex2}
\end{equation}
and $x_j=0, j=1,2,3,4$.
Notice that $t_1<t_3<t_4<t_2$.

Now the four time slices of the minimal surfaces $14,23,12,34$ are given by $t_{ij}=\frac{1}{2}(t_i+t_j)$
\begin{equation}
    t_{14}=0,\qquad
    t_{23}=L
    \frac{c+\tilde{c}}{(1-c)(1+\tilde{c})}
    \qquad
    t_{12}=L\frac{c}{1-c},\qquad
    t_{34}=L\frac{\tilde{c}}{1+\tilde{c}}
\end{equation}
And similarly the radii of the minimal surfaces are given by $r_{ij}=\frac{1}{2}|t_i-t_j|$. These are the following
\begin{equation}
    r_{14}=L,\qquad
    r_{23}=L\frac{1+\tilde{c}c}{(1-c)(1+\tilde{c})},\qquad
    r_{12}=L\frac{1}{1-c},\qquad
    r_{34}=L\frac{1}{1+\tilde{c}}
\end{equation}
Therefore all of the minimal surfaces and CFT subsystems are determined by the initial data $V_1,\,p_1,\,p_4$ and the interpolation 
ansatz.

In Section 3.3 we will conduct a detailed study to show that the same argument holds for the other direction of the duality as well. 
Namely we will prove that given the four tips of the causal diamonds one can explicitly construct the world sheet of the corresponding string segment.

For the time being let us summarize the causal constraints needed for
our family of string segments. The defining $AdS$ vectors $V_1,V_2,V_3,V_4$ and the null vectors $p_1,p_2,p_3,p_4$ need to satisfy:
\begin{enumerate}
    \item $V_i^->0$ for all $i=1,2,3,4$.
    \item These vectors satisfy the previously assumed boundary conditions and the interpolation ansatz,
    \item $p_i\cdot p_j<0$ and $p_i^-p_j^-<0$ for neighbouring null vectors,
    \item $p_i\cdot p_j>0$ and $p_i^-p_j^->0$ for antipodal null vectors,
    \item And finally $p_1^0/p_1^-<p_3^0/p_3^-<p_4^0/p_4^-<p_2^0/p_2^-$.
\end{enumerate}

The first assumption is necessary to be able to represent the vertices in the Poincaré patch. The motivation behind the first two conditions are clear. The third condition is a matter of choice, because the sign of $p_1\cdot p_4$ can be arbitrary. We have choosen it to be negative to be consistent with the literature. Condition three and four provides a set of pairwise timelike separated boundary points $x_i^\mu$ via \ref{fontos}. The final condition gives an ordering of these boundary points.

Hovewer some of these conditions are not independent and we can rewrite them to a physically more motivated form with the minimal number of assumptions. If we introduce the boundary points $x_1^\mu,x_2^\mu,x_3^\mu,x_4^\mu$ which are expressed from the original null vectors $p_i$ via \eqref{tips}, these conditions are the following:
\begin{enumerate}
    \item $V_i^->0$ for all $i=1,2,3,4$.
    \item These vectors satisfy the previously assumed boundary conditions and the interpolation ansatz,
    \item $p_1\cdot p_4<0$,
    \item $(x_i-x_j)\bullet(x_i-x_j)<0$ for all neighbouring $i,j$ edges,
    \item $t_1<t_3<t_4<t_2$.
\end{enumerate}
These are equivalent with our previous assumptions. For conditions one two and five the equivalence is obvious. For the other two it is not so evident.

The third condition is motivated by the reason that in the following we will connect the segment area and entanglement entropies of spacelike boundary CFT subsystems. These subsystems being spacelike they are surrounded by the cones $p_i\cdot X=0$ whose tips $x_i^\mu$ are timelik separated. It is not hard to see that taking into account conditions two to four actually not only the boundary points of neighbouring edges are timelike separated from each other but all of them pairwise. This can be seen by calculating the inner products $p_1\cdot p_2, p_4\cdot p_2\cdot p_3, p_1\cdot p_3$ and $p_4\cdot p_2$ and assuming that $p_1\cdot p_4<0$:
\begin{align}
    p_1\cdot p_2&=\frac{p_1\cdot p_4}{L^2-\frac{1}{2}p_1\cdot p_4}<0\\
    p_3\cdot p_4&=\frac{p_1\cdot p_4}{L^2-\frac{1}{2}p_1\cdot p_4}<0\\
    p_2\cdot p_3&=p_1\cdot p_4<0\\
    p_1\cdot p_3&=\frac{\frac{1}{2}(p_1\cdot p_4)^2}{L^2-\frac{1}{2}p_1\cdot p_4}>0\\
    p_2\cdot p_4&=\frac{\frac{1}{2}(p_1\cdot p_4)^2}{L^2-\frac{1}{2}p_1\cdot p_4}>0\\
\end{align}
Hence by supposing condition four it follows that the coordinates $p_i^-$ and $p_j^-$ for adjacent edges should have different signs. But then these coordinates of opposite edges $1-3$ and $2-4$ have the same signs hence their products are positive. As we have seen $p_1\cdot p_3>0$ and $p_2\cdot p_4>0$ therefore $(x_1-x_3)\bullet(x_1-x_3)<0$ and $(x_2-x_4)\bullet(x_2-x_4)<0$. Hence the new conditions three and four give back the third and fourth assumptions of the previous ones but more like in a sense of boundary points and intervals. This also means that the only allowed configuration motivated by the boundary field theory is the one where the string segment is timelike.

\begin{comment}
    \begin{figure}[!h]
    \centering
    \includegraphics[width=0.6\textwidth]{segment.png}
    \caption{A static world sheet of a string segment (green rectangle) in the $x=0$ plane of the Poincar\'e patch. The vertices of the segment are defined by the Poincar\'e representatives $V_i^{\hat{\mu}}$ of the vectors $V_i\in {\rm AdS}_3\subset {\mathbb R}^{2,2}$ represented by black dots with the respective numbering. On the other hand the edges are labelled by the null vectors $p_1,p_2,p_3,p_4$. The edges define light rays (dashed lines) intersecting the boundary (for the static configuration the $t$ axis) in the points $t_1,t_2,t_3,t_4$.}
    \label{fig:segment}
\end{figure}
\end{comment}
\begin{figure}[!h]
    \centering
    \includegraphics[width=0.6\textwidth]{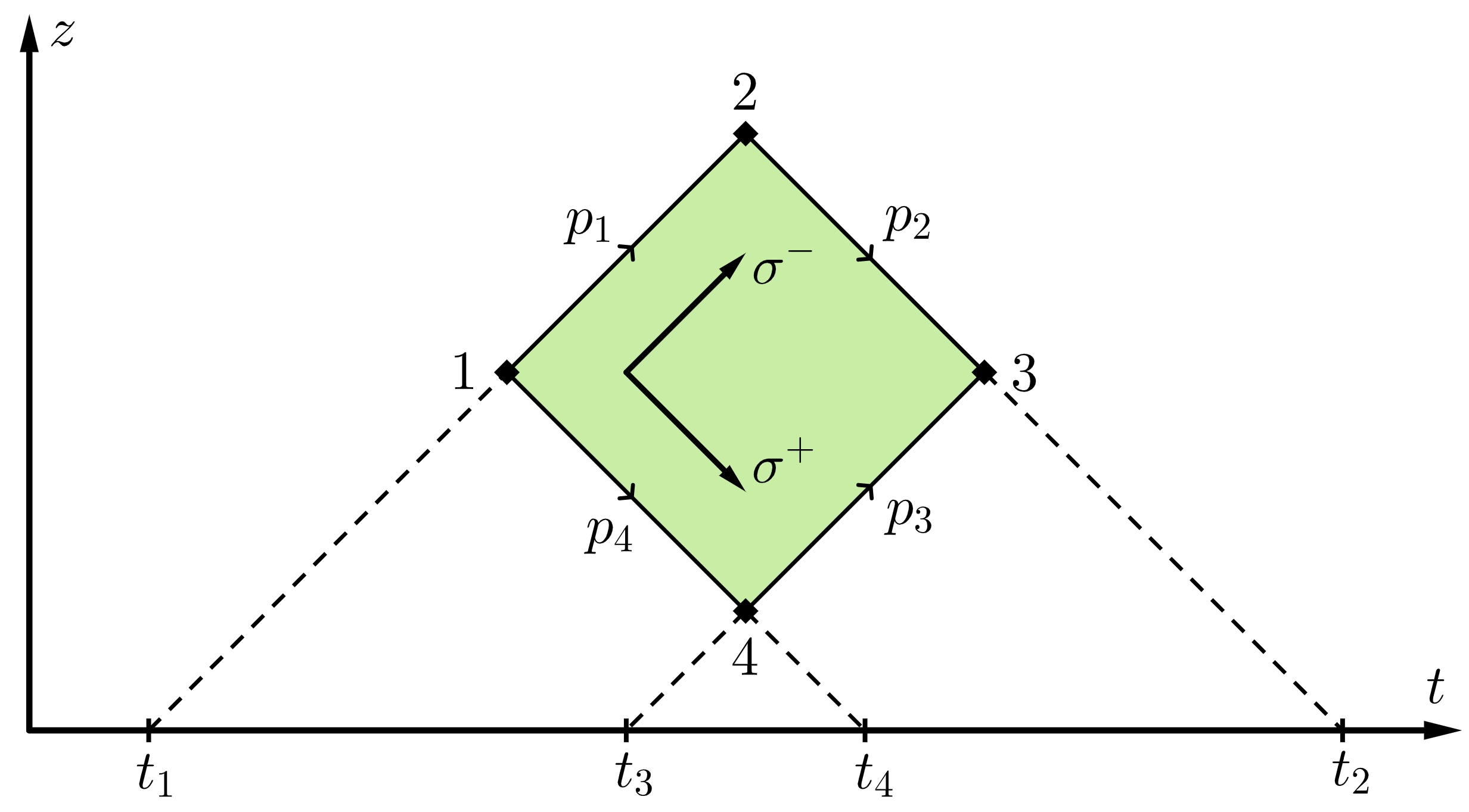}
    \caption{A static world sheet of a string segment (green rectangle) in the $x=0$ plane of the Poincar\'e patch. The vertices of the segment are defined by the Poincar\'e representatives $V_i^{\hat{\mu}}$ of the vectors $V_i\in {\rm AdS}_3\subset {\mathbb R}^{2,2}$ represented by black dots with the respective numbering. On the other hand the edges are labelled by the null vectors $p_1,p_2,p_3,p_4$. The edges define light rays (dashed lines) intersecting the boundary (for the static configuration the $t$ axis) in the points $t_1,t_2,t_3,t_4$.}
\end{figure}

\subsection{Boundary points provided by light rays}

According to Figure 3. via connecting the neighbouring vertices $V_i$ and $V_j$
by lines, the four edges of the string world-sheet give rise to four points of intersection on the boundary.
Let us first show that these lines are null geodesics
(representatives of light rays) of the (\ref{pmetric}) metric on the Poincar\'e patch, and then calculate the coordinates of these  points of intersection.

A calculation of the Christoffel symbols of our (\ref{pmetric}) metric shows that the geodesic equation for the affinely parametrized 
geodesic curve with coordinates $(t(\lambda),x(\lambda),z(\lambda))$ boils down to the following set
\begin{equation}
\ddot{t}=\frac{2\dot{t}\dot{z}}{z},\qquad
\ddot{x}=\frac{2\dot{x}\dot{z}}{z},\qquad
\ddot{z}=\frac{\dot{t}^2-\dot{x}^2+\dot{z}^2}{z}
\label{geodeq}
\end{equation}
For null geodesics the tangent vectors to this curve are lightlike hence we have 
$-\dot{t}^2+\dot{x}^2+\dot{z}^2=0$, then we get
\begin{equation}
\begin{pmatrix}\ddot{t}\\ \ddot{x}\\ \ddot{z}\end{pmatrix}=\frac{2\dot{z}}{z}\begin{pmatrix}\dot{t}\\ \dot{x}\\ \dot{z}\end{pmatrix}
\label{nullgeod}
\end{equation}
It is easy to check that the solutions give lines of the following two types

\begin{equation}
\begin{pmatrix}t(\lambda)\\ x(\lambda)\\ z(\lambda)\end{pmatrix}=\begin{pmatrix}a^t\\ a^x\\ 0\end{pmatrix}+
\frac{1}{\lambda}\begin{pmatrix}n^t\\ n^x\\ n^z\end{pmatrix},\qquad
-(n^t)^2+(n^x)^2+(n^z)^2=0
\label{solution1}
\end{equation}
\begin{equation}
\begin{pmatrix}t(\lambda)\\ x(\lambda)\\ z(\lambda)\end{pmatrix}=\begin{pmatrix}b^t\\ b^x\\ b^z\end{pmatrix}+{\lambda}\begin{pmatrix}n^t\\ n^x\\ 0\end{pmatrix},\qquad
-(n^t)^2+(n^x)^2=0
\label{solution2}
\end{equation}
For our considerations we need the line of Eq.(\ref{solution1})
which intersects the boundary at the boundary point $a^{\hat{\mu}},\hat{\mu}=(t,x,z)$ and can be written in the form
\begin{equation}
x^{\hat{\mu}}(\lambda)=a^{\hat{\mu}}+\frac{1}{\lambda}n^{\hat{\mu}}, \qquad g_{\hat{\mu}\hat{\nu}}n^{\hat{\mu}}
n^{\hat{\nu}}=0
\label{shortgeod1}
\end{equation}
Let us then consider one of the edges of our string world sheet,
e.g. the $12$ edge which is connecting the Poincar\'e patch representatives of $V_1$
and $V_2$. See Figure 3. for an illustration.
Let us call the Poincar\'e representatives of these points by $V_1^{\hat{\mu}}$ and 
$V_2^{\hat{\mu}}$.
Then we have
\begin{equation}
x^{\hat{\mu}}(\lambda_1)=V_1^{\hat{\mu}}=\frac{L}{V_1^-}\begin{pmatrix}V_1^0\\V_1^2\\L\end{pmatrix},\qquad
x^{\hat{\mu}}(\lambda_2)=V_2^{\hat{\mu}}=\frac{L}{V_2^-}\begin{pmatrix}V_2^0\\V_2^2\\L\end{pmatrix}
\label{vek}
\end{equation}
Now we can express the null tangent vector as
\begin{equation}
n^{\hat{\mu}}=\frac{\lambda_1\lambda_2}{\lambda_1-\lambda_2}\left(V_2^{\hat{\mu}}-V_1^{\hat{\mu}}\right)
\label{elso1}
\end{equation}
with
\begin{equation}
\frac{\lambda_1\lambda_2}{\lambda_1-\lambda_2}=\frac{n^z}{L^2}\left(\frac{V_1^-V_2^-}{V_1^--V_2^-}\right)
\end{equation}
From here one can see that the first two components of $n^{\hat{\mu}}$ are
\begin{equation}
L\frac{n^t}{n^z}=\frac{V_1^-V_2^{0}-V_2^-V_1^{0}}{V_1^--V_2^-},\qquad
L\frac{n^x}{n^z}=\frac{V_1^-V_2^{2}-V_2^-V_1^{2}}{V_1^--V_2^-}
\label{enkif}
\end{equation}
and the third component $n^z$ can be choosen arbitrarily.
Of course we still have to ensure that $-(n^t)^2+(n^x)^2+(n^z)^2=n^{\hat{\mu}}n_{\hat{\mu}}=0$.
But one can check that the fulfillement of this condition is guaranteed by the constraints of 
Eq. (\ref{propisegment})
needed for the very definition of the world sheet of our string segments.

At last we calculate the point of intersection of our null geodesic with the boundary. We calculate
\begin{equation}
a^{\hat{\mu}}=\frac{1}{2}\left(V_1^{\hat{\mu}}+V_2^{\hat{\mu}}-\frac{\lambda_1+\lambda_2}{\lambda_1\lambda_2}n^{\hat{\mu}}\right)
\end{equation} 
Since $a^z=0$ a calculation shows that this can be written as
\begin{equation}
a^{t}=L\frac{{p_1}^{0}}{p_1^-},\qquad
a^{x}=L\frac{{p_1}^{2}}{p_1^-}
\label{boundintersect}
\end{equation}
Hence for all of our null vectors assigned to the edges of the world sheet of our string segment 
we obtain for the coordinates of the corresponding boundary points
\begin{equation}\label{eq:zero_coord}
a_i^{t}=L\frac{p_i^{0}}{p_i^-},\qquad
a_i^{x}=L\frac{p_i^{2}}{p_i^-},\qquad
(a_i^t,a_i^x)=(t_i,x_i)
\end{equation}
with $i=1,2,3,4$.
In particular for our example of the previous subsection, one can form the points of intersection in the boundary (see Figure 3.).
\begin{equation}
   t_1= L\frac{p_1^0}{p_1^-}=-L,\qquad
   t_2= L\frac{p_2^0}{p_2^-}=L\frac{1+c}{1-c},\qquad
   t_3= L\frac{p_3^0}{p_3^-}=L\frac{\tilde{c}-1}{\tilde{c}+1},\qquad
   t_4= L\frac{p_4^0}{p_4^-}=L
\nonumber
\end{equation}
These coordinates satisfy the fourth constraint, namely $p_1^0/p_1^-<p_3^0/p_3^-<p_4^0/p_4^-<p_2^0/p_2^-$, i.e. $t_1<t_3<t_4<t_2$. Therefore for this example all of the four causality conditions are satisfied.

\begin{figure}[!h]
    \centering
    \includegraphics[width=0.6\textwidth]{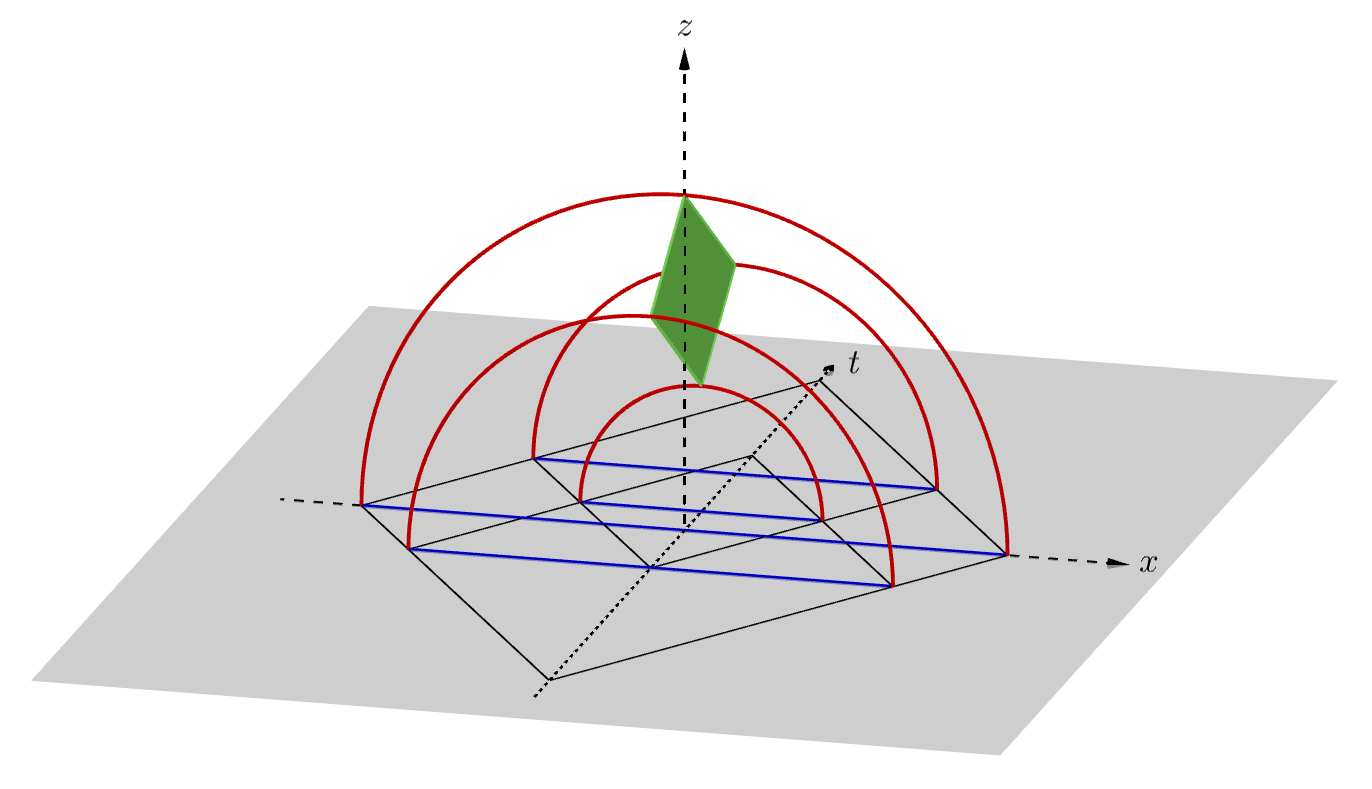}
    \caption{The correspondence between segmented strings, extremal (minimal) surfaces, and boundary subregions. The world sheet of a string segment as parametrized in Figure 3. is shown in green. The edges of the world sheet are parametrized by null vectors. The neighbouring null vectors give rise to extremal (minimal) surfaces (red). They in turn determine causal diamonds of subregions (blue) in the boundary. In the figure the special case where the boundary regions are parallel is shown. Here the tips of the causal diamonds are having varying $t$ coordinates but all have $x=0$, hence they are lying on a time axis of an inertial system.}
\label{Fig4}
\end{figure}

\subsection{String segments emerging from the data of causal diamonds}

\subsubsection{Momentum conservation emerging}

In the following we will consider quadruplets of points $x_j^{\mu}\in{\mathbb R}^{1,1}$,
$j=1,2,3,4$ representing causally ordered events giving rise to causal diamonds. Such time ordered events will be either lying on a time axis of some inertial system, or on a hyperbola describing some non inertial system exhibiting motion with constant acceleration.
Instead of the components of such events $t_j,x_j$ we will often use their light cone ones
$a_j=t_j+x_j$ and $\overline{a}_j=t_j-x_j$.
Here we would like to show, how the equation of momentum conservation for string segments in the bulk can be derived from the boundary data of diamonds.

We define our set of physical systems needed for this reconstruction to be the ones parametrized by  $(N^+,N,\overline{N},N^-)$ i.e. a quadruplet of real numbers 
satisfying the constraint
\begin{equation}
N^+N^--\overline{N}N=L^2
\label{normalconstr}
\end{equation}
Later we will see that such quadruplets correspond to the
(\ref{pm2}) components of the (\ref{normalis}) normal vectors for the world sheets of string segments. Our aim in this subsection is to construct such sheets of string segments from the boundary data $x_j^{\mu}\in{\mathbb R}^{1,1}$ in a holographic manner.

As a starting point we are fixing a quadruplet
$N^a$ in order to define a particular system with light cone coordinates $(a,\overline{a})$, by demanding that for such coordinates the following equation holds 
\begin{equation}
N^-\overline{a}a-N\overline{a}L-\overline{N}aL+N^+L^2=0   
\label{hypline}
\end{equation}
This equation is satisfied by a continuum set of points with coordinates $(t,x)$ 
with $a=t+x$ and $\overline{a}=t-x$.
In particular one can consider quadruplets from this particular set constrained as 
\begin{equation}
N^-\overline{a}_ja_j-N\overline{a}_jL-L\overline{N}a_jL+N^+L^2=0   
\label{hyplinespec}
\end{equation}
\begin{equation}
a_1<a_3<a_4<a_2,\qquad \overline{a}_1<\overline{a}_3<\overline{a}_4<\overline{a}_2   
\label{unusual}
\end{equation}

Now we consider two cases.
The first case is when $N^-=0$.
In this case we have 
\begin{equation}
N\overline{a}_j+\overline{N}a_j=N^+L
\label{unusual2}
\end{equation}
which is the equation of a line. Notice that due to (\ref{normalconstr}) we have $-\overline{N}N=L^2$ which says that the two component vector in ${\mathbb R}^{1,1}$ with light cone coordinates $(N, \overline{N})$ is spacelike.
Hence the vectors defined by point pairs with light cone coordinates $(a_i-a_j,\overline{a}_i-\overline{a}_j)$ are orthogonal to a spacelike vector, then they are timelike. The result is that the lines through such pairs of points constitute the time axis of some inertial frame in ${\mathbb R}^{1,1}$.

The second case is when $N^-\neq 0$.
In this case one has
\begin{equation}
N^+=\frac{L^2+\overline{N}N}{N^-}
\end{equation}
then
\begin{equation}
-(\overline{a}_j-\overline{a}_0)(a_j-a_0)=\varrho^2>0   
\label{hyphyphurra}
\end{equation}
with
\begin{equation}
a_0:=L\frac{N}{N^-},\qquad
\overline{a}_0=L\frac{\overline{N}}{N^-}
,\qquad \vert\varrho\vert=\frac{L^2}{\vert N^-\vert}
\label{adatok}
\end{equation}
Clearly Eq.(\ref{hyphyphurra}) defines a hyperbola centered at  
$(a_0,\overline{a}_0)$ with radius squared $\varrho^2$.
Notice that the vectors with light cone components $(a_j-a_0,\overline{a}_j-\overline{a}_0)$ are spacelike. Physically the four points are representing events lying on the world line of a system proceeding with constant acceleration 

\begin{equation}
g:=\frac{\vert N^-\vert}{L^2} 
\label{gyors}
\end{equation}
The somewhat unusually labelled points of Eq.(\ref{unusual})
are ordered according to the occurrence of the corresponding events with respect to the proper time of the accelerating system exhibiting hyperbolic motion.

Moreover, since $x_i-a_0+x_j-a_0$ is also spacelike and by virtue of
\begin{equation}
(x_i-a_0)\bullet(x_i-a_0)=(x_j-a_0)\bullet(x_j-a_0)=\varrho^2
\end{equation}
we have 
$(x_i-a_0)^2-(x_j-a_0)^2=0$
meaning that $x_i-a_0+x_j-a_0$
is orthogonal to 
$(x_i-x_j)$ then we learn that
$\Delta_{kj}=x_j-x_k$ is timelike.
Hence such point pairs lying on a hyperbola can also be used to form causal diamonds with future and past tips being $x_j$ and $x_i$ respectively.

Now for quadruplets of points representing events on noninertial frames exhibiting hyperbolic motion one can consider cross ratios like the ones
\begin{equation}
 \frac{(a_2-a_1)(a_4-a_3)}{(a_4-a_1)(a_2-a_3)}>0,\qquad
\frac{(a_2-a_1)(a_4-a_3)}{(a_3-a_1)(a_2-a_4)}>0
\label{crossfinger}
\end{equation}
Since by virtue of (\ref{hyphyphurra}) we have that 
\begin{equation}
a_j-a_k=(a_j-a_0)-(a_k-a_0)=\varrho^2\left(\frac{\overline{a}_k-\overline{a}_j}{(\overline{a_j}-\overline{a}_0)(\overline{a_k}-\overline{a}_0)}\right)    
\end{equation}
combining this with (\ref{unusual}) means that
\begin{equation}
\frac{(a_2-a_1)(a_4-a_3)}{(a_4-a_1)(a_2-a_3)}
=\frac{(\overline{a}_2-\overline{a}_1)(\overline{a}_4-\overline{a}_3)}{(\overline{a}_4-\overline{a}_1)(\overline{a}_2-\overline{a}_3)}>0
\label{cross1}
\end{equation}
Similar relation hold for the other cross ratio
\begin{equation}
\frac{(a_2-a_1)(a_4-a_3)}{(a_3-a_1)(a_2-a_4)}
=\frac{(\overline{a}_2-\overline{a}_1)(\overline{a}_4-\overline{a}_3)}{(\overline{a}_3-\overline{a}_1)(\overline{a}_2-\overline{a}_4)}>0
\label{cross2}
\end{equation}
For inertial frames one has 
using (\ref{unusual2}) the equation
$N(a_i-a_j)=-\overline{N}(\overline{a}_i-\overline{a}_j)$ which yields similar "reality conditions" for the (\ref{crossfinger}) cross ratios.

Now as a generalization of the second formula of Eq.(\ref{err}) we define
\begin{equation}
r_{jk}=\frac{1}{2}\sqrt{(a_j-a_k)(\overline{a}_j-\overline{a}_k)}
\label{errjk}
\end{equation}
Then we have
\begin{equation}
\frac{r_{12}r_{34}}{r_{14}r_{23}}=\frac{(a_2-a_1)(a_4-a_3)}{(a_4-a_1)(a_2-a_3)}
=\frac{(\overline{a}_2-\overline{a}_1)(\overline{a}_4-\overline{a}_3)}{(\overline{a}_4-\overline{a}_1)(\overline{a}_2-\overline{a}_3)}>0
\label{baromijo}
\end{equation}
and
\begin{equation}
\frac{r_{12}r_{34}}{r_{13}r_{24}}=\frac{(a_2-a_1)(a_4-a_3)}{(a_3-a_1)(a_2-a_4)}
=\frac{(\overline{a}_2-\overline{a}_1)(\overline{a}_4-\overline{a}_3)}{(\overline{a}_3-\overline{a}_1)(\overline{a}_2-\overline{a}_4)}>0
\label{baromijo2}
\end{equation}

Now we divide the identity
\begin{equation}
(a_1-a_2)(a_3-a_4)-(a_1-a_3)(a_2-a_4)+(a_1-a_4)(a_2-a_3)=0
\label{plucker}
\end{equation}
by $(a_1-a_4)(a_2-a_3)$ and use Eq.(\ref{baromijo}) to arrive at the formula
\begin{equation}
r_{12}r_{34}+r_{13}r_{24}=r_{14}r_{23}
\label{hasznalni}
\end{equation}
There are alternative ways of writing this (Plücker relation)
for example
\begin{equation}
    \frac{r_{14}}{r_{13}r_{34}}=\frac{r_{24}}{r_{23}r_{34}}+\frac{r_{12}}{r_{23}r_{13}},\qquad
    \frac{r_{23}}{r_{12}r_{13}}=\frac{r_{24}}{r_{12}r_{14}}+\frac{r_{34}}{r_{13}r_{14}}
\label{pl2}
\end{equation}

In the following we use the helicity formalism briefly summarized in Appendix B.
First we define a set of lightlike vectors $Z_j^{a}\in{\mathbb R}^{2,2}$ $j=1,2,3,4$
in terms of the $2\times 2$ matrices
in the helicity formalism we have
\begin{equation}
{\mathcal Z}_j:=z_j^T\otimes\bar{z}_j=
\begin{pmatrix}a_j\bar{a}_j&La_j\\L\bar{a}_j&L^2\end{pmatrix}
=\begin{pmatrix}
Z_j^+&Z_j\\\bar{Z}_j&Z_j^-    
\end{pmatrix},\qquad z_j=(a_j,L)
\label{zeee}
\end{equation}
In Appendix B
the following identities are proved
\begin{equation}
\frac{{\mathcal Z}_1-{\mathcal Z}_2}{r_{12}^2}r_{12}r_{34}+ \frac{{\mathcal Z}_1-{\mathcal Z}_3}{r_{13}^2}r_{13}r_{24}=\frac{{\mathcal Z}_1-{\mathcal Z}_4}{r_{14}^2}r_{14}r_{23}  
\label{matpluck}
\end{equation}
\begin{equation}
\frac{r_{13}}{r_{12}r_{23}}-\frac{r_{24}}{r_{12}r_{14}}=\frac{r_{13}}{r_{14}r_{34}}-\frac{r_{24}}{r_{23}r_{34}}
\label{important2}
\end{equation}
Let us now multiply (\ref{matpluck}) by $\frac{r_{13}}{r_{23}r_{34}}$ and then use in the coefficient of ${\mathcal Z_1}$ the identity (\ref{important2}).
Defining the  quantities
\begin{equation}
 \Lambda_{ij,k}:= \frac{r_{ij}}{r_{ik}r_{jk}}   
\label{lambdagauge}
\end{equation}
then we get the formula
\begin{equation}
-\Lambda_{24,1}{\mathcal Z}_1+\Lambda_{13,2}{\mathcal Z}_2=-\Lambda_{24,3}{\mathcal Z}_3+\Lambda_{13,4}{\mathcal Z}_4
\end{equation}
Defining
\begin{equation}
 {\mathcal P_1}:=-\Lambda_{24,1}{\mathcal Z}_1,\quad {\mathcal P}_2:=\Lambda_{13,2}{\mathcal Z}_2\quad {\mathcal P}_3:=-\Lambda_{24,3}{\mathcal Z}_3,\quad {\mathcal P}_4:=\Lambda_{13,4}{\mathcal Z}_4
\label{pek}
\end{equation}
this equation has the form
\begin{equation}
{\mathcal P}_1+{\mathcal P}_2={\mathcal P}_3+{\mathcal P}_4
\label{conserv}
\end{equation}
or using instead of $2\times 2$ matrices vectors $p_j^a$
the momentum conservation formula (\ref{momconv2})
follows.
Hence we completed our task of reconstructing momentum conservation for segmented strings from causal diamond data.

\subsubsection{The explicit form of the emerging stringy data}

Let us now define the vectors
\begin{equation}
     V_1=L^2\left(\frac{p_3+p_4}{p_3\cdot p_4}-
    \frac{p_1+p_4}{p_1\cdot p_4}\right)
\label{veegy}
\end{equation}
\begin{equation}
     V_2=L^2\left(\frac{p_2-p_1}{p_1\cdot p_2}-\frac{p_2-p_3}{p_2\cdot p_3}\right)
\end{equation}
\begin{equation}
     V_3=L^2\left(\frac{p_2+p_3}{p_2\cdot p_3}-\frac{p_1+p_2}{p_1\cdot p_2}\right)
\end{equation}
\begin{equation}
V_4=L^2\left(\frac{p_3-p_4}{p_3\cdot p_4}-\frac{p_1-p_4}{p_1\cdot p_4}\right)
\label{venegy}
\end{equation}
One can then check that these vectors are comprising the four vertices of the world sheet of a string segment.
Indeed these vectors taken together with the $p_j^a$ are satisfying the (\ref{interpol})
interpolation ansatz and the (\ref{scattt}) scattering equation.
Moreover, since we have 
$p_1+p_2=p_3+p_4$ and as can be easily checked 
$p_1\cdot p_2 =p_3\cdot p_4$ and $p_2\cdot p_3=p_1\cdot p_4$
these expressions for $V_j$ can have alternative appearances.
Explicitly one has the identities
\begin{equation}
p_1\cdot p_2=p_3\cdot p_4=-2L^2\frac{r_{13}r_{24}}{r_{14}r_{23}},\qquad
p_2\cdot p_3=p_1\cdot p_4=-2L^2\frac{r_{13}r_{24}}{r_{12}r_{34}}
\label{Liouvillhez}
\end{equation}
and one also has
$2L^2(p_2\cdot p_4)=2L^2(p_1\cdot p_3)=(p_1\cdot p_2)(p_2\cdot p_3)$.

One can simplify further the expressions of Eqs.(\ref{veegy})-(\ref{venegy})
by using the identity
\begin{equation}
(a_i-a_j)z_k=(a_i-a_k)z_j+(a_k-a_j)z_i,\qquad
(\bar{a}_i-\bar{a}_j)\bar{z}_k=(\bar{a}_i-\bar{a}_k)\bar{z}_j+(\bar{a}_k-\bar{a}_j)\bar{z}_i
\end{equation}
Then using this and the fact that according to Eq.(\ref{zeee}) the $z_j$ are row vectors, a straightforward calculation yields
the expressions
\begin{equation}
{\mathcal V}_1=\frac{(a_4-a_3)(\bar{a}_3-\bar{a}_1)z_1^T\otimes \bar{z}_4+(a_3-a_1)(\bar{a}_4-\bar{a}_3)z_4^T\otimes \bar{z}_1}{\sqrt{(a_4-a_1)(a_3-a_1)(a_4-a_3)(\overline{a}_4-\overline{a}_1)(\overline{a}_3-\overline{a}_1)(\overline{a}_4-\overline{a}_3)      }}
\label{veujegy}
\end{equation}
\begin{equation}
{\mathcal V}_2=\frac{(a_2-a_3)(\bar{a}_3-\bar{a}_1)z_1^T\otimes \bar{z}_2+(a_3-a_1)(\bar{a}_2-\bar{a}_3)z_2^T\otimes \bar{z}_1}{\sqrt{(a_2-a_1)(a_2-a_3)(a_3-a_1)(\overline{a}_2-\overline{a}_1)(\overline{a}_2-\overline{a}_3)(\overline{a}_3-\overline{a}_1)      }}
\label{veujketto}
\end{equation}
\begin{equation}
{\mathcal V}_3=\frac{(a_2-a_1)(\bar{a}_3-\bar{a}_1)z_3^T\otimes \bar{z}_2+(a_3-a_1)(\bar{a}_2-\bar{a}_1)z_2^T\otimes \bar{z}_3}{\sqrt{(a_2-a_1)(a_2-a_3)(a_3-a_1)(\overline{a}_2-\overline{a}_1)(\overline{a}_2-\overline{a}_3)(\overline{a}_3-\overline{a}_1)      }}
\label{veujharom}
\end{equation}
\begin{equation}
{\mathcal V}_4=\frac{(a_4-a_1)(\bar{a}_3-\bar{a}_1)z_3^T\otimes \bar{z}_4+(a_3-a_1)(\bar{a}_4-\bar{a}_1)z_4^T\otimes \bar{z}_3}{\sqrt{(a_4-a_1)(a_3-a_1)(a_4-a_3)(\overline{a}_4-\overline{a}_1)(\overline{a}_3-\overline{a}_1)(\overline{a}_4-\overline{a}_3)      }}
\label{veujnegy}
\end{equation}

 Next let us verify that the quadruplets of Eq.(\ref{normalconstr}) 
are indeed comprising the normal vectors of possible string segment world sheets.
A straightforward calculation with some details given in the Appendix A gives for
the components of the vector of Eq. (\ref{normalis}) 
\begin{equation}
N=-L\frac{\overline{a}_1a_1(a_4-a_3)+\overline{a}_3a_3(a_1-a_4)+\overline{a}_4a_4(a_3-a_1)}{\sqrt{(a_4-a_1)(a_3-a_1)(a_4-a_3)(\overline{a}_4-\overline{a}_1)(\overline{a}_3-\overline{a}_1)(\overline{a}_4-\overline{a}_3)      }}    
\label{egyeske}
\end{equation}
\begin{equation}
\overline{N}=L\frac{\overline{a}_1a_1(\overline{a}_4-\overline{a}_3)+\overline{a}_3a_3(\overline{a}_1-\overline{a}_4)+\overline{a}_4a_4(\overline{a}_3-\overline{a}_1)}{\sqrt{(a_4-a_1)(a_3-a_1)(a_4-a_3)(\overline{a}_4-\overline{a}_1)(\overline{a}_3-\overline{a}_1)(\overline{a}_4-\overline{a}_3)      }}    
\end{equation}
\begin{equation}
{N}^+=\frac{\overline{a}_1a_1(a_3\overline{a}_4-a_4\overline{a}_3)+\overline{a}_3a_3(a_4\overline{a}_1-a_1\overline{a}_4)+\overline{a}_4a_4(a_1\overline{a}_3-a_3\overline{a}_1)}{\sqrt{(a_4-a_1)(a_3-a_1)(a_4-a_3)(\overline{a}_4-\overline{a}_1)(\overline{a}_3-\overline{a}_1)(\overline{a}_4-\overline{a}_3)      }}    
\end{equation}
\begin{equation}
{N}^-=-L^2\frac{(a_3\overline{a}_4-a_4\overline{a}_3)+(a_4\overline{a}_1-a_1\overline{a}_4)+(a_1\overline{a}_3-a_3\overline{a}_1)}{\sqrt{(a_4-a_1)(a_3-a_1)(a_4-a_3)(\overline{a}_4-\overline{a}_1)(\overline{a}_3-\overline{a}_1)(\overline{a}_4-\overline{a}_3)      }}.    
\label{utolsocska}
\end{equation}
One can verify that this expression can be written in a very compact form as
\begin{equation}
{\mathcal N}=\frac{(a_4-a_3)(\bar{a}_3-\bar{a}_1)z_1^T\otimes \bar{z}_4-(a_3-a_1)(\bar{a}_4-\bar{a}_3)z_4^T\otimes \bar{z}_1}{\sqrt{(a_4-a_1)(a_3-a_1)(a_4-a_3)(\overline{a}_4-\overline{a}_1)(\overline{a}_3-\overline{a}_1)(\overline{a}_4-\overline{a}_3)}}
\label{enegy1}
\end{equation}
One can then realize that for example
\begin{equation}
{\mathcal V}_1=\alpha^3_{14}z_1^T\otimes \bar{z}_4+\bar{\alpha}^3_{14}z_4^T\otimes \bar{z}_1,\qquad
{\mathcal N}=\alpha^3_{14}z_1^T\otimes \bar{z}_4-\bar{\alpha}^3_{14}z_4^T\otimes \bar{z}_1
\end{equation}
where
\begin{equation}
\alpha^3_{14}=\sqrt{\frac{(a_4-a_3)(\bar{a}_3-\bar{a}_1)}{(a_4-a_1)(\bar{a}_4-\bar{a}_3)(a_3-a_1)(\bar{a}_4-\bar{a}_1)}}.
\end{equation}
Hence by virtue of the fact that in the helicity basis the metric is $\frac{1}{2}\epsilon\otimes \epsilon$ the well known equations $N\cdot N=-V_1\cdot V_1=L^2$ and $N\cdot V_1=0$ immediately follow.
Notice also that due to the (\ref{cross1}) "reality" condition there are alternative ways of expressing these vectors.
For example it is easy to check that
$\alpha^3_{14}=\alpha^2_{14}$    
yielding another expression for $V_1$ and $N$.

Notice that the normal vector is defined by an arbitrary triple from a quadruplet of points  $(a_j,\overline{a}_j)$ , $j=1,2,3,4$.
The expressions above singled out the triple with labels $(1,3,4)$.
It is also important to realize that (as can be verified by an explicit calculation) an arbitrary triple uniquely defines the (\ref{adatok}) parameters $(\varrho,a_0,\overline{a}_0)$ of the (\ref{hyphyphurra}) hyperbola.
Notice also that the constraint of Eq.(\ref{hyplinespec}) can be regarded as a version of the bulk equation 
$N\cdot p_j=0$ clearly valid for segmented strings.

As a special case one can consider a line through the origin with the four points having $a_j=\overline{a}_j$ i.e. $x_j=0$. In this case the points satisfy the equation $a_j(N+\overline{N})=0$. This yields $N=-\overline{N}$ hence the only nonvanishing component of the normal vector is $N^2=-L$. This is in accordance with Eq.(\ref{egyeske}).
In this case we have a line with points having only time components with $t_1<t_3<t_4<t_2$ localized on the time axis.
The four vertices of the string world sheet in this case is
\begin{align}
    \begin{pmatrix}
        V_1^-\\V_1^0\\V_1^+\\V_1^2
    \end{pmatrix}=\frac{1}{t_4-t_1}
    \begin{pmatrix}
        2L^2\\
        L(t_1+t_4)\\
        2t_1t_4\\
        0
    \end{pmatrix}&&
    \begin{pmatrix}
        V_2^-\\V_2^0\\V_2^+\\V_2^2
    \end{pmatrix}=\frac{1}{t_2-t_1}
    \begin{pmatrix}
        2L^2\\
        L(t_1+t_2)\\
        2t_1t_2\\
        0
    \end{pmatrix}\\
    \begin{pmatrix}
        V_3^-\\V_3^0\\V_3^+\\V_3^2
    \end{pmatrix}=\frac{1}{t_2-t_3}
    \begin{pmatrix}
        2L^2\\
        L(t_2+t_3)\\
        2t_2t_3\\
        0
    \end{pmatrix}&&
    \begin{pmatrix}
        V_4^-\\V_4^0\\V_4^+\\V_4^2
    \end{pmatrix}=\frac{1}{t_4-t_3}
    \begin{pmatrix}
        2L^2\\
        L(t_4+t_3)\\
        2t_4t_3\\
        0
    \end{pmatrix}
\label{visszateres}
\end{align}
And he vectors $p_i$
\begin{align}
    \begin{pmatrix}
        p_1^-\\p_1^0\\p_1^+\\p_1^2
    \end{pmatrix}=-2\frac{(t_2-t_4)}{(t_2-t_1)(t_4-t_1)}
    \begin{pmatrix}
        L^2\\
        Lt_1\\
        t_1^2\\
        0
    \end{pmatrix}&&
    \begin{pmatrix}
        p_2^-\\p_2^0\\p_2^+\\p_2^2
    \end{pmatrix}=2\frac{(t_3-t_1)}{(t_2-t_1)(t_2-t_3)}
    \begin{pmatrix}
        L^2\\
        Lt_2\\
        t_2^2\\
        0
    \end{pmatrix}\\
    \begin{pmatrix}
        p_3^-\\p_3^0\\p_3^+\\p_3^2
    \end{pmatrix}=-2\frac{(t_2-t_4)}{(t_2-t_3)(t_4-t_3)}
    \begin{pmatrix}
        L^2\\
        Lt_3\\
        t_3^2\\
        0
    \end{pmatrix}&&
    \begin{pmatrix}
        p_4^-\\p_4^0\\p_4^+\\p_4^2
    \end{pmatrix}=2\frac{(t_3-t_1)}{(t_4-t_1)(t_4-t_3)}
    \begin{pmatrix}
        L^2\\
        Lt_4\\
        t_4^2\\
        0
    \end{pmatrix}
\end{align}
One can also  verify that the momentum conservation
$p_1+p_2=p_3+p_4$ holds. This parametrization can be checked to boil down to the illustrative example of Eqs.(\ref{illex})-(\ref{illex2}).
This case is also illustrated on the left hand side figure of Figure 5. The remaining cases (boosted inertial frame, and non ineertial accelerated frame) are illustrated in the rest of Figure 5.
\begin{figure*}[!h]
        \centering
        \begin{subfigure}[b]{0.3\textwidth}
            \centering
            \includegraphics[width=\textwidth]{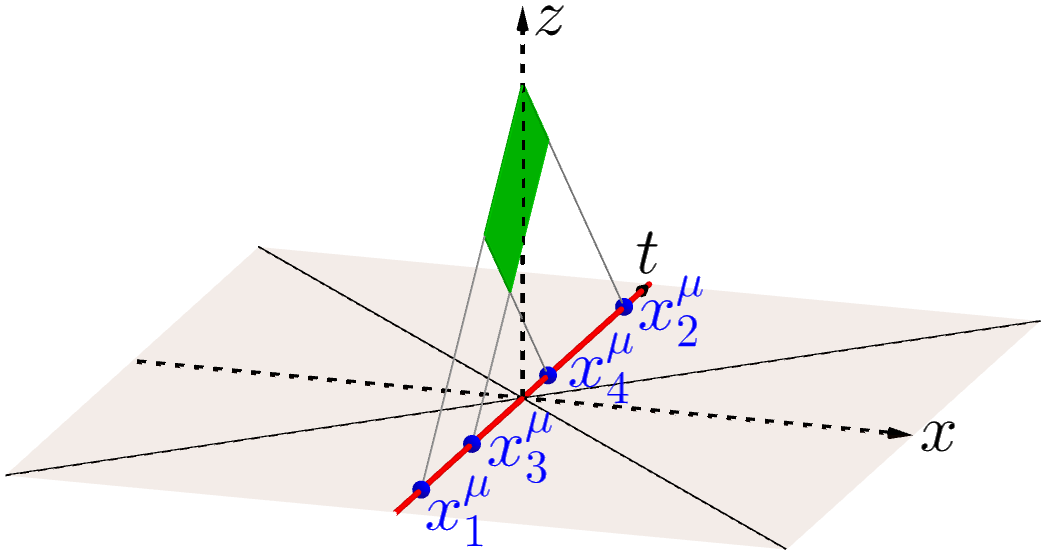}
        \end{subfigure}
        \hfill
        \begin{subfigure}[b]{0.3\textwidth}
            \centering
            \includegraphics[width=\textwidth]{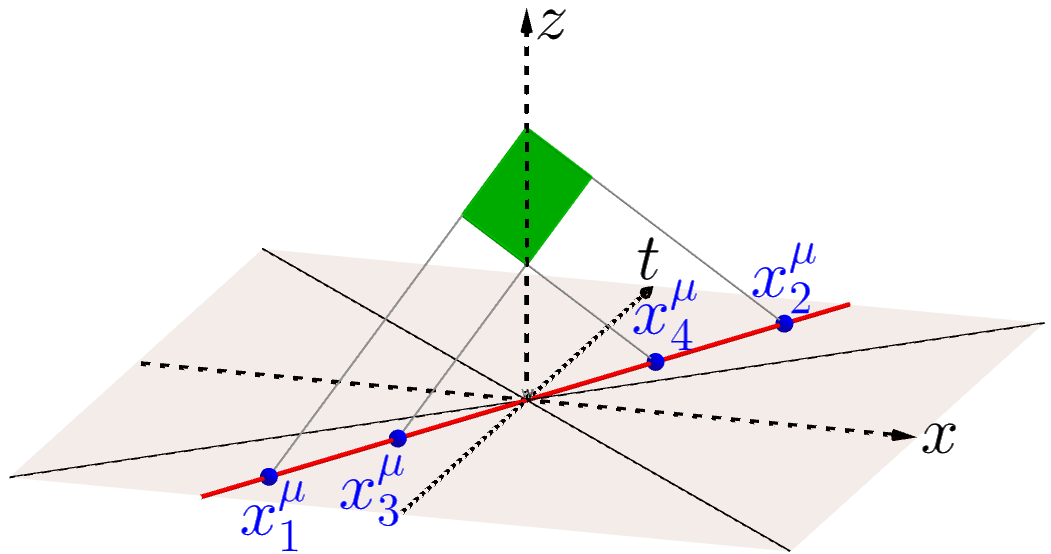}
        \end{subfigure}
        \hfill
        \begin{subfigure}[b]{0.3\textwidth}  
            \centering 
            \includegraphics[width=\textwidth]{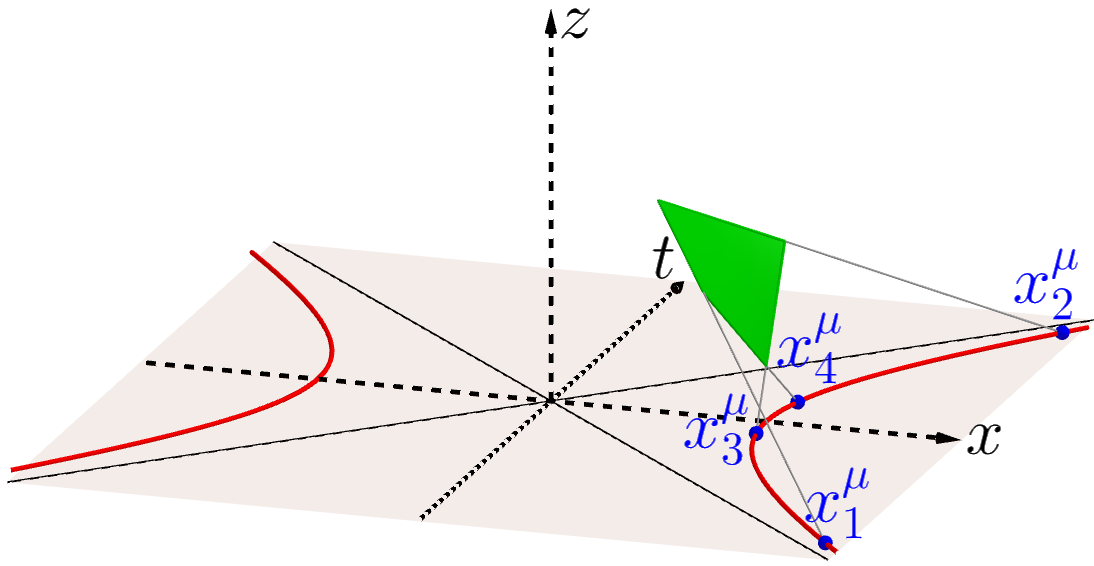}
        \end{subfigure}
        \vskip\baselineskip
        \centering
        \begin{subfigure}[b]{0.3\textwidth}
            \centering
            \includegraphics[width=\textwidth]{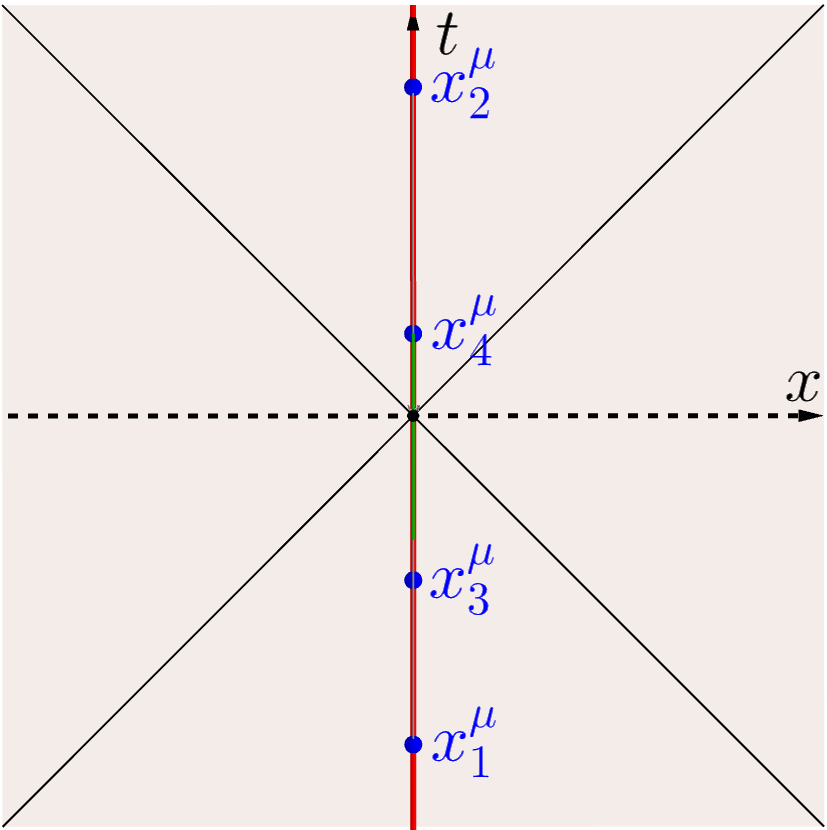}
        \end{subfigure}
        \hfill
        \begin{subfigure}[b]{0.3\textwidth}
            \centering
            \includegraphics[width=\textwidth]{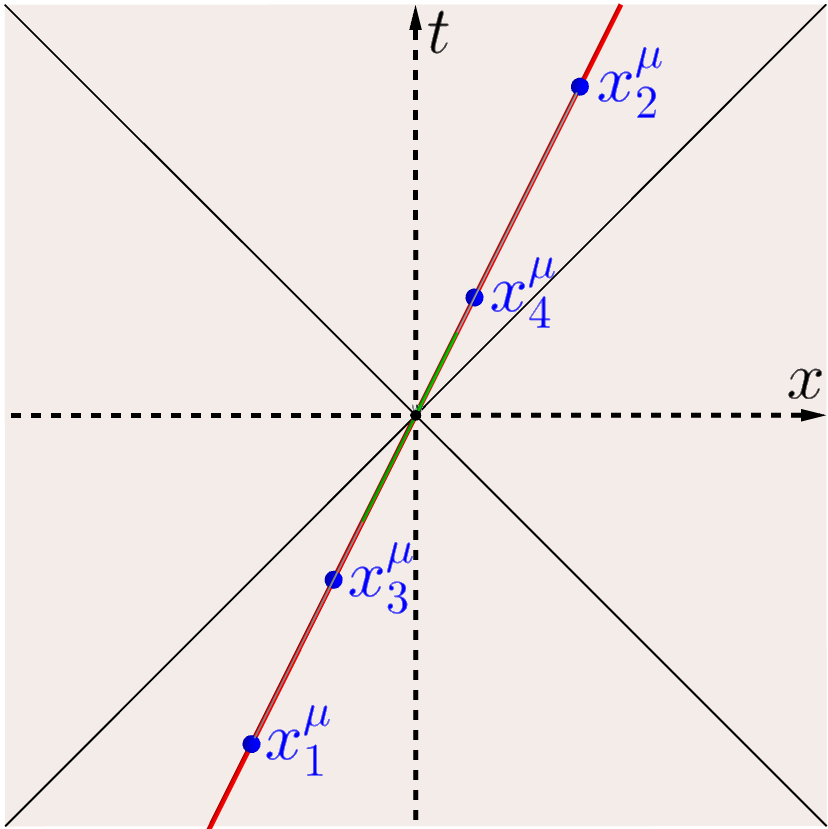}
        \end{subfigure}
        \hfill
        \begin{subfigure}[b]{0.3\textwidth}  
            \centering 
            \includegraphics[width=\textwidth]{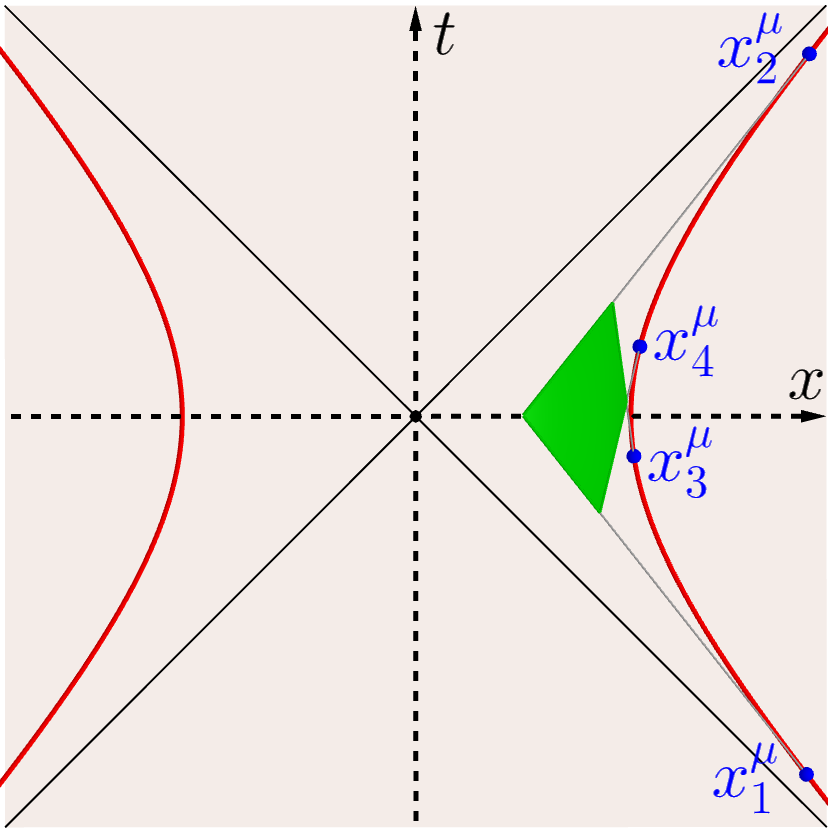}
        \end{subfigure}
        \caption{Causally ordered sets of boundary points representing consecutive events occurring in inertial, boosted inertial and noninertial  frames (frames with constant acceleration) exhibiting hyperbolic motion. According to Eq.(\ref{gyors}) the acceleration, corresponding to the situation represented by the rightmost figures, is connected to the normal vector of the world sheet (shown in green) of the string segment. In the upper part of these figures the bulk perspective displaying the lightlike geodesics needed for the reconstruction of the sheet in a holographic manner is also shown. For the corresponding causal diamonds of the leftmost figure see Figure 4. }  
        \label{fig:}
\end{figure*}
These figures nicely display that the green a string world sheet segment is emerging as a holographic image from an ordered set of boundary data provided by the geometry of the future and past tips of the corresponding causal diamonds (see also Figure 4.).
Amusingly precisely as in optics the information from the bulk to boundary and vice versa is propagated by lightlike geodesics, i.e. light rays!

Notice also that the geometry of momentum conservation in the bulk $p_1+p_2=p_3+p_4$ is connected to the "reality condition" for the cross ratios in the boundary. This latter one is ensuring that the corresponding projected points are representing an ordered set of events localized in inertial or noninertial constantly accelerated frames with acceleration determined by Eq.(\ref{gyors}).
Indeed from momentum conservation one has
\begin{equation}
    \det\begin{pmatrix}
     p_1^-&p_1^0&p_1^+&p_1^2\\  
     p_2^-&p_2^0&p_2^+&p_2^2\\  
     p_3^-&p_3^0&p_3^+&p_3^2\\  
     p_4^-&p_4^0&p_4^+&p_4^2
    \end{pmatrix}=0
\label{konzerv}
\end{equation}
Dividing each row by $p_i^-$ with the appropriate index $i$ and using the properties of the determinant one gets
\begin{equation}\label{eq:determinant}
    \det\begin{pmatrix}
     1&\frac{t_1}{L}&\frac{x_1}{L}&-\frac{x_1\bullet x_1}{L^2}\\  
     1&\frac{t_2}{L}&\frac{x_2}{L}&-\frac{x_2\bullet x_2}{L^2}\\  
     1&\frac{t_3}{L}&\frac{x_3}{L}&-\frac{x_3\bullet x_3}{L^2}\\  
     1&\frac{t_4}{L}&\frac{x_4}{L}&-\frac{x_4\bullet x_4}{L^2}
    \end{pmatrix}=0
\end{equation}
Which gives a constraint between the eight quantities $t_i$ and $x_i$ leaving seven degrees of freedom in all. It coincides with the number of parameters that were enough to specify a segment via the vectors $V_1$, $p_1$ and $p_4$. Therefore seven boundary coordinates - 3 causal cones and one remaining coordinate - are enough to define a bulk string segment that satisfies the Nambu equations of motion. It is important to note that after introducing the null coordinates $a_i=t_i+x_i$ and $\overline{a}_i=t_i-x_i$ the resulting equation gives precisely reality constraints like (\ref{baromijo}) for the cross ratio of these coordinates.

\subsection{Area of the world sheet of string segments}
As a first step in the direction of revealing the basic features of our correspondence let us calculate the area of the world sheet of a string segment.  It is achieved by determining first the induced metric $g_{\alpha\beta}=\partial_{\alpha} X\cdot \partial_{\beta} X$ on the surface, where $\alpha,\beta=\sigma^-,\sigma^+$. Using the interpolation ansatz \eqref{interpol} the components of the metric tensor are
\begin{equation}
    g_{\alpha\beta}=L^2\frac{p_1\cdot p_4}{(L^2-\sigma^-\sigma^-\frac{1}{2}p_1\cdot p_4)^2}
    \begin{pmatrix}
    0&1\\
    1&0
    \end{pmatrix}
\end{equation}
The area of the patch is then given by the formula
\begin{equation}
    A_{\diamond}=L^2\int_0^1 d\sigma^- \int_0^1 d\sigma^+\sqrt{-g}
\end{equation}
Where $g=\text{det}g_{ab}$. With a short calculation one gets for the area
\begin{equation}
    A_{\diamond}=L^2\log\left(\frac{L^2-\frac{1}{2}p_1\cdot p_4}{L^2}\right)^2
\label{shortarea}
\end{equation}
Now one can define the null vectors $p_2=V_3-(V_1+p_1)$ and $p_3=V_3-(V_1+p_4)$, where $V_3=X(1,1)$. As we have seen these vectors can be explicitly expressed as
\begin{align}
    p_2&=\frac{\frac{1}{2}p_1\cdot p_4}{L^2-\frac{1}{2}p_1\cdot p_4}\left(2V_1+p_1+\frac{2L^2}{p_1\cdot p_4}p_4\right)\\
    p_3&=\frac{\frac{1}{2}p_1\cdot p_4}{L^2-\frac{1}{2}p_1\cdot p_4}\left(2V_1+\frac{2L^2}{p_1\cdot p_4}p_1+p_4\right)
\end{align}
It is easy to show that
\begin{equation}
    p_1\cdot p_2=L^2\frac{p_1\cdot p_4}{L^2-\frac{1}{2}p_1\cdot p_4}
\end{equation}
Since $p_1\cdot p_1=p_2\cdot p_2=p_4\cdot p_4=0$ this formula can be rewritten as:
\begin{equation}
    \frac{L^2-\frac{1}{2}p_1\cdot p_4}{L^2}=-\frac{(p_1-p_4)^2}{(p_1+p_2)^2}
\end{equation}
Using momentum conservation the area of the patch can be written in the form
\begin{equation}
A_{\diamond}=L^2\log\frac{(p_1-p_4)^2(p_3-p_2)^2}{(p_1+p_2)^2(p_3+p_4)^2}=L^2\log\frac{(p_1\cdot p_4)(p_2\cdot p_3)}{(p_1\cdot p_2)(p_3\cdot p_4)}
\label{areaeq}
\end{equation}

Let us now use the (\ref{fontos}) identity for the null vectors $p_i$ and $p_j$ to arrive at
\begin{equation}
(a_i-a_j)(\bar{a}_i-\bar{a}_j)=2L^2\frac{p_ i\cdot p_j}{p_i^-p_j^-}
\end{equation}
Now for the static case we have $x_i=x_j=0$ hence we have $(t_i-t_j)^2$ for the left hand side.
Hence we obtain the alternative formula
for the area of a single string segment in the Poincaré patch as \cite{DV1}
\begin{equation}
A_{\diamond}=2L^2\log\frac{(t_4-t_1)(t_3-t_2)}{(t_4-t_3)(t_2-t_1)}
\label{boundalternative}
\end{equation}
where we have used the ordering in the causal constraints of Section 3.2 namely the one $t_1<t_3<t_4<t_2$.
Notice that by \eqref{eq:zero_coord} this expression is a combination of terms of type
\begin{equation}
A(ij)=2L^2\log\left|\frac{p_i^0}{p_i^-}-\frac{p_j^0}{p_j^-}\right|
\label{aij}
\end{equation}
of dimension area.
Recall that the area of the world sheet of the string segment of Eq.(\ref{areaeq}) is given in terms of bulk data ($p_i, i=1,2,3,4$).
On the other hand the dual formula (\ref{boundalternative}) displays this area in terms of boundary data ($t_i=Lp_i^0/p_i^-$).
The connection between the dual descriptions is effected by the holographic projection to the boundary as illustrated in Figure 3. and realized by the lines of the null geodesics of the form provided by Eq.(\ref{shortgeod1}).

\subsection{Conditional mutual information as the area of the world sheet}

Our $AdS_3$ spacetime is static meaning that it admits a timelike Killing field. Then this three dimensional spacetime has a canonical foliation with spacelike slices which in the Poincar\'e patch is given by the $t={\rm const}$ hypersurfaces. This foliation also naturally extends to the conformal boundary where a subregion ${\mathcal R}$ is singled out.
In Figure 4. one can easily identify such slices. They are the vertical hyperplanes containing both the blue regions (${\mathcal R}_{ij}$) of the boundary, and the red extremal surfaces ($X_{{\mathcal R}_{ij}}$)
of the bulk. Due to the Euclidean signature of the bulk spacelike slice an extremal surface $X_{\mathcal R}$ is a minimal one and it is guaranteed to exist.

We have seen that the area of a static, $AdS_3$ minimal surface $X_{{\mathcal R}_{ij}}$ defined by the null vectors $p_i$ and $p_j$ is
\begin{equation}
{\mathcal A}(ij):={\mathcal A}(X_{{\mathcal R}_{ij}})=2L\log\frac{L}{\delta}\left|\frac{p_i^0}{p_i^-}-\frac{p_j^0}{p_j^-}\right|
\label{jobb}
\end{equation}
 Now according to  Ryu and Takayanagi\cite{RT} this purely classical geometric calculation of the minimal surface $X_{{\mathcal R}_{ij}}$ on a constant time slice in the bulk has a boundary dual\footnote{Do not confuse our quantity ${\mathcal A}(ij)$ which is of dimension length due to the fact that our minimal "surface" is a geodesic "line", with $A(ij)$ of (\ref{aij}) which is of dimension length squared.}. Indeed, it is given by a calculation of the entanglement entropy $S({\mathcal R}_{ij})$ for the CFT vacuum quantum state of  a subsystem ${\mathcal R}_{ij}$, homologous to $X_{{\mathcal R}_{ij}}$. For a $2d$ CFT characterized by the value of the central charge $c$ we have\cite{Cardy}
\begin{equation}\label{eq:entropy_stat_3d}
S(ij):=S({\mathcal R}_{ij})=\frac{c}{3}\log\frac{\vert x_i^1-x_j^1\vert}{\delta}=\frac{c}{3}\log\frac{\vert t_i-t_j\vert}{\delta}
\end{equation}
where similar to the situation shown in Figure 2. here for the end points of ${\mathcal R}_{ij}$ we use the parametrization  
\begin{equation}
(x_i^0,x_i^1)=\left(\frac{t_i+t_j}{2},\frac{t_i-t_j}{2}\right),\qquad
(x_j^0,x_j^1)=\left(\frac{t_i+t_j}{2},\frac{t_j-t_i}{2}\right)
\end{equation}
We have $t_i=Lp_i^0/p_i^-$ yielding
\begin{equation}
S(ij):=S({\mathcal R}_{ij})=\frac{c}{3}\log\frac{L}{\delta}
\left|\frac{p_i^0}{p_i^-}-\frac{p_j^0}{p_j^-}\right|
\label{bal}
\end{equation}

\begin{figure}[!h]
    \centering
    \includegraphics[width=0.6\textwidth]{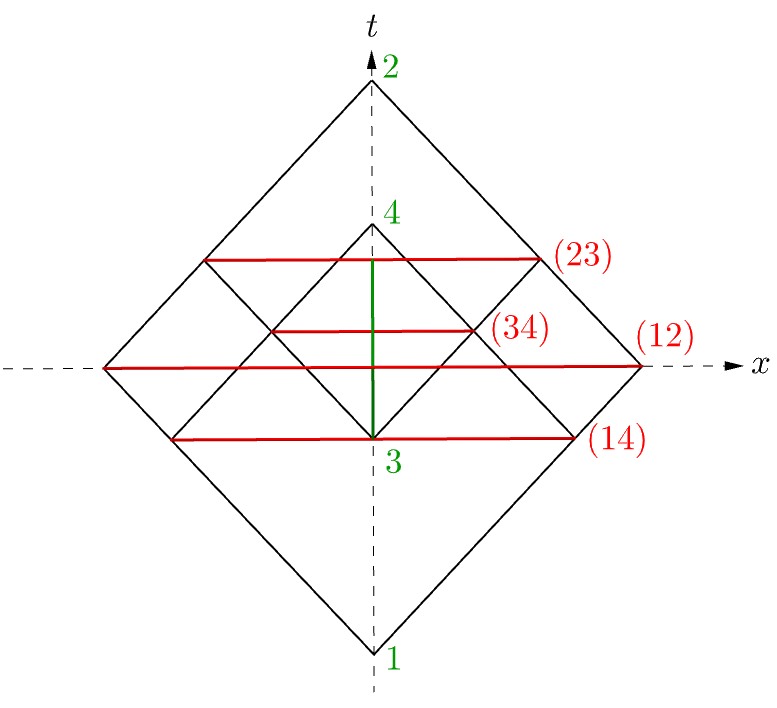}
    \caption{The boundary perspective of the string world sheet and the boundary regions hidden by minimal surfaces (red line segments) as seen from the bulk direction.
    The green line segment is the projection of the string world sheet segment. The red lines are the minimal surfaces homologous to the boundary regions: ${\mathcal R}_{12},{\mathcal R}_{34},{\mathcal R}_{23}, {\mathcal R}_{14}$. The corresponding causal diamonds (black) are also shown. For a 3D perspective see Figure 4.}
\end{figure}
Combining Eqs.(\ref{jobb}) and (\ref{bal}) and the Brown-Henneaux formula\cite{BH}
\begin{equation}
c=\frac{3L}{2G}
\label{BH}
\end{equation}
where $G$ is the three dimensional Newton constant, yields
\begin{equation}
S({\mathcal R}_{ij})=\frac{{\mathcal A}(X_{{\mathcal R}_{ij}})}{4G}
\label{RTkeplet}
\end{equation}
which is of course the Ryu-Takayanagi formula\cite{RT} for the static scenario.

Bearing in mind our considerations of segmented strings the slightly unusual parametrization provided by Eq.(\ref{RTkeplet}) of a well-known result yields some additional insight.
Indeed, let us also recall our (\ref{boundalternative}) formula
\begin{equation}
A_{\diamond}=L({\mathcal A}(14)+{\mathcal A}(23)-{\mathcal A}(12)-{\mathcal A}(34))
\label{justlike_area}
\end{equation}
for the area of the world sheet of a string segment
where ${\mathcal A}(ij)$ is given by Eq.(\ref{aij}).
Since according to Eq.(\ref{RTkeplet})
we have $S(ij)={\mathcal A}(ij)/4G$ one obtains
\begin{equation}
\frac{A_{\diamond}}{4GL}=S(14)+S(23)-S(12)-S(34)
\label{heureka}
\end{equation}
This result leads to a correspondence between the area formula for the world sheet of a string segment and a combination of entanglement entropies.

\begin{figure}[!h]
    \centering
    \includegraphics[width=0.6\textwidth]{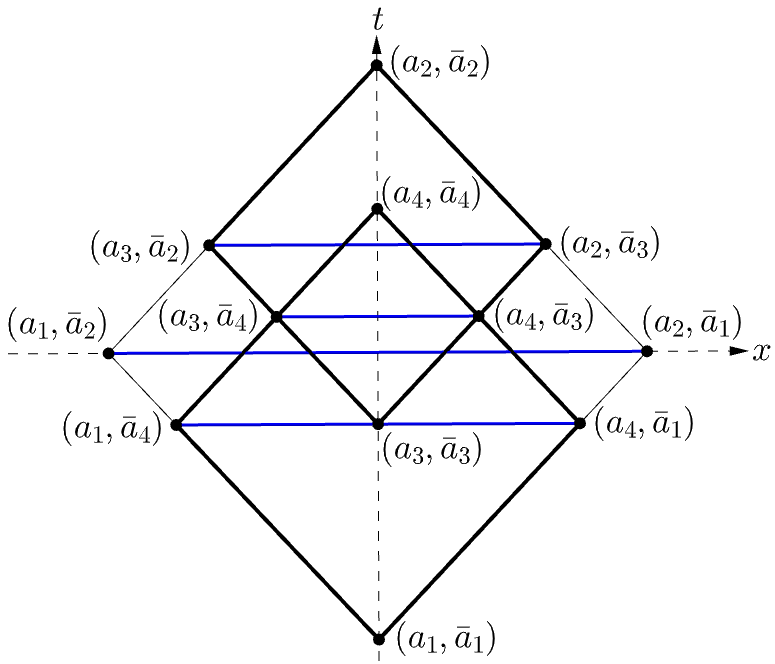}
    \caption{The parametrization of the vertices of the causal diamonds living on the boundary. It is the $z=0$ section of Figure 4. The coordinates are light cone ones.}
\end{figure}

\begin{figure}[!h]
    \centering
    \includegraphics[width=0.6\textwidth]{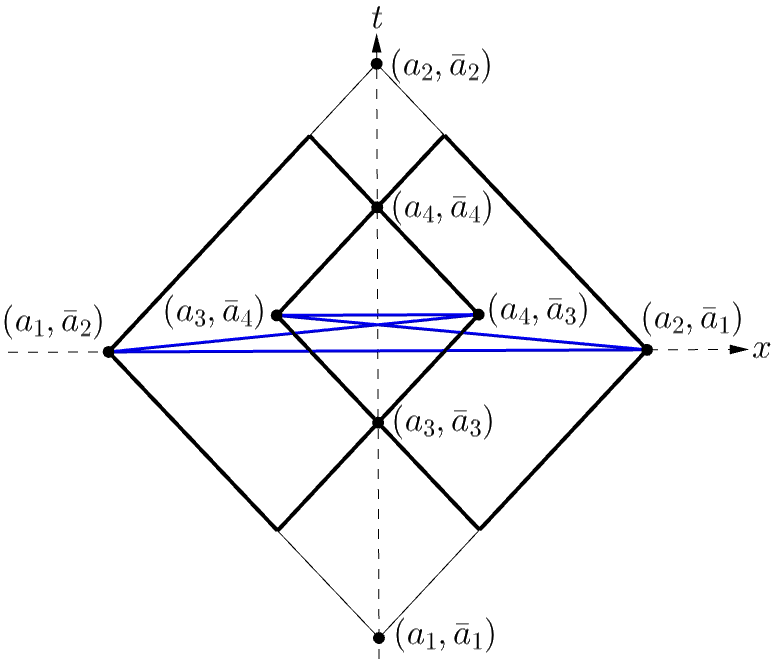}
    \caption{The arrangement of causal diamonds of the boundary needed for establishing a relation between the area of the world sheet of a string segment and conditional mutual information of  boundary subregions. The diamonds have the same intersection and union followed by causal completion as the ones of Figure 7.
    The four blue lines in this case contain two diagonal ones corresponding to the new regions ${\mathcal R}_{d_1}$ and ${\mathcal R}_{d_2}$. They are featuring the trapezoid configuration
    of Figure 9. and Eq.(\ref{justlike2}).}
\end{figure}

Now what is this combination of entanglement entropies showing up at the right hand side of (\ref{heureka})?
The combination featuring strong subadditivity (SSA) immediately
jumps into ones mind. 
SSA is the statement that
$S(14)+S(23)-S(12)-S(34)\geq 0$ for regions ${\mathcal R}_{ij}$ with the restrictions ${\mathcal R}_{12}={\mathcal R}_{14}\cup {\mathcal R}_{23}$, ${\mathcal R}_{34}={\mathcal R}_{14}\cap {\mathcal R}_{23}$ lying on the same time  slice. However, now our regions are not lying on the same time slice.  In any case if some version of this interpretation were consistent then by virtue of Eq.(\ref{heureka}) the nonnegativity of this quantity would be connected to the nonnegativity of area measured in units provided by $4GL$.

\begin{figure}[!h]
    \centering
    \includegraphics[width=0.6\textwidth]{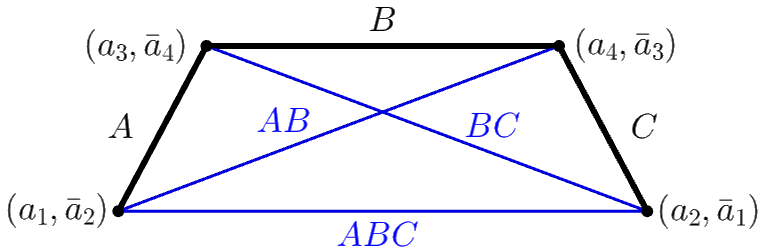}
    \caption{The trapezoid configuration living inside of Figure 8. The top ($B$) and bottom segments ($ABC$) are the ones ${\mathcal R}_{34}$ and ${\mathcal R}_{12}$ of Figure 6. However, the remaining segments ${\mathcal R}_{23}$ and ${\mathcal R}_{14}$ from Figure 6. are replaced by ${\mathcal R}_{d_1}$ and ${\mathcal R}_{d_2}$ representing the diagonals $AB$ and $BC$ of the trapezoid.} 
    \label{trapezoid}
\end{figure}

In order to gain some insight into these issues
and find such an interpretation let us recall that
\begin{equation}
x_i\bullet x_i=-t_i^2+x_i^2=- a_i{\overline{a}}_i
\label{minkover}
\end{equation}
Then for $x_i=0$ one has 
\begin{equation}
(a_i,\overline{a}_i)=(t_i,t_i),\qquad i=1,2,3,4
\label{kordlight}
\end{equation} 
Let us now consider the Minkowski lengths of the blue spacelike regions of Figures 7. and 8.
with the special points having coordinates as displayed in Eq.(\ref{kordlight}).
Clearly the horizontal regions ${\mathcal R}_{ij}$ of Figure 6. have length squares given by 
$(t_i-t_j)^2$.
On the other hand using (\ref{minkover}) both of the spacelike diagonal regions ${\mathcal R}_{d_1}$ and ${\mathcal R}_{d_2}$ have length squares
$(t_4-t_1)(t_2-t_3)>0$ due to the fourth causal constraint of Section 3.2 i.e.
$t_1<t_3<t_4<t_2$.
Moreover, the regularized areas 
for both of the minimal surfaces  $X_{d_1}$ and $X_{d_2}$
(regularized lengths of geodesics)
associated to these diagonals
are
\begin{equation}
{\mathcal A}(d_{j}):={\mathcal A}(X_{d_j})=L\log\left(\frac{(t_4-t_1)(t_2-t_3)}{\delta^2}\right),\qquad j=1,2
\label{diaglength}
\end{equation}
Notice that both ${\mathcal R}_{d_j}$ and $X_{d_j}$ are on the same "boosted" spacelike hyperplanes $H_j$ $j=1,2$ which are totally geodesic submanifolds of the Poincar\'e patch.
For these submanifolds the notions of extremal surfaces and minimal surfaces showing up in the covariant version of the holographic entanglement entropy proposal coincide\cite{HRT}. 

Now using Eq.(\ref{diaglength}) one can notice that just like in Eq.(\ref{justlike_area}) the combination
\begin{equation}
A_{\diamond}=L\left({{\mathcal A}(d_1)+{\mathcal A}(d_2)-\mathcal A}(12)-
{\mathcal A}(34)\right)
\label{justlike2}
\end{equation}
also gives the area of the world sheet of our string segment.
A comparison of the boundary causal diamonds of Figures 7. and 8. (answering the two different representations for the area $A_{\diamond}$ given by Eqs. (\ref{justlike_area})
and (\ref{justlike2})) reveals that
both of them has the same intersection and union followed by causal completion. 

However, the intersecting causal diamonds of Figure 7. are also featuring a trapezoid showing up in \cite{CasiniHuerta} and in Figure 10. in Ref.\cite{He} used for checking the strong subadditivity for the covariant 
holographic entanglement entropy proposal\cite{HRT}.
In the general case this version of the proposal was proved by Wall\cite{Wall} under the assumption of the null curvature condition which in our case is satisfied. 
The advantage of the occurrence of the trapezoid configuration is that in this case the three intervals $A,B,C$ showing up in the SSA
are adjacent, hence their unions $AB$,$BC$ and $ABC$ are also sensible. See Figure 9. Since by \cite{HRT}
the entropy of such an interval
is given by the shortest spacelike geodesic in the three dimensional bulk connecting its endpoints subject to the homology condition
the SSA interpretation of the combination on the right hand side of (\ref{heureka}) is legitimate. Notice also that
in the notation of Figure 8. the quantity of Eq.(\ref{heureka}) is the conditional mutual information  
\begin{equation}
I(A,C\vert B)= S(AB)+S(AC)-S(ABC)-S(B)
\label{condmutinf}
\end{equation}

Then we have our final result
\begin{equation}
    I(A,C\vert B)=\frac{A_{\diamond}}{4GL}
    \label{heureka2}
\end{equation}
i.e. the area formula for the world sheet of a string segment can indeed be  holographically related to strong subadditivity for certain boundary subsystems showing up as components of a trapezoid configuration of Figure 8.

\subsection{The area of the world sheet of a general string segment}

In the following we would like to generalize Eq.(\ref{heureka2}) for general string segments.
For this we have to consider further the extremal surfaces (spacelike geodesics) homologous to boosted time slices. 
In order to do this we rewrite Eq.(\ref{eq:min_line}), bearing in mind equation (\ref{er2}),
in the following form

\begin{equation}
    z^2+\frac{4r^2}{(\Delta_t)^2}(x-x_0)^2=r^2,\qquad t=\frac{\Delta_x}{\Delta_t}(x-x_0)+t_0
\end{equation}
Hence the parametric equations of the extremal surfaces (half ellipses) are
\begin{equation}
\left.\begin{aligned}
    z&=r\cos{\Theta}\\
    x&=\frac{1}{2}\Delta_t\sin{\Theta}+x_0\\
    t&=\frac{1}{2}\Delta_x\sin{\Theta}+t_0
\end{aligned}\right\}    
\end{equation}
 By calculating the induced metric and integrating it with respect to $z$ between the cutoff $\delta$ and $r$ the area of the extremal surface (the regularized length of the geodesic) is given by (see also Eq.(\ref{moregeneral}) for a more general calculation)
\begin{equation}
    {\mathcal A}(ij)=2L\log\frac{2r_{ij}}{\delta}=L\log\frac{L^2}{\delta^2}\left\{\left(\frac{p_i^0}{p_i^-}-\frac{p_j^0}{p_j^-}\right)^2-\left(\frac{p_i^2}{p_i^-}-\frac{p_j^2}{p_j^-}\right)^2\right\}
\end{equation}
This gives back our previous (\ref{jobb}) formula in the  $x=0$ case.

\begin{comment}
Introducing:
\begin{equation}
    \frac{a}{L}=\frac{\lambda^1}{2\lambda^2}+\frac{\tilde{\lambda^2}}{2\tilde{\lambda^1}},\qquad \frac{b}{L}=\frac{\lambda^1}{2\lambda^2}-\frac{\tilde{\lambda^2}}{2\tilde{\lambda^1}},\qquad
    \frac{\alpha}{L}=\frac{\chi^1}{2\chi^2}+\frac{\tilde{\chi^2}}{2\tilde{\chi^1}},\qquad
    \frac{\beta}{L}=\frac{\chi^1}{2\chi^2}-\frac{\tilde{\chi^2}}{2\tilde{\chi^1}},\qquad
\end{equation}
\end{comment}

One can rewrite this result by using the spinor formalism reviewed in Appendix A. Then instead of the four vector $p^{\mu}$ we have the $2\times 2$ matrix
\begin{equation}
    {\mathcal P}=
    \begin{pmatrix}
    p^1+p^{-1}&p^0+p^{2}\\
    p^0-p^{2}&p^1-p^{-1}
    \end{pmatrix}=\begin{pmatrix}p^+&p\\\bar{p}&{p}^-\end{pmatrix}
\label{Pp}
\end{equation}.
Since $\text{det}(P)=-p^2=0$ we can write $P$ in the following form
\begin{equation}
P=L\lambda^T\otimes\tilde{\lambda}^T=L\begin{pmatrix}\lambda^1\tilde{\lambda}^1
&\lambda^1\tilde{\lambda}^2\\
\lambda^2\tilde{\lambda}^1&\lambda^2
\tilde{\lambda}^2\end{pmatrix}=\lambda^2\tilde{\lambda}^2\begin{pmatrix}\overline{a^\lambda}a/L&a\\\overline{a}&L\end{pmatrix}
\label{dek}
\end{equation}
where $\lambda$ is the row vector
$(\lambda^1,\lambda^2)$.
We call $\lambda^a$ as a left-moving, $\tilde{\lambda}^{\dot{a}}$ as a right-moving spinor. Then we have
\begin{equation}
\frac{a}{L}=\frac{\lambda^1}{\lambda^2}=\frac{p}{p^-},\qquad \frac{\overline{a}}{L}=\frac{\tilde{\lambda^1}}{\tilde{\lambda^2}}=\frac{\overline{p}}{p^-}
\label{celest}
\end{equation}
Using the spinor notation the radius $r_{ij}$ in terms of spinor variables of the null vectors $p_i$ and $p_j$ is
\begin{equation}
    r_{ij}^2=\frac{L^2}{4}\left(\frac{\lambda^1_i}{\lambda^2_i}-\frac{\lambda^1_j}{\lambda^2_j}\right)\left(\frac{\tilde{\lambda}^1_i}{\tilde{\lambda}^2_i}-\frac{\tilde{\lambda}^1_j}{\tilde{\lambda}^2_j}\right)
\end{equation}
Therefore the area of the surface defined by the null vectors $p_i$ and $p_j$ is
\begin{equation}
\begin{aligned}
    {\mathcal A}(ij)&=L\log\frac{L}{\delta}\left(\frac{\lambda_i^1}{\lambda_i^2}-\frac{\lambda_j^1}{\lambda_j^2}\right)+L\log\frac{L}{\delta}\left(\frac{\tilde{\lambda}_i^1}{\tilde{\lambda}_i^2}-\frac{\tilde{\lambda}_j^1}{\tilde{\lambda}_j^2}\right)=\\
    &=L\log\frac{1}{\delta}|a_i-a_j|+L\log\frac{1}{\delta}|\overline{a}_i-\overline{a}_j|
\end{aligned}
\end{equation}
Which gives back our previous result for the case of $x=\text{const}$. Notice that the left and right moving parts separate, namely one can write ${\mathcal A}(ij)={\mathcal A}^L(ij)+{\mathcal A}^R(ij)$. 

It is important to note that the $SO(2,2)$ symmetry of the $AdS_3$ space splits up to an $SL(2)_L\times SL(2)_R$ symmetry in the spinor formalism where $SL_L(2)$ acts on $\lambda$ and the other acts on $\tilde{\lambda}$ respectively. An $SL(2)$ transformation of that kind can be written in the form
\begin{equation}
    \begin{pmatrix}
    \lambda'^1\\
    \lambda'^2
    \end{pmatrix}=
    \begin{pmatrix}
    \alpha&\beta\\
    \gamma&\delta
    \end{pmatrix}
    \begin{pmatrix}
    \lambda^1\\
    \lambda^2
    \end{pmatrix},\qquad \alpha\delta-\beta\gamma=1
\end{equation}
Hence $a$ transforms as
\begin{equation}
    a'=L\frac{\frac{\alpha}{L} a+\beta}{\frac{\gamma}{L} a+\delta}
\end{equation}
which is a Möbius transformation. The same holds for $\overline{a}_i$. Therefore the $SO(2,2)$ symmetries generate Möbius transformations of the coordinates $a$ and $\overline{a}$.  
In Section 3.3. we have also seen that these coordinates are arising as the projections of the edges of the world sheet of the string segment via null geodesics, i.e. light rays.
These observations make it possible to elucidate the holographic meaning of ${\mathcal A}_{\diamond}$ in the most general context, the task we turn to in the next section.

\subsection{General string segments and entropies}
We have seen that the area of a string segment defined by the null vectors $p_1,p_2,p_3,p_4$ in general is given by
\begin{equation}
A_{\diamond}=L^2 \log\left[\frac{(p_1-p_4)^2}{(p_1+p_2)^2}\right]^2=L^2 \log\frac{(p_1-p_4)^2(p_2-p_3)^2}{(p_1+p_2)^2(p_3+p_4)^2}
\end{equation}
which is $SO(2,2)$ invariant therefore it holds for any arbitrary patch arrangement. 

In order to gain some additional insight into the structure of this formula one can rewrite it using  the spinor formalism of the previous section. Let us consider again the (\ref{Pp}) matrix $\mathcal P$ associated to a null vector $p^{\mu}$ and its (\ref{dek}) decomposition.
Let us moreover define the following inner products
\begin{equation}
\langle\lambda,\chi\rangle=\lambda^T\epsilon \chi,\qquad
[\tilde{\lambda},\tilde{\chi}]=\tilde{\lambda}^T\epsilon\tilde{\chi}
\end{equation}
The patch area can then be written as
\begin{equation}
A_{\diamond}=L^2\log\left(\frac{\langle\lambda_1,\lambda_4\rangle[\tilde{\lambda}_1,\tilde{\lambda}_4]}{\langle\lambda_1,\lambda_2\rangle[\tilde{\lambda}_1,\tilde{\lambda}_2]}\right)^2
\end{equation}
Using the constraint $p_1+p_2=p_3+p_4$ and that $p_i^2=0$ the patch area can be written in the form
\begin{equation}
A_{\diamond}=L^2\log\left|\frac{\left\langle\lambda_{1}, \lambda_{4}\right\rangle\left\langle\lambda_{2}, \lambda_{3}\right\rangle}{\left\langle\lambda_{1}, \lambda_{2}\right\rangle\left\langle\lambda_{3}, \lambda_{4}\right\rangle}\right|+
L^2\log\left|\frac{[\tilde{\lambda}_{1}, \tilde{\lambda}_{4}][\tilde{\lambda}_{2}, \tilde{\lambda}_{3}]}{[\tilde{\lambda}_{1}, \tilde{\lambda}_{2}][\tilde{\lambda}_{3}, \tilde{\lambda}_{4}]}\right|
\end{equation}
Where the lower index indicates the label of the given null vector. Using the definition of the spinor inner products and dividing the numerator and the denominator by $\lambda_1^2\lambda_2^2\lambda_3^2\lambda_4^2$ and $\tilde{\lambda}_1^1\tilde{\lambda}_2^1\tilde{\lambda}_3^1\tilde{\lambda}_4^1$ in the first and the second term respectively the patch area is
\begin{equation}
    A_{\diamond}=L^2\log\left|\frac{(a_1-a_4)(a_2-a_3)}{(a_1-a_2)(a_3-a_4)}\right|+L^2\log\left|\frac{(\tilde{a}_1-\tilde{a}_4)(\tilde{a}_2-\tilde{a}_3)}{(\tilde{a}_1-\tilde{a}_2)(\tilde{a}_3-\tilde{a}_4)}\right|
\label{realityentropy}
\end{equation}
Where
as usual Eq.(\ref{celest}) holds.
Notice that the left and right terms again separate.

Now it is easy to see that in the general $AdS_3$ case the string patch area $A_{\diamond}$ is connected to the areas of the extremal surfaces via
\begin{equation}
    A_{\diamond}=L({\mathcal A}(14)+{\mathcal A}(23)-{\mathcal A}(34)-{\mathcal A}(12))
\end{equation}
As in the previously discussed special case this result can be recast in a form elucidating the connection between string segments and CFT entanglement

\begin{equation}
    \frac{A_{\diamond}}{4GL}=S(14)+S(23)-S(34)-S(12)
\label{simplenice}
\end{equation}
Therefore the duality between the extremal surfaces, string segments and boundary subsystems holds in general. 

We also remark
that due to the conditions
\begin{align}
    p_1\cdot V_2=p_1\cdot V_1=0&&p_2\cdot V_3=p_2\cdot V_2=0\\
    p_3\cdot V_3=p_3\cdot V_4=0&&p_4\cdot V_4=p_4\cdot V_1=0
\end{align}
the tips of the world sheet of the segment lie on the extremal surfaces in the general case as well.
Moreover,
thanks to the separation of left and right components our dualities hold between the two pieces separately.

In order to obtain another justification for our formula between areas of world sheets and combinations of entanglement entropies for general configurations we can also proceed as follows.
Consider a spacelike CFT interval which is giving rise to the causal diamond whose past and future tips are $(t_i,x_i)$ and $(t_j,x_j)$. Define $(a_i,\overline{a}_i)=(t_i+x_i,t_i-x_i)$ and $(a_j,\overline{a}_j)=(t_j+t_j,x_j-x_j)$. The boundary causal diamond is bordered by two null cones defined by the equations $p_i\cdot X= p_j\cdot X=0$. Therefore the coordinates $x_i,t_i,x_j,t_j$ can be expressed as
    $t_i=\frac{p_i^0}{p_i^-}, x_i=\frac{p_i^2}{p_i^-}$
First if we assume that $x_i=x_j=0$ then 
$a_i=\overline{a}_i=t_i$ and $a_j=\overline{a}_j=t_j$ and 
the entanglement entropy of the CFT subsystem is 
\begin{equation}
S(ij)=\frac{c}{6}\log \frac{|a_i-a_j|}{\delta}+\frac{c}{6}\log \frac{|\overline{a}_i-\overline{a}_j|}{\delta}=\frac{c}{3}\log \frac{|a_i-a_j|}{\delta}=\frac{c}{3}\log \frac{|t_i-t_j|}{\delta}
\end{equation}

This formula transforms\cite{Myers} under the maps $z\to f(z)$, $\overline{z}\to \overline{f}(\overline{z})$ as
\begin{equation}
S(ij)=\frac{c}{12} \log \frac{(f(a_i)-f(a_j))^{2}}{\delta^{2} f^{\prime}(a_i) f^{\prime}(a_j)}+\frac{c}{12} \log \frac{(\overline{f}(\overline{a_i})-\overline{f}(\overline{a_j}))^{2}}{\delta^{2} \overline{f}^{\prime}(\overline{a_i}) \overline{f}^{\prime}(\overline{a_j})}
\label{schwarz}
\end{equation}
$SO(2,2)$ transformations of the $AdS_3$ space generate global conformal transformations of the form $a_i\to\frac{aa_i+b}{ca_i+d}$, $ad-bc=1$ and $\overline{a_i}\to\frac{\overline{a}\overline{a_i}+\overline{b}}{\overline{c}\overline{a_i}+\overline{d}}$, $\overline{a}\overline{d}-\overline{b}\overline{c}=1$ and similarly for $a_j,\overline{a}_j$ in the asymptotic limit. It can be easily shown, that $S$ is invariant under these maps. Then any general causal diamond can be transformed into a special case $x_i=x_j=0$ by a pair of conformal transformations. Therefore the entanglement entropy of a generally alligned CFT subsystem is given by
\begin{equation}
    S(ij)=\frac{c}{6}\log \frac{|a_i-a_j|}{\delta}+\frac{c}{6}\log \frac{|\overline{a}_i-\overline{a}_j|}{\delta}
\end{equation}
Where
Eq.(\ref{celest}) holds
and the null vectors $p_i$ and $p_j$ define the usual null cones in the Poincaré patch and their projections to the boundary.
Hence the (\ref{simplenice}) formula between the area of a string segment stretched by the null vectors $p_1,p_2,p_3,p_4$ and the entanglement entropies of the corresponding subsystems holds for any string segment with arbitrary normal vector. This gives us the opportunity to build up any string solution from small segments whose geometry is captured by the entanglement of timelike separated CFT subsystems.

\begin{comment}
\subsection{General arrangement}

From the special case with normal vector $N=(0,0,0,1)$ we can generate any string segment with arbitrary normal vector by a global $SO(2,2)$ transformation. Consider a global transformation $\Lambda\in SO(2,2)$. The defining null vectors  $p_1,p_2,p_3,p_4$  of the string segment transform as $p_i'=\Lambda p_i$. Similarly the normal vector $N$ transforms as $N'=\Lambda N$. Which is now arbitrary.

Since $\Lambda$ is a global transformation the points $X$ of the string segment stretched by the vectors $p_i'$ are connected to the original ones by this global transformation. The Nambu-Goto action
\begin{equation}
A=\int d\sigma d\tau \sqrt{-g}
\end{equation}

is constructed in such a way that it is invariant under the symmetry transformations of the underlying geometry. Therefore the area $A$ of the string segment is invariant under $\Lambda$.

The same argument holds for the minimal surfaces defined by the intersections of any pair of $p_i\cdot X=0$ and $p_j\cdot X=0$ cones. Their area (length):
\begin{equation}
\ell=\int dS\sqrt{h}
\end{equation}
Is invariant under $\Lambda$, therefore the area $\ell'$  of the surface determined by $p_i'\cdot X=0$ and $p_j'\cdot X=0$ is equal to $\ell$.

\end{comment}

\subsection{The physical meaning of trapezoids related to string world sheets}

Clearly our general result displayed in Eq.(\ref{simplenice}) can also be given a quantum information theoretic meaning. This can be seen via repeating the argument we made in Section 3.5. This time one has to use a generalized trapezoid configuration with the corresponding set of points $(a_i,\bar{a}_i)$ lying on a hyperboloid  rather than on a line. See the rightmost part of Figure 5. 
The net result then will be of the same form as the one of Eq.(\ref{heureka2}). Namely, we will have in this most general case a nice correspondence between the area of the string world sheet segment and the conditional mutual information calculated for the subsystems featuring a distorted trapezoid i.e. a distortion of the one of Figure 9.

Providing the basic mathematical objects for relating world sheets of the bulk to 
quantum information theoretic quantities of the boundary,
it is worth exploring the physical meaning of these trapezoids as objects connected to segmented strings.
In order to shed some light on the meaning of trapezoids we use the helicity formalism briefly summarized in Appendix A.
Since the points of the trapezoids are lying on spacelike lines or hyperbolas which are dual to the timelike ones related to systems of inertial and noninertial observers, we first turn to a Theorem encapsulating a generalization of the reality conditions of Eqs.(\ref{cross1})-(\ref{cross2}).

\subsubsection{Spacelike duals of timelike lines and hyperbolas}

Let us consider the following Theorem\footnote{This result is a reformulation   
Proposition 3.4. of Ref.\cite{Brewer}}
\begin{theorem}
For a set of distinct points  
$(a_i,\bar{a}_i), i=1,3,4$ and $(a,\bar{a})$ in ${\mathbb R}^{1,1}$ satisfying the 
(\ref{unusual})
conditions with $(a_2,\bar{a}_2)\equiv (a,\bar{a})$ let us define
\begin{equation}
\zeta:=\frac{(a_3-a_1)(a-a_4)}{(a-a_1)(a_4-a_3)}>0,\qquad    
\bar{\zeta}:=\frac{(\bar{a}_3-\bar{a}_1)(\bar{a}-\bar{a}_4)}{(\bar{a}-\bar{a}_1)(\bar{a}_4-\bar{a}_3)}>0
\label{chichi}
\end{equation}
then $\zeta =\bar{\zeta}$ if and only if
\begin{equation}
N\cdot Z=0.
\end{equation}
Here  $Z,N\in{\mathbb R}^{2,2}$ are null and spacelike vectors with their components written in the helicity formalism as
\begin{equation}
{\mathcal Z}:=z^T\otimes\bar{z}=
\begin{pmatrix}a\bar{a}&La\\L\bar{a}&L^2\end{pmatrix},\qquad {\mathcal N}:=\psi_1^T\otimes \overline{\psi}_4-\psi_4^T\otimes\overline{\psi}_1=\begin{pmatrix}N^+&N\\\bar{N}&N^-\end{pmatrix}
\end{equation}
\end{theorem}
Notice that here ${\mathcal N}$ is the matrix analogue of the normal vector $N$ of the string world sheet segment familiar from Eq.(\ref{enegy1}). The explicit form for the components of $\psi_1$ and $\psi_4$ can be written as
\begin{equation}
\psi_1=\frac{1}{\sqrt{{\mathcal D}}}(a_4-a_3)z_1,\qquad
\psi_4=\frac{1}{\sqrt{{\mathcal D}}}(a_3-a_1)z_4,
\label{vesszo12}
\end{equation}
\begin{equation}
\overline{\psi}_1=    
\frac{1}{\sqrt{\overline{\mathcal D}}}(\bar{a}_4-\bar{a}_3)\bar{z}_1,\qquad
\overline{\psi}_4= 
\frac{1}{\sqrt{\overline{\mathcal D}}}(\bar{a}_3-\bar{a}_1)\bar{z}_4
\label{vesszo22}
\end{equation}
where we have the row vectors
\begin{equation}
z_i=(a_i,L),\qquad
\bar{z}_i=(\bar{a}_i,L),\qquad i=1,4
\end{equation}
and
\begin{equation}
{\mathcal D}=(a_4-a_1)(a_3-a_1)(a_4-a_3)>0,\qquad \bar{\mathcal D}=(\overline{a}_4-\overline{a}_1)(\overline{a}_3-\overline{a}_1)(\overline{a}_4-\overline{a}_3)>0
\label{dek22}
\end{equation}

The Theorem states that a sufficient and necessary condition for the "reality" condition ${\zeta}=\bar{\zeta}$ for the ${\mathbb R}^{1,1}$ cross ratios to hold is the orthogonality condition $N\cdot Z=0$ for the null and spacelike vectors $Z,N\in {\mathbb R}^{2,2}$.
But our condition $N\cdot Z=0$ is precisely the familiar one of Eq.(\ref{hypline}) and (\ref{hyphyphurra}) which is defining our lines and hyperbolas in the boundary. The coordinates of the three points (this time $(a_i,\bar{a}_i)$ with $i=1,3,4$) characterize a line or a hyperbola. This data is encapsulated in the vector ${\mathcal N}$ in the helicity representation. One can regard the fourth point $(a,\bar{a})$ as a one moving on the particular line or hyperbola determined by ${\mathcal N}$.
The Theorem shows that a sufficient and necessary condition of these four points to be related to events on world lines of inertial observers or ones moving with constant acceleration is the reality condition for cross ratios. 
In Section 3.3. we have already given half of the proof of this theorem. For a full short proof see Appendix A. 
It is also important to realize that our reality condition is related to momentum conservation for segmented strings, see Eq.(\ref{konzerv}).

In order to obtain insight on trapezoids we have to consider an important generalization to Theorem 1.
Notice that Theorem 1 can easily be generalized even for the case when one is replacing the (\ref{unusual}) causality constraint for a set of four points $(b_i,\bar{b}_i)$ satisfying
\begin{equation}
b_1<b_3<b_4<b_2,\qquad \bar{b}_1>\bar{b}_3>\bar{b}_4>\bar{b}_2
\label{causality2}
\end{equation}
i.e. when the left moving coordinates have a reversed causal ordering.
A particularly important realization of this physical situation is given by the choice
\begin{equation}
(b_1,\bar{b}_1)=(a_1,\bar{a}_2),\quad
(b_2,\bar{b}_2)=(a_2,\bar{a}_1),\quad
(b_3,\bar{b}_3)=(a_3,\bar{a}_4),\quad
(b_4,\bar{b}_4)=(a_4,\bar{a}_3)
\label{bes}
\end{equation}
corresponding to the four points of our trapezoid labeled from left to to right (see Figure 9.).

Now it is easy to see that reality condition $\zeta=\bar{\zeta}$ is still satisfied. Then  from (\ref{egyeske})-(\ref{utolsocska}) one can see that
under the replacement
 $(a_i,\bar{a}_i)\mapsto (b_i,\bar{b}_i)$
and $N\mapsto M$ (with the sign under the square root changed) our  Theorem still holds.
It is crucial now to realize that in this new case one has $M\cdot M<0$, i.e. now the $M\in {\mathbb R}^{2,2}$ vector is timelike.
Notably one can also choose
$M\in AdS_3$ i.e. $M\cdot M=-L^2$.
Then in this dual situation the spacelike lines and hyperbolas are having spacelike tangent vectors. 
Hence the events represented by points of our trapezoids are characterized by the geometric property of either being localized on spacelike lines or spacelike hyperbolas.
This dual situation clearly describes  the spacelike analogue of the previous timelike case.

To verify this claim with an explicit calculation let us consider
\begin{equation}
\psi_1^{\prime}=\frac{1}{\sqrt{{\mathcal D}^{\prime}}}(b_4-b_3)z_1^{\prime},\qquad
\psi_2^{\prime}=\frac{1}{\sqrt{{\mathcal D}^{\prime}}}(b_3-b_1)z_4^{\prime},
\label{vesszo1}
\end{equation}
\begin{equation}
\overline{\psi}_1^{\prime}=    
\frac{1}{\sqrt{-\overline{\mathcal D}^{\prime}}}(\bar{b}_4-\bar{b}_3)\bar{z}_1^{\prime},\qquad
\overline{\psi}_2^{\prime}= 
\frac{1}{\sqrt{-\overline{\mathcal D}^{\prime}}}(\bar{b}_3-\bar{b}_1)\bar{z}_4^{\prime}
\label{vesszo2}
\end{equation}
where
\begin{equation}
z_i^{\prime}=(b_i,L),\qquad
\bar{z}_i^{\prime}=(\bar{b}_i,L),\qquad i=1,4
\end{equation}
and
\begin{equation}
{\mathcal D}^{\prime}=(b_4-b_1)(b_3-b_1)(b_4-b_3)>0,\qquad \bar{\mathcal D}^{\prime}=(\overline{b}_4-\overline{b}_1)(\overline{b}_3-\overline{b}_1)(\overline{b}_4-\overline{b}_3)<0
\label{dek2}
\end{equation}
Clearly by virtue of Eq.(\ref{bes}) one can express these in terms of the original coordinates, and then we have
$\psi_1^{\prime}=\psi_1$ and 
$\psi_2^{\prime}=\psi_2$, however 
\begin{equation}
\overline{\psi}_1^{\prime}=    
\frac{1}{\
\sqrt{-\overline{\mathcal D}^{\prime}}}    
(\bar{a}_3-\bar{a}_4)\bar{z}_2,\qquad
\overline{\psi}_2^{\prime}= 
\frac{1}{\sqrt{-\overline{\mathcal D}^{\prime}}}(\bar{a}_4-\bar{a}_2)\bar{z}_3
\label{vesszo3}
\end{equation}
Now the image of the $N\mapsto M$ 
($\mathcal{N}\mapsto \mathcal{M}$) transformation can be described as follows
\begin{equation}
\mathcal{M}=\psi_1^T\otimes\bar{\psi}_2^{\prime}-
\psi_2^T\otimes\bar{\psi}_1^{\prime}
\end{equation}
We would like to express this quantity in terms of some familiar quantities related to segmented strings.
One can show that
\begin{equation}
\mathcal{M}=\sqrt{\frac{1+\zeta^2}{\zeta^2}}\frac{1}{2}\left(\mathcal{V}_3-\mathcal{V}_1\right)    
\label{trapmeaning}
\end{equation}
Since the overall normalization is meaningless it is the bulk vector ${V}_3-{V}_1$ which determines the boundary hyperbola on which the events corresponding to the points of our trapezoid are located.
As we already know these vectors are the ones showing up in the (\ref{veujegy})-(\ref{veujnegy}) list of points comprising the vertices of a world sheet in ${\rm AdS}_3$.
The vector defining the trapezoid configuration is the one pointing from the initial point $V_1$ to the opposite one $V_3$ along the diagonal of the sheet. Notice that though the normalized vector
$M$ is a vector lying in ${\rm AdS}_3$, however it is easy to see that it is not lying on the world sheet itself.

To put things into a physical context it is worth considering the simplest situation of illustrative value. It is the case
where the four points representing the future and past tips of the causal diamonds are located on the $t$-axis symmetrically to the $x$-axis. (See the left hand side of Figure 5. modified appropriately.)
\begin{equation}
t_1=-t_2,\qquad t_3=-t_4
\label{nagyonspeci}
\end{equation}
Then we have 
\begin{equation}
{\mathcal M}=
   \frac{t_2-t_4}{t_2+t_4}
   \begin{pmatrix}
   0&-L\\-L&0\end{pmatrix}
\end{equation}
This means that the $AdS_3$ vector $M^a$ determines a line in the boundary, namely the one with equation $a+\bar{a}=0$ i.e. $t=0$, that is the $x$-axis.
Hence the four points in this case are localized on the $x$-axis.

\subsubsection{Flow lines for trapezoid points and the modular flow}

The upshot of these considerations is the following.
We have a conformal field theory (CFT) on ${\mathbb R}^{1,1}$ and we consider the vacuum state of this CFT.
Then we choose a causal diamond $\bf D$ with the future and past tips of it given by the light cone coordinates $(a_2,\bar{a}_2)$ and 
$(a_1,\bar{a}_1)$.
In a special case (which we can always obtain by performing a conformal transformation)  for an $R\in{\mathbb R}^+$ one can have $(a_2,\bar{a}_2)=(R,R)$ and
$(a_1,\bar{a}_1)=(-R,-R)$ i.e. in $(t,x)$ coordinates 
we have $x_2^{\mu}=(R,0)$ for the future tip and
$x_1^{\mu}=(-R,0)$ for the past one. This is the Universe of an inertial observer with a finite life time $2R$. This lifetime can be regarded as the proper time measured by the observer.
$\bf D$ is the region of events with which the observer can exchange signals in his/her lifetime via sending signals and receiving response\cite{Rovelli}.

Then one can consider a Cauchy slice of 
$\bf D$ , e.g. one can choose the following part of the $x$ axis: ${\mathcal R}=[-R,R]\subset {\mathbb R}$. 
If we integrate out the degrees of freedom in the complement $\overline{\mathcal R}$ we are left with the reduced density matrix $\rho$ describing the remaining degrees of freedom in ${\mathcal R}$. The entanglement entropy across the two endpoints of the interval, which is an $S^0={\mathbb Z}_2$, is our von Neumann entropy $S({\mathcal R})$.
The reduced density matrix can be expressed as $\rho=e^{-H}$ where $H$ is the modular Hamiltonian known from axiomatic quantum field theory\cite{Haag}.
The unitary operator $U(s)=e^{-sH}$ generates a symmetry (a flow called the modular flow) of the Universe based on $\bf D$. This means that the symmetry transforms the algebra $\mathcal A$ of observables of $\bf D$ into itself.

$H$ is generically a nonlocal operator, but it is known that for the diamond $\bf D$ it is a local one\cite{Hislop,Towards}. This can be proved by starting from the Rindler wedge ${\bf W}$ where the modular Hamiltonian is known to be a local operator which is just a representation of the usual Lorentz boost\cite{Bisognano}. Then one can employ the well-known map\cite{Hislop,Towards} from ${\bf W}$ to ${\bf D}$ and obtain the local modular Hamiltonian for ${\bf D}$.
What we need in the following is merely the explicit form\cite{Towards} of the modular flow on $\bf D$ starting from a point $(a,\bar{a})=(a(0),\bar{a}(0))$. It is given by the formula
\begin{equation}
a(s)=R\frac{\left({R}+a\right)-e^{-2\pi s}\left({R}-a\right)}
{\left({R}+a\right)+e^{-2\pi s}\left({R}-a\right)},\qquad
\bar{a}(s)=-{R}\frac{\left({R}-\bar{a}\right)-e^{2\pi s}\left({R}+\bar{a}\right)}
{\left({R}-\bar{a}\right)+e^{2\pi s}\left({R}+\bar{a}\right)}
\end{equation}
Indeed, we have $(a(-\infty),\bar{a}(-\infty))=(a_1,\bar{a_1})$
and 
$(a(\infty),\bar{a}(\infty))=(a_2,\bar{a_2})$.
Alternatively one can write ($a=t+x,\bar{a}=t-x$)
\begin{equation}
t(s)=R\frac{(R^2+t^2-x^2)\sinh(2\pi s)+2Rt\cosh (2\pi s)}
{(R^2-t^2+x^2)+(R^2+t^2-x^2)\cosh(2\pi s)+2Rt\sinh(2\pi s)}
\end{equation}
\begin{equation}
x(s)=R\frac{2Rx}
{(R^2-t^2+x^2)+(R^2+t^2-x^2)\cosh(2\pi s)+2Rt\sinh(2\pi s)}
\label{egyeske1}
\end{equation}

Now we wish to show that the flow lines of the modular flow give the accelerated and inertial frames of reference we used in Section  3.3.
There we demonstrated that a connection exist between the world lines of such observers and the world sheets of string segments.

In order to do this we recall that the accelerated frames of reference fitting into the causal diamond $\bf D$ exhibiting hyperbolic motion,  in proper time parametrization, should have the form
\begin{equation}t(\tau)=\frac{1}{g}\sinh(g\tau),\qquad
    x(\tau)=\frac{1}{g}\cosh(g\tau)+x_0
\label{ketteske2}
\end{equation}
where $g$ is the acceleration. $g$ and $\tau$ have to be subjected to some constraints to be specified below.
In order to identify these
constraints we use the boundary conditions that at $\tau=0$ and $s=0$ the two different parametrizations, namely (\ref{egyeske1}) and (\ref{ketteske2}) match\cite{Rovelli}. 
This means that\footnote{The somewhat unusual convention of shifting $x_0$ by $1/g$ was introduced to be also in accord with our previous parametrization of hyperbolas in Eq.(\ref{hyphyphurra}) by $a_0=t_0+x_0$ and $\bar{a}_0=t_0-x_0$.}
\begin{equation}
x(0)=x:=\frac{1}{g}+x_0,\qquad t(0)=t:=t_0=0     
\label{ixnull}
\end{equation}
An accelerated observer in the diamond universe $\bf D$ is having a finite lifetime $2\tau_0$ i.e. $-\tau_0\leq \tau\leq \tau_0$. Since we should have $t(\tau_0)=R$ and $x(\tau_0)=0$ we get using (\ref{ketteske2})
\begin{equation}
gR=\sinh(g\tau_0), \qquad gx_0=-\cosh(g\tau_0)    
\label{confuse}
\end{equation}
Combining this with Eq.(\ref{ixnull}) one gets
\begin{equation}
(gR)^2-(gx)^2=-2(gx),\qquad
(gR)^2+(gx)^2=2(gx_0)(gx)
\label{jajdecuki}
\end{equation}
Since we have $t=0$ in Eq.(\ref{egyeske1})
we get using (\ref{jajdecuki})
\begin{equation}
t(s)=R \frac{\sinh(2\pi s)}{\cosh(2\pi s)-gx_0},\qquad
x(s)=R \frac{gR}{gx_0-\cosh(2\pi s)}
\label{upperhalfcoord}
\end{equation}
hence we see that
\begin{equation}
(x(s)-x_0)^2-t(s)^2=\frac{1}{g^2}   
\end{equation}
This is of the (\ref{hyphyphurra}) form
hence now one can finally make contact with the results of Section 3.3. as
\begin{equation}
\varrho^2 =\frac{1}{g^2},\qquad a_0=-\bar{a}_0=x_0,\qquad (a_1,\bar{a}_1)=(-R,-R),\qquad
(a_2,\bar{a}_2)=(R,R)
\label{niceconnect}
\end{equation}
These are quantities that can be expressed in terms of the components of the normal vector of the string world sheet as displayed in Eqs. (\ref{adatok})  and (\ref{egyeske})-(\ref{utolsocska}).
(In order to fit our data to the diamond universe $\bf D$ it is worth rewriting the components of the normal vector $N$ in terms of $(a_i,\bar{a}_i)$ with $i=1,2,4$ rather than with the ones of Section 3.3., namely $i=1,3,4$.)
In this case we also have the formulae
\begin{equation}
x_0=\frac{R^2-a_4\bar{a}_4}{a_4-\bar{a}_4},\qquad
g=\sqrt{\frac{(a_4-\bar{a}_4)^2}{(R^2-a_4^2)(R^2-{\bar{a}}_4^2)}}
\label{gejee}
\end{equation}
Of course these expressions for $x_0$ and $g$ work for any point on the worldline of the accelerated observable.

In order to get back to trapezoids notice that one can write (\ref{upperhalfcoord}) in the form
\begin{equation}
    t(u,v)=R \frac{\sinh(v )}{\cosh(v)+\cosh(u)},\qquad
x(u,v)=R \frac{\sinh(u)}{\cosh(v)+\cosh(u)}
\label{parapara}
\end{equation}
where $v=2\pi s$ and $u=-g\tau_0$.
We see that the tips of the causal diamonds in the boundary are on the modular curve with $u$ fixed. The parameter $v$ changes along the modular curve which is a timelike hyperbola representing an accelerating observer in ${\bf D}$.
The four tips are related to the four edges of the string world sheet segment in the bulk. These four tips are parametrized as $(a_i,\bar{a}_i)$ with $i=1,2,3,4$ in the boundary.

However, we also have another set of hyperbola.
These are parametrized according to Eq.(\ref{causality2}) i.e. $(b_i,\bar{b}_i)$ with $i=1,2,3,4$.
These four points give rise to the trapezoid configurations where the subsystems needed for establishing a connection with conditional mutual informations live. 
From Eq.(\ref{parapara}) we also see that these dual hyperbolas
are the ones with $v$ fixed. Since $gR=-\sinh u$ the change in the parameter $u$ maps worldlines of observers with different acceleration.
Comparing Figures 7. and 8. one can see that in both of these dual two cases one has two intersecting causal diamonds with intersections and causal completions being the same. However, in the first case the diamonds are timelike separated in the second they are spacelike separated. 

In closing this section one can take this line of reasoning one step further.
For the diamond universe characterized by $R$  one can also write
\begin{equation}
t_i\pm x_i=R\frac{\sinh v_i \pm \sinh u}{\cosh v_i +\cosh u}=R\tanh\left(\frac{v_i\pm u}{2}\right),\qquad i=1,2,3,4
\label{ad1}
\end{equation}
with
$v_1=-\infty$ and $v_2=\infty$ corresponding to the (\ref{niceconnect}) tips of the diamond.
Then it is instructive to introduce
\begin{equation}
v_i:=\tilde{\tau}_i/R,\qquad i=1,2,3,4
\label{confuse2}
\end{equation}
On the other hand one can recall that since the four points are on the same hyperbola and for this hyperbola we have
\begin{equation}
\sinh u=-gR
\label{accelerate11}
\end{equation}
where $g$ is the constant acceleration characterizing the relevant hyperbolic motion.
Using now the reality condition for cross ratios and  Eq.(\ref{realityentropy}) one gets
\begin{equation}
e^{A_{\diamond}/2L^2}=\frac{(a_4-a_1)(a_2-a_3)}{(a_2-a_1)(a_4-a_3)}
\end{equation}
We have for example
\begin{equation}
a_4-a_3=R\left[\tanh\left(\frac{v_4+u}{2}\right)-\tanh\left(\frac{v_3+u}{2}\right)\right]=R    
\frac{\sinh\left(\frac{v_4-v_3}{2}\right)}{\cosh\left(\frac{v_4+u}{2}\right)\cosh\left(\frac{v_3+u}{2}\right)}
\end{equation}
Then a calculation shows that
\begin{equation}
e^{A_{\diamond}/2L^2}-1=\frac{1}{e^{(\tilde{\tau}_4-\tilde{\tau}_3)/R}-1}=\frac{1}{e^{(v_4-v_3)}-1}
\label{jaj1}
\end{equation}
Hence the area segment of the string world sheet is related to the time elapsed 
according to the modular parameter $s_i$ where $v_i=2\pi s_i=\tilde{\tau}_i/R$ with $i=3,4$ labeling the other two points on the timelike hyperbola not corresponding to the tips. \footnote{Do not confuse the proper time $-\tau_0\leq\tau\leq \tau_0$ of Eq.(\ref{confuse}) associated with a particular accelerating observer with the modular time $-\infty\leq\tilde{\tau}\leq \infty$ of Eq.(\ref{confuse2}). They are related to each other by the formula
\begin{equation}
\tanh(g\tau)=
\frac{gR \sinh\left( \frac{\tilde{\tau}}{R}  \right)}{1+\sqrt{1+g^2R^2}\cosh\left( \frac{ \tilde{\tau}}{R}  \right) }
\nonumber
\end{equation}
As is well-known the interplay between $\tau$ and $\tilde{\tau}$ is related to the thermal time hypothesis see Ref.\cite{Rovelli}.}

However, according to the other (spacelike) hyperbola where now $v$ is constant and $u_i$ with ($i=1,2,3,4$) is changing and related to the four points of the trapezoid, an alternative interpretation can also be given. It is encapsulated in the dual expression
\begin{equation}
e^{A_{\diamond}/2L^2}-1=\frac{1}{e^{(u_4-u_3)}-1}
\qquad
\sinh u_i=-g_iR
\label{jaj2}
\end{equation}
Hence in this case the area can be expressed by the accelerations $g_3$ and $g_4$
of the observers going through the points with coordinates $(a_3,\bar{a}_4)$
and $(a_4,\bar{a}_3)$. It can be shown that these points are on a spacelike hyperbola characterized by a new constant $v$  with modular parameter related to the old ones by $s=(s_3+s_4)/2$.

Comparing Eqs.(\ref{jaj1}) and (\ref{jaj2}) one can conclude that going from timelike hyperbolas to spacelike ones (corresponding to trapezoids) amounts to either relating the area for string segments
to a change in modular time $\tilde{\tau}$
or to a change in acceleration $g$.

How these considerations are related to the entanglement structure of the CFT vacuum?
In order to answer this question just notice that according to Eq.(\ref{heureka2}) and the (\ref{BH}) Brown-Henneaux relation one can also express Eq.(\ref{jaj1}) via $I\equiv I(A,C\vert B)$ as
\begin{equation}
e^{\frac{3}{c}I}-1=
\frac{1}{e^{(\tilde{\tau}_4-\tilde{\tau}_3)/R}-1}
\end{equation}
Since our considerations are valid only for the central charge $c$ of the CFT being large one can write
\begin{equation}
I\simeq\frac{c/3}{e^{(\tilde{\tau}_4-\tilde{\tau}_3)/R}-1}=
\frac{c/3}{e^{({u}_4-{u}_3)}-1}
\label{jajdeklassz}
\end{equation}
where according to Eqs.(\ref{jaj1})-(\ref{jaj2})
we can also look at this formula in a dual manner.
It is also known that one can associate a temperature with our causal diamond. This is the temperature an inertial observer of the diamond universe with finite lifetime $2R$ observes. The result is \cite{Rovelli}
\begin{equation}
T=\frac{1}{\pi R} 
\end{equation}
calculated using the thermal time hypothesis. This temperature is twice as big as the one showing up in Eq.(3.19) of Ref.\cite{Towards} after identifying a thermal density matrix as a mixed state of the diamond.

\subsection{Duality between strings and entanglement, Toda equation}
In this section we show how the duality between segmented strings and subsystems for the $CFT_2$ vacuum can be applied to patch together different world sheet segments.
 
First let us observe that there is a gauge degree of freedom in the definition of the extremal  surfaces defined by 
$p_i\cdot X=p_j\cdot X=0$ 
arising from a rescaling degree of freedom for the defining null vectors. Notice that the rescaling $p_1\rightarrow \Lambda_1 p_1$ and $p_4\rightarrow \Lambda_4 p_4$ generates a rescaling $p_1\cdot p_4\rightarrow \Lambda_1\Lambda_4 p_1\cdot p_4$ hence different null vectors $p_2$ and $p_3$ due to the interpolation ansatz. However, the normal vector of the segment given by
\begin{equation}
    N_a=\frac{\epsilon_{a b c d} V_1^{b} p_1^{c} p_4^{d}}{p_1\cdot p_4}
\end{equation}
 stays invariant.
Notice in this respect that in Eq.(\ref{pek}) the relationship between $p_i$ and $Z_i$ is featuring precisely such local rescaling factors.
Indeed, in Section 3.3 a careful fixing of this gauge
degree of freedom was crucial for establishing momentum conservation $p_1+p_2=p_3+p_4$ hence arriving at a unique reconstruction of the stringy data from the boundary one.

\begin{figure}[t]
    \centering
    \includegraphics[width=0.8\textwidth]{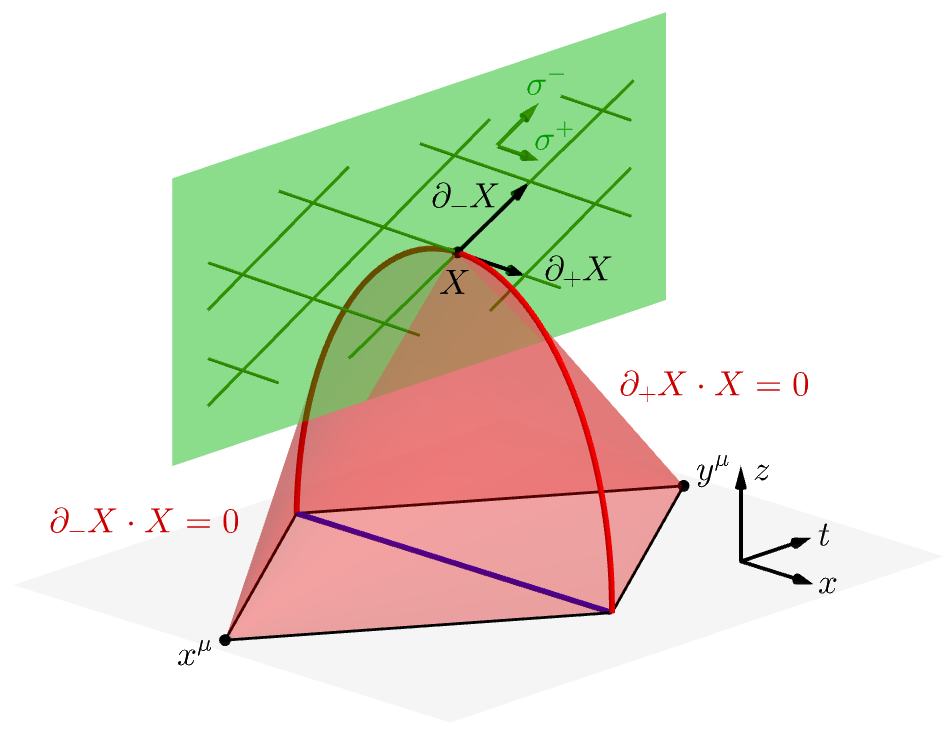}
    \label{fig:lattice}
    \caption{An illustration of the connection between the discretized variables of segmented strings to the ones showing up in the continuous one\cite{Maldacena}. In the notation of Section 3.1 clearly we have $V_1\leftrightarrow X$, $\partial_-X\leftrightarrow p_1$,
    $\partial_+X\leftrightarrow p_4$.
    The equations in red define the cones with future ($y$) and past ($x$) tips. Their intersections give rise to the extremal surface (red line) which is now a spacelike geodesic. This Figure should be compared to Figure 1. 
    }
\end{figure}

It is also important to note that by defining the following quantities \cite{Alday}
\begin{equation}
    e^{2 \alpha}=\frac{1}{2} \partial X \cdot \bar{\partial} X,\qquad \pi=-\frac{1}{2}N \cdot \partial^2 X, \qquad \overline{\pi}=\frac{1}{2} N \cdot \bar{\partial}^2 X
\end{equation}
$\alpha$ satisfies the generalized sinh-Gordon equation:
\begin{equation}
    \partial \bar{\partial} \alpha-e^{2 \alpha}+\pi\bar{\pi} e^{-2 \alpha}=0
\end{equation}
Comparing Eq.(\ref{Ncont}) with (\ref{normalis}) for strings with constant normal vectors $\pi=\bar{\pi}\equiv 0$ which gives the Liouville equation. For a string segment given by the initial data $p_1$ and $p_4$
using equation (\ref{Liouvillhez}) we get
\begin{equation}
    e^{2\alpha}=\frac{1}{2}p_1\cdot p_4=
-L^2\frac{r_{13}r_{24}}{r_{12}r_{34}}
\end{equation}
Which connects the $p_1\rightarrow \Lambda_1 p_1$, $p_4\rightarrow \Lambda_4 p_4$ gauge degree of freedom in the minimal surface theory to the different values of the variable $\alpha$ in the discretized Liouville equation.
Notice that had we chosen
instead of the pair $(p_1,p_4)$ the much simpler pair of lightlike vectors $(Z_1,Z_4)$ with helicity representatives $({\mathcal Z}_1,{\mathcal Z}_2)$ of Eq.(\ref{zeee}) we would have obtained
$Z_1\cdot Z_4=2L^2r_{14}^2$ which is not featuring a cross ratio.
Then Eq.(\ref{shortarea}) would not have yielded the correct cross ratio for the area of the world sheet segment.
Hence the gauge fixing
$Z_1\mapsto -\Lambda_{24,1}Z_1=p_1$ and
$Z_4\mapsto \Lambda_{13,4}Z_4=p_4$ with the (\ref{lambdagauge}) factors is crucial for obtaining a unique lift from boundary data to bulk one reconstructing the world sheet of a string segment.

\begin{figure}[!h]
    \centering
    \includegraphics[width=0.6\textwidth]{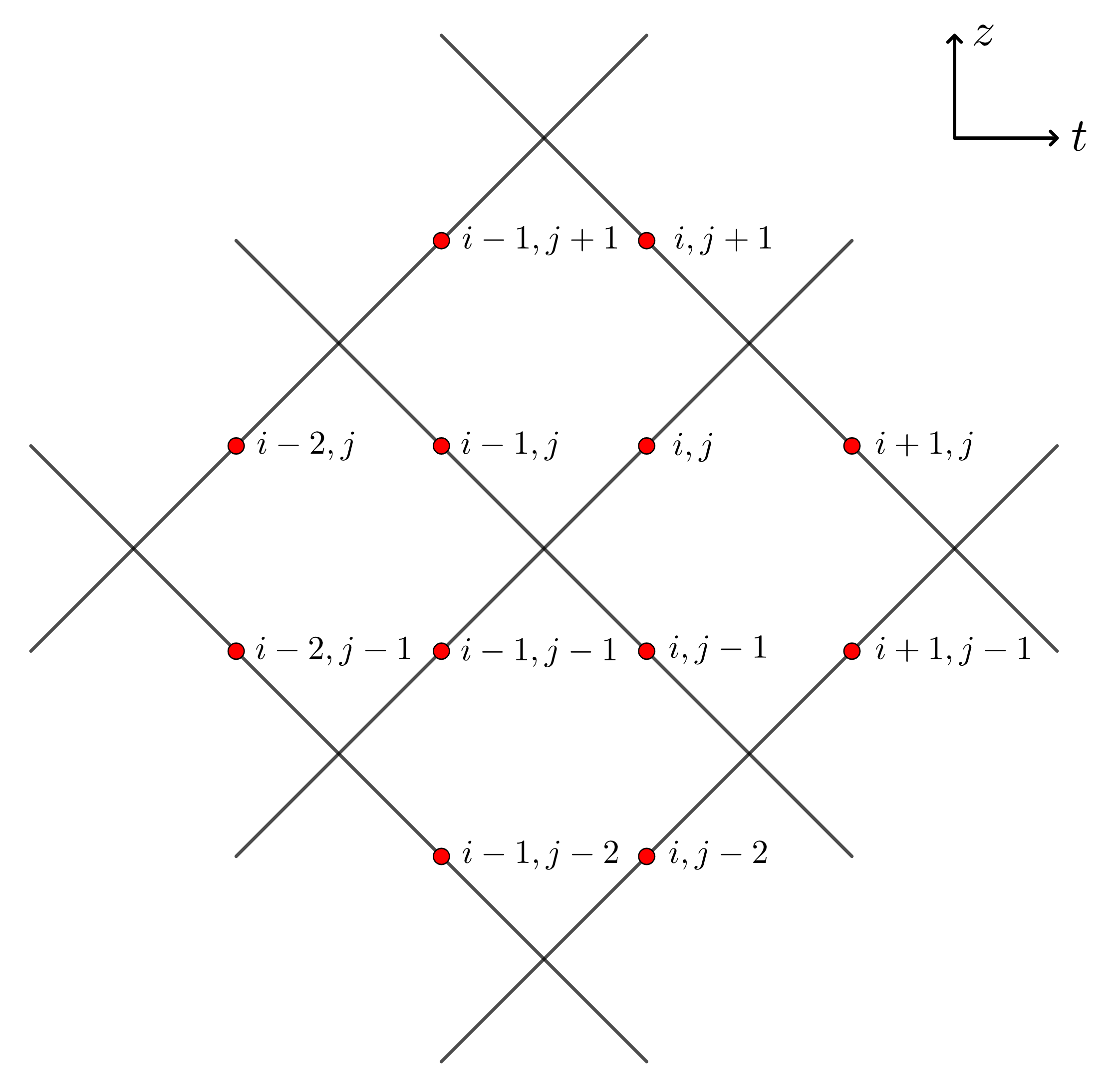}
    \caption{A worldsheet built up from string segments. Each side of a given segment is labeled by its defining lightlike vector $p_{ij},p_{i+1j},p_{i+1,j-1}$ or $p_{ij-1}$.}
\end{figure}

After these observations let us address the problem of patching together the pieces of information provided by different string segments.
Consider a lattice of string segments. The segments are determined by four null vectors $p_{ij}$, $p_{i-1j}$, $p_{i-1j+1}$ and $p_{ij+1}$. The area of a given segment is:
\begin{equation}
    A=L^2\log\left|\frac{(a_{i-1j}-a_{i-1j+1})(a_{ij}-a_{ij+1})}{(a_{i-1j}-a_{ij})(a_{ij+1}-a_{i-1j+1})}\right|+L^2\log\left|\frac{(\tilde{a}_{i-1j}-\tilde{a}_{i-1j+1})(\tilde{a}_{ij}-\tilde{a}_{ij+1})}{(\tilde{a}_{i-1j}-\tilde{a}_{ij})(\tilde{a}_{ij+1}-\tilde{a}_{i-1j+1})}\right|
\end{equation}
Where $a_{ij}=L\frac{p_{ij}^0+p_{ij}^2}{p_{ij}^1-p_{ij}^{-1}}$ and $\tilde{a}_{ij}=L\frac{p_{ij}^0-p_{ij}^2}{p_{ij}^1-p_{ij}^{-1}}$. The area can be split into two parts one with and one without a tilde on the variables. From now on we only considering only the left moving part, however the following argument holds for the other one as well. The total area of the string can be written in the form:
\begin{equation}
    A_{\text{tot}}=\sum_{i, j} \log \left|\frac{a_{i, j}-a_{i, j+1}}{a_{i, j}-a_{i+1, j}}\right|+\sum_{i, j} \log \left|\frac{\tilde{a}_{i, j}-\tilde{a}_{i, j+1}}{\tilde{a}_{i, j}-\tilde{a}_{i+1, j}}\right|
\end{equation}
The terms contain $a_{ij}$ are:
\begin{equation}
    A_{\text{tot}}=\dots-2\log|a_{ij}-a_{i+1j}|-2\log|a_{i-1j}-a_{ij}|+2\log|a_{ij}-a_{ij+1}|+2\log|a_{ij-1}-a_{ij}|+\dots
\end{equation}
Which is the discretized version of the Nambu-Goto action. The equation of motion is then given by the variation $\delta A_{\text{tot}}/\delta a_{ij}=0$. Hence the discretized equation of motion is \cite{DV1}:
\begin{equation}\label{eq:toda}
    \frac{1}{a_{i j}-a_{i j+1}}+\frac{1}{a_{i j}-a_{i j-1}}=\frac{1}{a_{i j}-a_{i+1 j}}+\frac{1}{a_{i j}-a_{i-1 j}}
\end{equation}

Now assume that the defining null vectors of each string segment satisfy the condition 4. from Section 5.4. Then it follows that $a_{ij}<a_{ij+1},a_{ij-1},a_{i-1j},a_{i+1j}$ for positively oriented $ij$ edge and $a_{ij}>a_{ij+1},a_{ij-1},a_{i-1j},a_{i+1j}$ for negatively oriented $ij$ edge (and similarly for $\tilde{a}_{ij}$). This simply means that for example in case of positively oriented setting the future part of the null cone of $p_{ij}$ intersects the past part of the $p_{ij+1},p_{ij-1},p_{i+1j},p_{i-1j}$ cones. Therefore one can write:
\begin{equation}
    \frac{1}{|a_{i j}-a_{i j+1}|}+\frac{1}{|a_{i j}-a_{i j-1}|}=\frac{1}{|a_{i j}-a_{i+1 j}|}+\frac{1}{|a_{i j}-a_{i-1 j}|}
\end{equation}

In the boundary theory the left component of the entanglement entropy of a CFT interval determined by two $p_{ij}$ and $p_{kl}$ null vectors can be written in the form:
\begin{equation}
    S^L(ij,kl)=\frac{c}{6}\log\frac{|a_{ij}-a_{kl}|}{\delta}
\end{equation}
Using this expression one can rewrite \eqref{eq:toda} 
\begin{equation}
    e^{-\frac{c}{6}S^L(ij,ij+1)}+e^{-\frac{c}{6}S^L(ij,ij-1)}=e^{-\frac{c}{6}S^L(ij,i+1j)}+e^{-\frac{c}{6}S^L(ij,i-1j)}
\end{equation}
Therefore the Toda equation generates a relation between entanglement entropies. By varying $A_{\text{tot}}$ with respect to $\tilde{a}_{ij}$ it can be shown that the same equation holds for the right component $S^R$.

\section{Correspondence in even dimensions}

Now we turn to the higher dimensional case. Our aim is to show that correspondences similar to the ones discussed in the previous sections hold even in the $AdS_{d+1}/CFT_d$ scenario if $d$ is even.
We point out that after a careful reconsideration of our results a nice quantum information theoretic interpretation for the area of the string world sheet segment emerges. 
These results can naturally be connected to existing ones in the literature.

\subsection{$AdS$ space and the Poincaré patch}

The $d$ dimensional anti-de Sitter space is the locus of points in $\mathbb{R}^{2,d}$ whose points satisfy the constraint
\begin{equation}
\begin{aligned}
    X\cdot X:&=-(X^{-1})^2-(X^0)^2+(X^1)^2+\dots+(X^d)^2=\\
    &=-(X^0)^2+(X^2)^2+\dots+(X^d)^2+X^+X^-=-L^2
\end{aligned}
\end{equation}
Where 
\begin{equation}
X^{\pm}=X^1\pm X^{-1}
\label{pm}
\end{equation}
and
$L$ is the AdS radius. The $d$ dimensional asymptotic boundary of the $AdS_{d+1}$ space is defined by the set:
\begin{equation}
    {\partial}_{\infty} AdS_{d+1}\coloneqq {\mathbb P}\{U\in{\mathbb R}^{2,d}\vert U\cdot U=0\}
\end{equation}
where ${\mathbb P}$ means projectivization.

Following our previous conventions we define the Poincaré patch representation of the $AdS_{d+1}$ in the following way:
\begin{equation}
\left(X^{-1},X^0,X^1,X^2,\dots,X^d\right)=
\left(\frac{-z^{2}-x\bullet x-L^2}{2 z}, L\frac{t}{z}, \frac{-z^{2}-x\bullet x+L^2}{2 z}, L\frac{x^1}{z},\dots,L\frac{x^{d-1}}{z}\right)
\label{patchpoin}
\end{equation}
And $z>0$. We have defined the Minkowski vector $x^\mu=(t,x^1,\dots,x^{d-1})$, $\mu=0,1,\dots,d-1$ and Minkowski inner product $x\bullet x=-t^2+(x^1)^2+\dots+(x^{d-1})^2$.
The line element in these coordinates is:
\begin{equation}\label{eq:metric_patch}
ds^2=L^2\frac{dz^2+dx\bullet dx}{z^2}
\end{equation}
The boundary of the $AdS$ space in the Poincaré patch is obtained by taking the $z\to 0$ limit. Notice that in this limit the metric is conformally equivalent to the $d$ dimensional Minkowki space. The Poincaré patch coordinates $x$ ond $z$ of an AdS point can be expressed by the global coordinates $X$ in the following way
\begin{equation}
    z=\frac{L^2}{X^-},\quad t=x^0=L\frac{X^0}{X^-},\quad x^1=L\frac{X^2}{X^-},\qquad\dots,\qquad x^{d-1}=L\frac{X^d}{X^-}
\end{equation}
Notice that in this patch those $AdS$ points are represented that satisfy the condition $X^->0$. The coordinates of a null vector $U$ representing a boundary point are given by:
\begin{equation}
    x_u^\mu=(x_u^0,x_u^1,\dots,x_u^{d-1})=\frac{L}{U^-}(U^0,U^2,\dots,U^d)\\
\label{bdycoord}
\end{equation}

Now by repeating the same steps as in Section 2.1. one can prove that Eq.(\ref{fontos}) holds in this general case hence we have
\begin{equation}
(x_u-x_v)\bullet(x_u-x_v)=-2L^2\frac{U\cdot V}{U^-V^-}
\label{veryimport}
\end{equation}
We notice that if $U\cdot V<0$ and $U^-V^-<0$ or $U\cdot V>0$ and $U^-V^->0$ then $x_u$ and $x_v$ are timelike separated.

\subsection{$AdS_{d+1}$ minimal surfaces}

The codimension two minimal surfaces $X_{\mathcal R}$ of the $AdS_{d+1}$ space homologous to a boundary region ${\mathcal R}$, with $\partial {\mathcal R}\simeq S^{d-2}$, can be defined by null vectors of the embedding space $\mathbb{R}^{2,d}$. Let $U$ and $V$ with $U\cdot U=V\cdot V=0$ be two null vectors such that
\begin{equation}
    U\cdot V<0\qquad \text{and}\qquad U^-V^-<0
\label{sepcond}
\end{equation} 
Then the minimal surface is the intersection of the following two submanifolds \cite{Myers}:
\begin{equation}
    U\cdot X=0,\qquad V\cdot X=0
\end{equation}Where $X\in AdS_{d+1}$. Using the Poincaré coordinates of Eq. (\ref{patchpoin}) these equations define null cones of the form
\begin{align}
    &z^2+(x-x_u)\bullet(x-x_u)=0 \label{cones_higher1}\\
    &z^2+(x-x_v)\bullet(x-x_v)=0 \label{cones_higher2}
\end{align}
where
\begin{equation}
    x_u^\mu=\frac{L}{U^-}(U^0,U^2,\dots,U^d),\qquad x_v^\mu=\frac{L}{V^-}(V^0,V^2,\dots,V^d)
\end{equation}
These null cones are residing in the Poincaré patch. The minimal surface in question is the surface given by their intersection. 

Subtracting the two equations it turns out that the surface lies in the subspace
\begin{equation}
    2x\bullet(x_v-x_u)+x_u\bullet x_u-x_v\bullet x_v=0
\end{equation}
Introducing the quantities
\begin{equation}
\Delta=x_u-x_v,\qquad x_0=\frac{1}{2}(x_u+x_v)
\label{tkp}
\end{equation}
this subspace can alternatively be described as
\begin{equation}
\Delta\bullet (x-x_0)=0
\label{subspacecond}
\end{equation}

From the equations of the cones one infers that
\begin{equation}
    (x-x_u)\bullet(x-x_u)<0,\qquad (x-x_v)\bullet(x-x_v)<0
\end{equation}
Hence the points of the minimal surface projected to the boundary are timelike separated from the centers of the cones.
These points are comprising a spherical region\cite{Myers} for which the entanglement entropy of the reduced density matrix of the boundary CFT state is calculated.
Moreover from Eq.(\ref{veryimport}) it is clear that by virtue of the (\ref{sepcond}) conditions
$x_u$ and $x_v$ are timelike separated too, hence we have $\Delta\bullet\Delta<0$.
One can check that, as explained in \cite{Myers}, the two cones define a causal diamond in the boundary. 
The centers of the cones define the upper and lower tips of this diamond.\footnote{This derivation can similarly be done for null vectors that satisfying $U\cdot V>0$. However, then their patch vectors are timelike separated if and only if $U^-V^->0$.}

Let us now write
\begin{equation}
x-x_u=x-x_0-\frac{\Delta}{2},\qquad x-x_v=x-x_0+\frac{\Delta}{2}
\label{masalak}
\end{equation}
Then both of Eqs.(\ref{cones_higher1} and \ref{cones_higher2}) gives
\begin{equation}
z^2+(x-x_0)\bullet (x-x_0)=-\frac{\Delta\bullet\Delta}{4}\equiv r^2
\label{minsurf}
\end{equation}
where we have used the fact that $x_u$ and $x_v$ are timelike separated for defining the positive quantity $r^2$.
The detailed form of Eq.(\ref{minsurf}) is
\begin{equation}
z^2-(t-t_0)^2+\vert\vert{\bf x}-{\bf x}_0\vert\vert^2=r^2,\qquad 4r^2=-\vert\vert\mathbf{\Delta}\vert\vert^2+(\Delta^0)^2
\label{detailedform}
\end{equation}
This equation is to be used together with the following version of Eq.(\ref{subspacecond})
\begin{equation}
\Delta^0(t-t_0)=\mathbf{\Delta}({\bf x}-{\bf x}_0)
\label{skalar}
\end{equation}
where on the right hand side we have an ordinary scalar product of two vectors in ${\mathbb R}^{d-1}$.
Expressing $t-t_0$ from this equation and using it in Eq.(\ref{detailedform}) gives
\begin{equation}
z^2+\vert\vert{\bf x}-{\bf x}_0\vert\vert^2 - \left(\frac{\mathbf{\Delta}({\bf x}-{\bf x}_0)}{\Delta^0}\right)^2=r^2
\label{verynice}
\end{equation}

Notice that there is a gauge degree of freedom given by rescalings of the form
\begin{equation}
U\mapsto \Lambda_uU,\qquad
V\mapsto \Lambda_vV,\qquad \Lambda_u\Lambda_v>0
\label{gaugedeg}
\end{equation}
These new null vectors define the same null cones $U\cdot X=0$ and $V\cdot X=0$ hence the same minimal surface.

Similarly to the $d=3$ case one can determine the area of a spherical $AdS_{d+1}$ minimal surface determined by the equations \eqref{cones_higher1} and \eqref{cones_higher2}. The result is well-known from the literature\cite{Towards,RT2}, however for the convenience of the reader a calculation is presented in Appendix B. To summarize the derivation, after some algebraic manipulation and reparametrization of the minimal surface the area can be determined via evaluating the following integral
\begin{equation}\label{eq:integral}
    {\mathcal A}(X_R)=L^{d-1}\Omega_{d-2}\int_{\delta/r}^1 dy\frac{\left(1-y^2\right)^{\frac{d-3}{2}}}{y^{d-1}}
\end{equation}
Where we have introduced again a $z>\delta$, $\delta\gg r$ cutoff and as before
\begin{equation}
    r^2=-\frac{1}{4}(x_u-x_v)\bullet(x_u-x_v)
\end{equation}
and we also referred the well-known formula
\begin{equation}
\Omega_{d-2}=2\frac{\pi^{(d-1)/2}}{\Gamma\left(\frac{d-1}{2}\right)}
\label{omegawell}
\end{equation}

During the calculations it turns out that the final result is significantly different when $d=\text{even}$ and $d=\text{odd}$. In the following we restrict our examinations to the case when $d=\text{even}$. In this case the area can be written in the following form
\begin{equation}\label{eq:min_area_gen}
\begin{aligned}
    {\mathcal A}(X_R)&=L^{d-1}\Omega_{d-2}\frac{1}{d-2}\frac{r^{d-2}}{\delta^{d-2}}+\\
    &+L^{d-1}\Omega_{d-2}\frac{(-1)^{(d-2)/2}}{2}\frac{(d-3)!!}{(d-2)!!}\log{\left(\frac{r^2}{\delta^2}\right)}+\\
    &+L^{d-1}\Omega_{d-2}F_d(1)+\\
    &+L^{d-1}\Omega_{d-2}\sum_{\substack{k=1\\k\neq(d-2)/2}}^\infty
    \frac{(-1)^{k+1}}{2^kk!}\frac{1}{(2k-d+2)}\frac{(d-3)!!}{(d-(2k+3))!!}\left(\frac{r}{\delta}\right)^{d-(2k+2)}
\end{aligned}
\end{equation}
The first term of the expression is prortional to the area of $\partial A$ that is with $r^{d-2}\Omega_{d-2}$. The second term is logarithmic similar to the $d=2$ case. The third term is a constant that comes from the evaluation of the integrand in \eqref{eq:integral} at the upper limit. The explicit form of this term is:
\begin{equation}
    F_d(1)=\sum_{\substack{k=0\\k\neq(d-2)/2}}^\infty
    \frac{(-1)^{k+1}}{2^kk!}\frac{1}{(2k-d+2)}\frac{(d-3)!!}{(d-(2k+3))!!}
\end{equation}
Finally the last term includes different powers of $r/\delta$. These are divergent and vanishing terms in the limit $\delta\to 0$.

The main point is that in the $AdS_{d+1}$ scenario, when $d=\text{even}$ there is a logarithmic term in the expression for the area of a minimal surfaces. Via the higher dimensional generalization of the Ryu-Takayanagi conjecture these terms are connected to the entanglement entropies of boundary regions. In the following we show that this establishes a connection between string segments and entanglement similarly to the $d=2$ case.

\subsection{String segments and the area/entropy relation in even dimensions}

In general two dimensional strings embedded into $AdS_{d+1}$ space are determined by the following equation of motion
\begin{equation}\label{eq:eom_gen}
    \partial_+\partial_- X-(\partial_-X\cdot\partial_+X)X=0
\end{equation}
which can be derived via the variation of the Nambu-Goto action. We have parametrized the string by the parameters $\sigma^-$ and $\sigma^+$. The Virasoro constraints for string are
\begin{equation}
\partial_- X\cdot \partial_- X=\partial_+ X\cdot\partial_+ X=0
\end{equation}
Let us now again define an $AdS_{d+1}$ string segment by the vectors of its vertices $V_i$ and the lightlike vectors of its edges
\begin{align}
p_1=V_2-V_1&&p_2=V_3-V_2\\
p_3=V_3-V_4&&p_4=V_4-V_1
\end{align}
The initial data set as usual is $(V_1,p_1,p_4)$ and
\begin{align}
    X(\sigma^-,0)&=V_1+\sigma^- p_1\\
    X(0,\sigma^+)&=V_1+\sigma^+ p_4
\end{align}
The remaining vertex and null vectors can be calculated by the interpolation ansatz
\begin{equation}
    X(\sigma^-,\sigma^+)=\frac{L^2+\sigma^-\sigma^+\frac{1}{2}p_1\cdot p_4}{L^2-\sigma^-\sigma^+\frac{1}{2}p_1\cdot p_4}V_1+L^2\frac{\sigma^-p_1+\sigma^+p_4}{L^2-\sigma^-\sigma^+\frac{1}{2}p_1\cdot p_4}
\end{equation}
 satisfying the equation of motion \eqref{eq:eom_gen}. Assume further that the causality conditions mentioned in Section 3.2 still holds for the vectors $V_i$ and $p_i$.

The area of the string segment can be calculated just like in the $AdS_{3}$ case. The final result is again
\begin{equation}
A_{\diamond}=L^2 \log\left[\frac{(p_1-p_4)^2}{(p_1+p_2)^2}\right]^2=L^2\log\frac{(p_1-p_4)^2(p_3-p_2)^2}{(p_1+p_2)^2(p_3+p_4)^2}
\end{equation}
Therefore by defining the usual quantity
\begin{equation}
    r_{ij}^2=-\frac{1}{4}(x_i-x_j)\bullet(x_i-x_j)
\end{equation}
by virtue of Eq.(\ref{veryimport}) 
the area of the string segment can be written in the form
\begin{equation}
    A_{\diamond}=L^2\log\frac{r_{14}^2r_{23}^2}{r_{34}^2r_{12}^2}
\label{vesd}
\end{equation}

Now consider the $AdS_{d+1}/CFT_{d}$ scenario for $d=\text{even}$. Choose an entangling surface $\partial {\mathcal R}$ in the boundary field theory. The general form of the Ryu-Takayanagi conjecture in the $AdS_{d+1}/CFT_{d}$ regime reads as
\begin{equation}
    S=\frac{{\mathcal A}(X_{\mathcal R})}{4G^{(d+1)}}
\end{equation}
Where $S$ is the entanglement entropy inside $\partial {\mathcal R}$, ${\mathcal A}(X_{\mathcal R})$ is the area of $X_{\mathcal R}$ the $AdS_{d+1}$ minimal surface homologous to $\partial {\mathcal R}$ and $G^{(d+1)}$ is the $d+1$ dimensional Newton's constant. According to \cite{RT2} the entanglement entropy inside $\partial {\mathcal R}$ can be written in the form
\begin{equation}
    S=\frac{\gamma_1}{2} \cdot \frac{\text{Area}(\partial {\mathcal R})}{\delta^2}+\gamma_2 \log \frac{l}{\delta}+S^{\text {others }}
\end{equation}
Where $\text{Area}(\partial {\mathcal R})$ is the area of $\partial {\mathcal R}$ and $S^{\text{others}}$ depends on the detailed shape of $\partial {\mathcal R}$. The constant $\gamma_2$ is universal in the sense, that it does not depend on the cutoff $\delta$. From now on we denote the universal logarithmic term as $S^{\text{uni}}$.

From the calculation of the previous subsection, using Eq.(\ref{eq:min_area_gen}) and Eq.(\ref{eq:area_gen2}) and the Ryu-Takayanagi formula the universal term of the entanglement for a $d-2$ dimensional, spherical entangling surface is
\begin{equation}\label{eq:entanglement_gen}
    S^{\text{uni}}(ij)=\frac{(-2\pi)^{(d-2)/2}L^{d-1}}{4G^{(d+1)}(d-2)!!}\log{\left(\frac{r_{ij}^2}{\delta^2}\right)}
\end{equation}
Where the homologous minimal surface is determined by the $\mathbb{R}^{2,d}$ null vectors $p_i$ and $p_j$ via the equations $p_i\cdot X=p_j\cdot X=0$. Therefore $r_{ij}^2$ is
\begin{equation}
    r_{ij}^2=-\frac{1}{4}(x_i-x_j)\bullet(x_i-x_j)
\end{equation}
where
\begin{equation}
    x_i^\mu=\frac{L}{p_i^-}(p_i^0,p_i^2,\dots,p_i^d)
\end{equation}
We remark that an alternative formula for this universal part of the entropy in even dimensions is given by the formula coming from the CFT side\cite{Towards}
\begin{equation}
    S^{\text{uni}}(ij)=
    4(-1)^{(d-2)/2}a_d^{\ast}\log{\left(\frac{r_{ij}}{\delta}\right)}
\label{hogyuni}
\end{equation}
where
$a_d^{\ast}$ is the coefficient of the A-type trace anomaly in the boundary CFT. For $d=2$ in accordance with the Brown-Henneaux formula we have $a_d^{\ast}=c/12$.

Let us now choose two $\mathbb{R}^{2,d}$ null vectors $p_1,p_4$ and an $AdS_{d+1}$ vector $V_1$. They determined a string segment in the $AdS_{d+1}$ space. The other two edge null vectors $p_2$ and $p_3$ of the segment are given by the interpolation ansatz. The area of the segment is
\begin{equation}
    A_{\diamond}=L^2\log\frac{r_{14}^2r_{23}^2}{r_{12}^2r_{34}^2}
    \label{diamondstring}
\end{equation}
The tips of the string segment are lying on the cones defined by the equations $p_i\cdot X=0$. The intersections of these cones define minimal surfaces $X_{\mathcal R}$ hence boundary entangling surfaces as well. Inside these entangling surfaces the entanglement entropy is given by \eqref{eq:entanglement_gen}. Now one can see that between the area of the string segment and the universal terms of the entanglement entropies
the following relation holds
\begin{equation}
S^{\text{uni}}(14)+S^{\text{uni}}(23)-S^{\text{uni}}(12)-S^{\text{uni}}(34)=
\left[\frac{(-2\pi)^{(d-2)/2}}{(d-2)!!}L^{d-2}\right]\left[\frac{
    A_{\diamond}}{4G^{(d+1)}L}\right]
\label{hogyan}
\end{equation}
One can also notice that the first term on the right
hand side is the volume $V(B_L^{d-2})$ of the $d-2$ dimensional Euclidean ball
with radius $L$ multiplied by a sign factor.
Then the general form of our formula for even $d$ that connects the geometry of segmented strings and the entanglement of the boundary CFT is
\begin{equation}
S^{\text{uni}}(14)+S^{\text{uni}}(23)-S^{\text{uni}}(12)-S^{\text{uni}}(34)=
\left[\frac{(-1)^{(d-2)/2}V(B_L^{d-2})
    A_{\diamond}}{4G^{(d+1)}L}\right]
\label{hogyan1}
\end{equation}
Notice that for the special case of $d=2$ the first term on the right hand side is one and then we are back to Eq.(\ref{simplenice}).

Since $p_i\cdot V_j=0$ for the segment tip and null vectors the correspondence geometrically is interpreted in the following way. If we choose a spacelike, spherical domain in the boundary CFT theory, its causal development is dual to cones in the bulk geometry. Their intersection defines an $AdS$ minimal surface. Taking multiple subsystems they will define timelike AdS string segments whose tips will lie on these surfaces. Their area will be proportional to the combination of entanglement entropies of the CFT subsystems seen on the left hand side of our formula.

Now from Section 3.7 we also know that the left hand side of (\ref{hogyan1}) has the nice quantum information theoretic interpretation of Eq.(\ref{heureka2}) in terms of conditional mutual information. Then the question arises, how can we reinterpret the left hand side of our formula of Eq. (\ref{hogyan1})  
in terms of similar quantities for $d>2$ with $d$ even?

\subsection{Reinterpretating the  area formula for $d$ arbitrary.}

\subsubsection{Trapezoids for $d>2$ $d$ even.}

Our result of Eq. (\ref{hogyan1}) shows that for even dimensions the area of the string world sheet segment can be expressed in terms of a combination 
of the universal terms of entanglement entropies
calculated for four different spheres.
These spheres are determined by the tips of causal diamonds arising from the intersection of four light cones.
In the $d=2$ case via Eq.(\ref{heureka2}) we managed to obtain a nice quantum information theoretic interpretation of this area as conditional mutual information based on dual regions comprising trapezoids. 
In the higher dimensional case, due to lack of explicit results for entanglement entropies for unions and intersections of spherical regions, we did not manage to find such an interpretation.
However, in the next subsection we will point out that there is a very interesting  alternative interpretation.

But first in this subsection we would like to offer some general
considerations to be used later.
First of all for simplicity let us rename the $d+2$ component momentum null vectors $p_i, i=1,2,3,4$ labeling the edges of our string world sheet segment as $A,B,C,D$. 
We use the parametrization similar to the one used in the rows of the matrix of Eq.(\ref{eq:determinant})
\begin{equation}
A=\lambda_1  \begin{pmatrix}
1\\-R/L\\{\bf 0}\\R^2/L^2\end    
{pmatrix},\quad
B=\lambda_2 \begin{pmatrix}
1\\R/L\\{\bf 0}\\R^2/L^2\end    
{pmatrix},\quad
C=\lambda_3 \begin{pmatrix}
1\\ \gamma/L\\{\bf c}/L\\-c\bullet c/L^2\end    
{pmatrix},\quad
D=\lambda_4 \begin{pmatrix}
1\\ \delta/L\\{\bf d}/L\\-d\bullet d/L^2\end    
{pmatrix}
\nonumber
\end{equation}
Here $\lambda_i=p_i^-$ (see also Eq.(\ref{konzerv})), and due to the causality conditions found at the end of Section 3.1 one must have $\lambda_i\lambda_j<0$ for neighbouring null vectors and 
$\lambda_i\lambda_j>0$ for antipodal ones.

We also introduce the notation
\begin{equation}
c\bullet c =-{(c^0)}^2+\vert\vert{\bf c}\vert\vert^2=-\gamma^2+\vert\vert{\bf c}\vert\vert^2,\qquad
d\bullet d=-{(d^0)}^2+\vert\vert{\bf d}\vert\vert^2=-\delta^2+\vert\vert{\bf d}\vert\vert^2
\end{equation}
Clearly $A,B,C,D\in{\mathbb R}^{d,2}$ and
$a,b,c,d\in{\mathbb R}^{d-1,1}$ with the latter labeling the tips of our causal diamonds. With the special parametrization for $A$ and $B$ (giving rise to the past and future tips of our basic causal diamond) we transformed our diamond to the arrangement familiar from the end of Section 3.8. Namely we have
\begin{equation}
a=\begin{pmatrix}\alpha\\{\bf a}\end{pmatrix}=
\begin{pmatrix}-R\\{\bf 0}\end{pmatrix},\qquad
b=\begin{pmatrix}\beta\\{\bf b}\end{pmatrix}=
\begin{pmatrix}R\\{\bf 0}\end{pmatrix}
\label{ab1}
\end{equation}
In terms of these new quantities the causality conditions of Section 3.1
to be satisfied are
\begin{equation}
\gamma<\delta< R
\end{equation}
The remaining set of causality conditions are stating that the Minkowski vectors: $a-b, b-c,c-d,a-d$ are all timelike.

Now in this notation stringy momentum conservation $A+B=C+D$ in the bulk ${\rm AdS}_{d+1}$
boils down to the following set of constraints for the four points $a,b,c,d$ in the boundary 
${\mathbb R}^{d-1,1}$

\begin{equation}
\lambda_1+\lambda_2=\lambda_3+\lambda_4,\qquad
R^2(\lambda_1+\lambda_2)=-c\bullet c\lambda_3-d\bullet d\lambda_4
\end{equation}

\begin{equation}
\lambda_3{\bf c}+\lambda_4{\bf d}=0,\qquad
R(\lambda_2-\lambda_1)=\gamma\lambda_3+\delta\lambda_4
\end{equation}

From the first three of these equations one gets
\begin{equation}
\frac{\bf c}{R^2+c\bullet c}=
\frac{\bf d}{R^2+d\bullet d}
\end{equation} 
This shows that the ${\mathbb R}^{d-1}$ vectors ${\bf c}$ and ${\bf d}$ are parallel.

Let us now define a vector ${\bf x_0}\in{\mathbb R}^{d-1}$ as the space component of the vector $x_0=(0,{\bf x}_0)^T\in{\mathbb R}^{1,d-1}$
as the one satisfying
\begin{equation}
\frac{2{\bf c}{\bf x}_0}{R^2+c\bullet c}=
\frac{2{\bf d}{\bf x}_0}{R^2+d\bullet d}=1
\end{equation}
Clearly this vector is defined up to adding an arbitrary vector orthogonal to ${\bf c}$.
Hence in this way we have defined a class of vectors having both orthogonal ${{\bf x}_0}^{\perp}$ and parallel 
${{\bf x}_0}^{\vert\vert}$ components
to both ${\bf c}$ and ${\bf d}$ (that are collinear).
Then we have 
\begin{equation}
-\gamma^2+\vert\vert {\bf c}-{\bf x}_0\vert\vert^2 =
-\delta^2+\vert\vert {\bf d}-{\bf x}_0\vert\vert^2=\varrho^2,\qquad \varrho^2=R_0^2-R^2,\qquad R_0^2=\vert\vert{\bf x}_0\vert\vert^2
\end{equation}
Hence for $\varrho^2>0$ the four points $a,b,c,d$ are on a class of single sheeted hyperboloids with centers parametrized by the vectors $x_0$ and the radius is  $\varrho$.
Via a ${\bf T}\in SO(d-1)$ transformation one can achieve that ${\bf x}_0=R_0{\bf e}_1$ where ${\bf e}_1=(1,0,0,\dots, 0)^T$. Under this transformation we have  $\tilde{\bf {c}}={\bf T}{\bf c}$.
In this basis we have 
\begin{equation}
-\gamma^2+(\tilde{c}_1-R_0)^2+\tilde{c}_2^2+\dots+\tilde{c}_{d-1}^2=\varrho^2    
\end{equation}
\begin{equation}
-\delta^2+(\tilde{d}_1-R_0)^2+\tilde{d}_2^2+\dots+\tilde{d}_{d-1}^2=\varrho^2    
\end{equation}
Clearly for $d=2$ according to Eqs.(\ref{niceconnect})-(\ref{gejee}) one has $a_4=d=\delta+d_1$ and $\bar{a}_4=\bar{d}=\delta-d_1$ hence 
\begin{equation}
x_0=\frac{R^2+d\bullet d}{2d_1}    
\end{equation}
which is the first of Eq.(\ref{gejee}).

Choosing $(\tilde{c}_2,\dots,\tilde{c}_{d-1})=(\tilde{d}_2,\dots,\tilde{d}_{d-1})=(0,\dots,0)$
one can achieve that the points $c$ and $d$ are again on a hyperbola
with the usual interpretation as events for observers executing hyperbolic
motion.
For a causal diamond placed at the origin one can then introduce coordinates of Jacobson and Visser\cite{Jacobson,Banks,Towards}
such that the observers of hyperbolic motion are moving along the orbits of conformal Killing vectors as follows.

We fix the diamond $\mathcal D$ as the intersection of the future light cone of $a$ and
the past light cone of $b$ of Eq.(\ref{ab1}). Then  the edge of the diamond is the boundary of a d-1-dimensional ball-shaped region $\Sigma$ which is a sphere $S^{d-2}$ of radius $R$.
The line element of $\mathcal D$ is in polar coordinates
\begin{equation}
ds^2=-dt^2+dr^2+r^2d\Omega^2_{d-2}=-dad\bar{a}+\frac{1}{4}(a-\bar{a})^2d\Omega^2_{d-2}
\end{equation}
with $a=t+r$ and $\bar{a}=t-r$.
Then the Jacobson-Viesser coordinates are just $0\leq u<\infty $ and  $-\infty<\tilde{\tau}<\infty$ related to $t$ and $r$
as
\begin{equation}
t=R\frac{\sinh\left(\frac{\tilde{\tau}}{R}\right)}{\cosh u+\cosh\left(\frac{\tilde{\tau}}{R}\right)},\qquad
r=R\frac{\sinh u}{\cosh u+\cosh\left(\frac{\tilde{\tau}}{R}\right)}
\label{JVcoord}
\end{equation}
with the corresponding line element
\begin{equation}
ds^2=C^2(u,\tilde{\tau})(-d\tilde{\tau}^2+R^2(du^2+\sinh^2ud\Omega^2_{d-2})),\qquad C^{-1}(u,\tilde{\tau})=\cosh\left(\frac{\tilde{\tau}}{R}\right)+\cosh u   
\end{equation}
Notice that the (\ref{JVcoord}) formula gives a conformal mapping from $\mathcal D$ to $\mathcal H\simeq {\mathbb R}\times{\mathbb H}^{d-1}$
where ${\mathbb H}^{d-1}$ is a d-1 dimensional hyperbolic space\cite{Towards}.

The use of these coordinates is in accord with the result
that the orbits of the conformal Killing field 
\begin{equation}
\xi=\frac{1}{2R}\left((R^2-a^2)\partial_a+(R^2-\bar{a}^2)\partial_{\bar{a}}\right)
=R\partial_{\tilde{\tau}}
\end{equation}
(i.e. the field generating evolution in modular time and preserving $\mathcal D$) have uniform acceleration inside the diamond\cite{Jacobson}. 
The flow lines of this Killing field coincide at the past and future tip of the diamond, but
never cross inside the diamond.
It can be also shown (see Appendix F of \cite{Jacobson}) that the proper acceleration depends only on $u$ , which is constant
and has the value
\begin{equation}
g(u)=\frac{1}{R}\sinh u
\label{acci}
\end{equation}on the conformal
Killing orbits. This is to be compared with (\ref{accelerate11}) of the $d=2$ case.
The central orbit at $u=0$ corresponding to an inertial observer is unaccelerated, and at the edge where $u\to\infty$
the acceleration diverges. These results
boils down to our findings for the $d=2$ case where the edge of the diamond is a $S^0\simeq {\mathbb Z}_2$ and
$\mathcal H\simeq {\mathbb R}\times{\mathbb R}^+$ i.e. the upper half plane.
Notice that in this special case an extra ${\mathbb Z}_2$ symmetry $u\mapsto -u$ is relating the right half and the left half of the causal diamond.

One is particularly interested in constant $\tilde{\tau}$ slices of our causal diamond. From the $d=2$ case we know that these slices correspond to dual spacelike hyperbolas giving rise to trapezoids.
We are expecting that the dual regions forming some analogue to these trapezoids  should live on such slices. 
The slices in question are foliating our diamond with ${\mathbb H}^{d-1}$ spaces hence the leaves of this foliation are having the induced metric
$h_{\hat{\mu}\hat{\nu}}= R^2C^2\sigma_{\hat{\mu}\hat{\nu}}$ with
$\sigma_{\hat{\mu}\hat{\nu}}$ being the metric on ${\mathbb H}^{d-1}$.
Then a calculation\cite{Jacobson} shows that $K$, the trace of the extrinsic curvature of these slices, is given by the formula
\begin{equation}
K(\tilde{\tau})=\frac{1-d}{R}\sinh\left(\frac{\tilde{\tau}}{R}\right)    
\label{extrinsic}
\end{equation}
hence it is independent of $u$ and vanishing on the $\tilde{\tau}=0$ slice.
Notice that for $d=2$
the formula $RK_i=-\sinh\left(\frac{\tilde{\tau}_i}{R}\right)=-\sinh(v_i)$ taken together with $Rg_i=-\sinh(u_i)$ for $i=3,4$
gives an alternative interpretation of the world sheet area  Eqs.(\ref{jaj1})-(\ref{jaj2}) of the string segment.
According to this interpretation the change in modular time $\tilde{\tau}_i$ is related to the change in the constant extrinsic curvatures
$K_i$ of the corresponding leaves.

For the higher dimensional case since we have  ${\bf c}\vert\vert{\bf d}$ and 
in this case like in Eq. (\ref{errjk}) one still has $4r_{jk}^2=(a_j-a_k)(\bar{a}_j-\bar{a}_k)$ a calculation using Eq.(\ref{diamondstring}) shows that this interpretation survives even for $d>2$.
Then expressions (\ref{jaj1})-(\ref{jaj2}) are also valid for the $d>2$ case. However, in the case of (\ref{jaj1}) modular time is related to the trace of the extrinsic curvature by formula (\ref{extrinsic}), moreover in the case of (\ref{jaj2}) in order to relate the coordinate $u$ to acceleration Eq.(\ref{acci}) is to be used.

Now we can again arrange the tips of the respective causal diamonds to form dual situations  as depicted in Figures 7. and 8. 
In order to see this notice that for $d>2$
 the "left and right tips"  of causal diamonds can be defined by the constraint 
$(\tilde{c}_2,\dots,\tilde{c}_{d-1})=(\tilde{d}_2,\dots,\tilde{d}_{d-1})=(0,\dots,0)$ coming from momentum conservation for segmented strings.
Hence we still have dual situations where the future and past, or the left and right tips of causal diamonds are arranged on timelike or spacelike hyperbolas corresponding to the coordinate lines of Jacobson Visser coordinates. 
This time however, unlike in the $d=2$ case , we have spherical subregions not merely intervals.  
Recall also here, that clearly in two dimensions,
a causal diamond can be defined either in terms of a pair of timelike separated points or
a pair of spacelike separated points.
Hence in the $d=2$ the moduli space of causal diamonds\cite{Myers} can equivalently be defined as
the space of timelike separated pairs of points
or as the space of spacelike separated pairs of points.
Due to a rather conterintuitive result (see Appendix A.3. of\cite{Myers}) the equivalence of these moduli spaces survives even for $d>2$. 
The upshot of these considerations is that the set of left and right tips can again be arranged to lie on a hyperbola forming a trapezoid.
They are coming from intersecting light cones of other spacelike separated points.
For a graphical illustration of intersecting light cones of two spacelike sparated points giving rise to a spacelike hyperbola 
lying on a codimension one hyperplane see Figure 11. of Ref.\cite{Myers}.
Unfortunately this argumentation runs out of steam, since the unions and intersections of the associated spherical regions are not spherical regions anymore.
For such configurations we were unable to find results in the existing literature that are checking strong subadditivity or calculating the corresponding conditional mutual informations\footnote{See however, Ref.\cite{Torroba} for recent results with calculations for mutual information for arbitrary spherical regions with large separation.} .
Hence arriving at a quantum information theoretic interpretation of (\ref{hogyan1}) of the (\ref{heureka2}) form (along this line of reasoning)  for $d>2$  $d$ remains a future challenge. 

\subsubsection{Fidelity susceptibility, $d$ even}

There is however, yet another interesting possibility for finding a quantum information theoretic meaning for expressions like Eq.(\ref{hogyan1}).
In order to see this just recall that the natural metric on the moduli space of causal diamonds\cite{Myers} 
can be connected to a quantity known as fidelity susceptibility or quantum information metric in quantum information\cite{Gu, Miyaji,Kinaiak}.
For a parameter dependent quntum state this latter quantity is defined via considering an infinitesimal change of some parameter  for a parameter dependent state
$\psi(\lambda)$.
Then the quantity
\begin{equation}
\vert\langle \psi(\lambda)\vert\psi(\lambda+d\lambda)\rangle \vert =1-G_{\lambda\lambda}(\delta\lambda)^2+\dots    
\label{fidelity}
\end{equation}
is called the fidelity, and $G_{\lambda\lambda}$ is called fidelity susceptibility or quantum information metric.

In order to reveal the relevance of $G$ to our considerations first recall that the $2d$ dimensional moduli space of causal diamonds $\mathcal K$, also called the {\it kinematic space}, 
can be represented as the coset space
\begin{equation}
{\mathcal K}=SO(d,2)/SO(d-1,1)\times SO(1,1)
\label{Kin}
\end{equation}
This space
can be endowed with the following metric\cite{Myers}
\begin{equation}
    d s_{\mathcal K}^2=\frac{4 {\ell}^2}{\left(x-y\right)^2}\left[-\eta_{\mu \nu}+\frac{2\left(x_\mu-y_\mu\right)\left(x_\nu-y_\nu\right)}{\left(x-y\right)^2}\right] d x^{\mu} d y^{\nu}:=h_{\mu\nu} dx^\mu dy^\nu
\label{modulimetric}
\end{equation}
where ${\ell}$ is an arbitrary length scale and the timelike separated points $x$ and $y$ label the past and future
tips of a causal diamond ${\bf D}$.
$ds^2$ then corresponds to the distance between ${\bf D}$ and its cousin where the new tips $(x+dx,y+dy)$ are slightly displaced ones.

It is then proved\cite{Myers} that ${\mathcal K}$ has $d$ spacelike and $d$ timelike directions. 
In particular if we move the centre of ${\bf D}$ with coordinates $(x^{\mu}+y^{\mu})/2$ by an infinitesimal amount in any of
the $d$ directions ($\mu=0,1,\dots d-1$) one has $ds^2>0$, i.e. for such deformations we are moving in a spacelike direction in ${\mathcal K}$.
Notice in particular that moving the centre of a causal diamond in the {\it timelike} direction in ${\mathbb R}^{1,d-1}$ produces 
a spacelike displacement in ${\mathcal K}$.
On the other hand a constant spacelike shift of the center in ${\mathbb R}^{1,d-1}$ {\it also} yields
a spacelike displacement\footnote{Timelike displacements in kinematic space correspond to deformations of diamonds keeping their center invariant\cite{Myers}.} in ${\mathcal K}$. 

Now based on the intuition of Figures 7. and 8. let us imagine specially displaced pairs of causal diamonds ${\bf D}_1$ and ${\bf D}_2$
in ${\mathbb R}^{1,d-1}$.
These displaced ones are of two dual kinds. They are displaced either along timelike or spacelike hyperbolas or lines.
For example for two infinitesimally displaced diamonds one can choose
${\bf D}\leftrightarrow (x_1^{\mu},x_4^{\mu}):=(x^{\mu},y^{\mu})$
and
${\bf D}+{\bf \delta D}\leftrightarrow (x_3^{\mu},x_2^{\mu}):=(x^{\mu}+dx^{\mu},y^{\mu}+dy^{\mu})$ such that the points should lie on hyperbolas of either kind.
Then due to the causality conditions like (\ref{unusual}) and (\ref{causality2}) one can show that
$0<\frac{r_{12}^2r_{34}^2}{r_{14}^2r_{23}^2}<1$.
Hence for $\delta A_{\diamond} \ll L^2$ according to (\ref{diamondstring}) we get
\begin{equation}
e^{-\delta A_{\diamond}/L^2}=1-\frac{\delta A_{\diamond}}{L^2}+\dots=
\frac{r_{12}^2r_{34}^2}{r_{14}^2r_{23}^2}:=r({\bf D},{\bf D}+{\bf \delta D})=1-\frac{1}{2{\ell}^2}ds_{\mathcal K}^2+\dots
\label{ddots}
\end{equation}
Here we used Eq.(2.25) of Ref.\cite{Myers} and for the definition of $r_{ij}$ see (\ref{errjk}). This formula connects the infinitesimal area element of the string world sheet segment and the (\ref{modulimetric}) line element
on kinematic space.
For these diamonds we have an associated string world sheet segment of infinitesimal area : $\delta A_{\diamond}$.
Since  $\delta A_{\diamond}>0$ then we are in accord with the fact that both of these configurations give {\it spacelike} separation in kinematic space hence one should have  $ds^2>0$. These are the cases when the tips of the diamonds are localized on timelike hyperbolas or dually
when
we have spacelike hyperbolas
on codimension one hyperplanes featuring trapezoids.

Now generally for ${\bf D}_1$ and
${\bf D}_2$ there are reduced density matrices
$\rho_1$ and $\rho_2$ calculated for the corresponding ball-shaped regions. They are obtained after tracing out the relevant parts of the vacuum state of the CFT. We then have the modular flows inside ${\bf D}_1$ and
${\bf D}_2$ with the corresponding modular Hamiltonians $H_1=-\log\rho_1$ and 
$H_2=-\log\rho_2$ at our disposal.
We can then calculate the Bures distance\cite{Bures} between the corresponding density operators, or taking instead of $\rho_1$ and $\rho_2$ the infinitesimally separated ones $\rho$ and 
$\rho+d\rho$:
for the calculation of the Bures metric.
It is known that in this case the fidelity susceptibility can be extracted from the overlap  $\vert\langle\psi_1\vert\psi_2\rangle\vert$ between certain normalized states $\psi_1$ and $\psi_2$ that form {\it parallel purifications}\cite{Uhlmann,Kirklin} of our density matrices\footnote{Two purifications are are parallel when the transition probability $\vert\langle\psi_1\vert\psi_2\rangle\vert$ is maximised.}. 
Then we expect that\footnote{The first equality of the next equation is Uhlmann's Theorem. For the holographic context see Ref.\cite{Kirklin}.}
\begin{equation}
\vert\langle\psi_1\vert\psi_2\rangle
\vert={\rm Tr}\left(\sqrt{\sqrt{\rho_1}\rho_2\sqrt{\rho_1}}\right) \simeq 
r({\bf D}_1,{\bf D}_2)=e^{-A_{\diamond}/L^2}
\label{Buresconjecture}
\end{equation}
Here the $\simeq$ symbol is referring to a correspondence whose precise form is yet to be established.
According to Eq.(\ref{ddots}) the infinitesimal version of the (\ref{Buresconjecture}) correspondence using Eq.(\ref{fidelity}) shows that the area $\delta A_{\diamond}$ of an infinitesimal string world sheet segment should be dual to fidelity susceptibility.
Of course in order to make this correspondence sound the explicit form of the purifications should be found
and their meaning should be clarified. 
Note in this respect that though the density matrices
$\rho_1$ and $\rho_2$ are acting on different Hilbert spaces, but according
to Ref.\cite{Kirklin},  appropriate maps can be found from each of these Hilbert spaces to a common one.
So this construction in principle can be done.

Luckily however, one can use an alternative and more explicit method for arriving at such a fidelity susceptibility interpretation. It can be achieved by
considering the variation of the modular flow via regarding the coordinates of the future and past tips of the causal diamonds as parameters\cite{Kinaiak}. In this approach kinematic space is playing the role of a parameter space in a usual Berry's Phase scenario\cite{Berry}.
One can then calculate the so called quantum geometric tensor\cite{Provost,BerryShapere} whose symmetric part gives rise to the Provost-Val\'ee metric\cite{Provost}) on kinematic space. 
There is a map provided by the spectral projector (associated to the vacuum)  coming from the spectral resolution of the modular Hamiltonian.
Then the
 Provost-Vallee metric is the pullback of the Fubini-Study metric\cite{LevayGeo} (which is now living on the space of rays of the Hilbert space associated with the causal diamond) with respect to this map to kinematic space.

In order to elaborate on these concepts note that
a useful generalization of Eq.(\ref{fidelity}) is to consider a family of Hamiltonians $H(\lambda)$  
with set of parameters $\lambda_I$ with $I=1,2,\dots $ forming a manifold $M$.
Then the family of spectral projectors $P_n({\lambda})$
with $H(\lambda)P_n(\lambda)=E_n(\lambda)P_n(\lambda)$
gives rise to a mapping $f^{(n)}:M\to P$ where $P$ is the space of rays of the Hilbert space $\mathcal H$ on which $H$ acts.
It can be shown that the pullback of the Fubini-Study metric on $P$
to $M$ by $f^{(n)}$ is of the Provost-Valee form\cite{Provost}
\begin{equation}
ds_{n}^2=\sum_{m\neq n}{\rm Re}\langle\partial_I n(\lambda)\vert m(\lambda)\rangle\langle m(\lambda)\vert\partial_J n(\lambda)\rangle d\lambda^Id\lambda^J=
G_{IJ}^{(n)}(\lambda)d\lambda^Id\lambda^J
\label{provi}
\end{equation}
where $P_n=\vert n(\lambda)\rangle\langle n(\lambda)\vert$.
In our case we have $M={\mathcal K}$ and $I,J=1,2,\dots 2d$ and parameter dependent Hamiltonian is the modular Hamiltonian. The corresponding calculations have been carried out in Ref.\cite{Kinaiak}. There it was also argued that
the vacuum state is also an eigenstate of the modular Hamiltonian for the spherical regions. Therefore, we can directly use the vacuum state in our calculations meaning that in Eq.(\ref{provi}) one can use $n=0$ corresponding to the vacuum.
Then from the analogue of Eq.(\ref{fidelity})
the result from Ref.\cite{Kinaiak}) for $d$ even is 
\begin{equation}
ds^2_0\simeq (-1)^{(d-2)/2}a_d^{\ast}\left(\frac{ds_{\mathcal K}}{2\pi\ell}\right)^2\log(\tilde{\varepsilon})
\label{china}
\end{equation}
i.e. the Provost-Valee metric for the vacuum state is related to 
the metric on kinematic space
\footnote{The $\log(\tilde{\varepsilon})$ term comes from a regularization factor\cite{Kinaiak}.} 
. 
We remark that this is the result one should be expecting based on the general considerations of Refs.\cite{Levaymod,LevayLand}
for parameter spaces having a coset space structure $G/H$ like ${\mathcal K}$. Indeed, the Maurer-Cartan form part on $G/H$ is representation independent and fixing the form of the line element 
$ds^2_{\mathcal K}$. On the other hand there are terms related to representations of $G$ on the Hilbert space of the theory\footnote{See for example Eqs. 5.5a of Ref.\cite{LevayLand}.} fixing the prefactors in this case to $a_d^{\ast}$.

Let us now recall Eq.(\ref{hogyuni}) and use our result of Eq.(\ref{hogyan1}) to obtain
\begin{equation}
a_d^{\ast}\left(\frac{ds_{\mathcal K}}{\ell}\right)^2=
\frac{V(B_L^{d-2})
    \delta A_{\diamond}}{4G^{(d+1)}L}
\label{hogyhogy}
\end{equation}
Taken together Eqs.(\ref{china}) and (\ref{hogyhogy}) we see that the Provost-Valee quantum information metric (fidelity susceptibilty)
is indeed related to the combination of entanglement entropies which is found on the left hand side of Eq.(\ref{hogyan1}).
Moreover, thanks to (\ref{hogyhogy}) the fidelity susceptibility is in turn related to the area of the infinitesimal segmented string world-sheet.

\subsubsection{Fidelity susceptibility, $d$ odd}

Surprisingly our fidelity susceptibility interpretation for the area of the stringy world-sheet survives even for the $d$ odd case.
In order to see this recall that it is possible to extend the relation between central charges and entanglement entropy to higher dimensions as far as the spacetime dimension is even. When we consider odd dimensional spacetime, we do not have any clear definition of central charges due to the absence of the Weyl anomaly\cite{RT2}. Hence the meaning of an analogous term $a_d^{\ast}$ showing up in a formula similar to (\ref{hogyhogy}) for $d$ odd at first sight is not clear.
Luckily, using the ideas of Refs.\cite{Sinha1,Perlmutter}
one can still generalize $a_d^{\ast}$ to $d$ odd.
According to these results for even dimensions $a_d^{\ast}$ is the usual coefficient of the $d$-dimensional Euler density in the conformal anomaly, and in odd dimensions $a_d^{\ast}\simeq \log Z_{S^d}$ with $Z_{S^d}$ being the partition function of the CFT on the sphere. 

With this generalization for $a_d^{\ast}$ at hand one can still ask the question whether there is a corresponding relation between the metric on kinematic space and the Provost-Vallee metric. The answer is yes\cite{Kinaiak} and for odd $d$ it is of the form
\begin{equation}
ds^2_0\simeq (-1)^{(d+1)/2}\pi a_d^{\ast}\left(\frac{ds_{\mathcal K}}{2\pi\ell}\right)^2
\label{china2}
\end{equation}
Now since according to Eq.(\ref{ddots}) we have $\delta A_{\diamond}/L^2= \left(ds_{\mathcal K}/\ell\right)^2/2$ valid for $d$ arbitrary then combining this with (\ref{china2}) we see that $ds_0^2$ still corresponds to $\delta A_{\diamond}$, hence the fidelity susceptibility interpretation of the area holds.

Notice however, that according to Eqs.(\ref{unihyp1})-(\ref{unihyp2}) 
the universal term of the area of the extremal surface $X_{\mathcal R}$ is of the form
${\mathcal A}_{uni}(X_{\mathcal R})=L^{d-1}V({\mathbb H}^{d-1})$
where  $V({\mathbb H}^{d-1})$ is the regularized volume of the $d-1$ dimensional hyperbolic space
with metric $ds^2_{{\mathbb H}^{d-1}}=du^2+\sinh^2ud\Omega^2_{d-2}$ calculated in Appendix C.
Now $V({\mathbb H}^{d-1})$ for $d$ odd is containing a constant term and not a logarithmic
one. For $d$ even we were able to directly relate the universal logarithmic term to the world sheet area of the string segment which was also logarithmic.
But although for $d$ arbitrary the world sheet segments are connected to $X_{\mathcal R}$ in a direct geometric manner, their areas are related  to the areas of such $X_{\mathcal R}$s only for $d$ even. Hence for $d$ odd the world sheet area cannot be reinterpreted in holographic entanglement entropy terms.  Regardless of this here we were able to confirm that the  {\it world sheet area/ fidelity susceptibility} duality is still valid
for $d$ arbitrary.

\subsubsection{Complexity}

Looking at the right hand side of Eq.(\ref{hogyhogy})
valid for $d$ even one can see that (for not necessarily infinitesimal world sheets) it is of the form
\begin{equation}
\frac{Vol(\Diamond\times B_L^{d-2})}{4G^{(d+1)}L}
\end{equation}
where $V_d=Vol(\Diamond\times B_L^{d-2})$ is the volume of our string world sheet segment $\Diamond$ times a $d-2$ dimensional ball with radius equals the $AdS$
length. Moreover, relating the relevant $a_d^{\ast}$ generalization\cite{Sinha1,Perlmutter} valid also for $d$ odd to similar terms we expect a similar relation to hold for general $d$. Now the volume $V_d$ is a codimension one object in ${\rm AdS}_{d+1}$.
According to the complexity equals volume conjecture\cite{Stanford}
(see also the related complexity equals action conjecture\cite{Suss2}) one should interpret this quantity as computational complexity $\mathcal C (\psi_1,\psi_2)$ in quantum information
i.e.
\begin{equation}
 \mathcal C(\psi_1,\psi_2)=\frac{Vol(\Diamond\times B_L^{d-2})}{4G^{(d+1)}L}   
\end{equation}

It is well-known that there is no consensus in the holographic community on how to define computational complexity in quantum field theory.
There is Nielsen's geometric method\cite{Nielsen,Nielsen2,Jeff}, Fubini-Study method\cite{Pastawski} and the path integral complexity proposal\cite{Caputa}.
In our context the proposal put forward in Ref.\cite{Pastawski} is of relevance. According to this proposal the circuit complexity ${\mathcal C} (\psi_1,\psi_2)$ calculated for a pure reference state $\psi_1$ and a pure target state $\psi_2$ is just the length of a geodesic connecting these states. 
If the geodesics are not unique, then one should choose the one minimizing the distance between the reference state and the target state.
The length of the  geodesic is calculated with respect to the metric of Eq.(\ref{provi}) and is given by\footnote{Here a factor of $2$ is added to be consistent with Nielsen's geometric method\cite{Ruan1}.} the formula
\begin{equation}
\mathcal C(\psi_1,\psi_2)=
\int_0^1 ds\sqrt{2G^{(0)}_{IJ}(\lambda(s))\dot{\lambda}^I\dot{\lambda}^J}
\label{complex1}
\end{equation}
where we $\lambda^I\leftrightarrow (x^\mu,y^{\mu})$.
An alternative definition of complexity is related to the energy of the geodesic
\begin{equation}
\mathcal C^{(\kappa=2)}(\psi_1,\psi_2)=
2\int_0^1 dsG^{(0)}_{IJ}(\lambda(s))\dot{\lambda}^I\dot{\lambda}^J=\mathcal C^2(\psi_1,\psi_2)
\label{complex2}
\end{equation}
here $\kappa$ is a parameter labeling the so called cost function in Nielsen's metod\cite{Jeff}.

Now since we also know that the line element of this metric showing up in these proposals is related
to a fidelity susceptibility
of the previous subsection it is natural to conjecture that the reference and target states in our context should be the parallel purifications of Eq.(\ref{Buresconjecture}). They are the ones purifying the density matrices $\rho_1$ and $\rho_2$ of the causal diamonds ${\bf D}_1$ and ${\bf D}_2$ in a special manner.
In order to prove this conjecture one should be able to relate the results of \cite{Kinaiak} we used in our considerations based on the Provost-Vallee metric of the previous subsection, to existing ones based on the Bures geometry\cite{Ruan2, Kirklin}. 
If this conjecture is true then there is an alternative interpretation of Eq. (\ref{complex1}) featuring the pullback of the Bures metric (also called the quantum Fisher metric) to $\mathcal K$  as the Provost-Valee one calculated in Ref.\cite{Kinaiak}.  
Note that the relevance of the Bures metric for calculating complexity was first put forward in the proposal (see Eq. 2.20 there) of Ref.\cite{Ruan1}.

After these considerations, and adopting the (\ref{complex2}) definition for complexity using the result
\begin{equation}
2ds_0^2\simeq \frac{\delta V_d}{4G^{(d+1)}L},\qquad {\delta V_d}=Vol(B_L^{d-2}\times \delta A_{\diamond})
\label{lenyegveg}
\end{equation}
one obtains for the infinitesimal version of (\ref{complex2}) the formula 
\begin{equation}
\mathcal C^{(\kappa=2)}(\psi,\psi +\delta\psi)\simeq \frac{\delta V_d}{4G^{(d+1)}L}     
\label{complexlenyeg}
\end{equation}
This formula relates the complexity between two states $\psi$ and $\psi +\delta\psi$ (parallel purifying the two density matrices $\rho$ and $\rho +\delta\rho$ corresponding to the causal diamonds 
${\bf D}$ and ${\bf D}+{\bf \delta D}$), to the infinitesimal area of the corresponding string worlds sheet.
It is natural to conjecture that a similar formula holds for arbitrary pairs of causal diamonds having a spacelike separation with respect to the metric in kinematic space.
According to this conjecture the geodesic distance calculated by (\ref{complex2}) should be dual to the $d$ dimensional volume containing the world sheet area segment.

\section{Conclusions}

In this paper we have revealed an interesting connection between segmented strings propagating in an $AdS_{d+1}$ background and ${\rm CFT}_d$ subsystems in
Minkowski spacetime  characterized by quantum information theoretic quantities
calculated for the vacuum state.
We have shown that the area of the world sheet of a string segment measured in appropriate units on the AdS side can be connected to fidelity susceptibility (quantum information metric) on the CFT side if $d$ arbitrary.
More precisely: we have shown that for the vacuum state the line element $ds^2_0$ of the quantum information metric can be related to the infinitesimal area of the string world sheet in the form of Eq.(\ref{lenyegveg}).
This quantity has another (see Eq.(\ref{complexlenyeg}))  interpretation as the computational complexity for infinitesimally separated states corresponding to causal diamonds
that are displaced in a spacelike manner (according to the metric of kinematic space).
In Eq.(\ref{Buresconjecture}) for the general case of not necessarily infinitesimally (but still spacelike) separated intersecting causal diamonds
we formulated the following conjecture.  The Bures fidelity calculated for the states $\rho_1$ and $\rho_2$ answering the causal diamonds ${\bf D}_1$ and ${\bf D}_2$ corresponds to 
$e^{-A_{\diamond}/2L^2}$ where $A_{\diamond}$
is the area of the worlds sheet segment of the string.

For the special case of $AdS_3$ the mathematical form of our fidelity susceptibility coincides with the combination of entanglement entropies showing up in proofs of strong subadditivity for the covariant holographic entanglement entropy proposal.
More precisely: in this case the segmented stringy area in units of $4GL$ is the conditional mutual information $I(A,C\vert B)$ calculated for a trapezoid configuration arising from boosted spacelike intervals $A$,$B$ and $C$.
For the special case of $AdS_{d+1}$ with $d$ even
the fidelity susceptibility interpretation again coincides with a one using certain combinations of entanglement entropies.
For $d$ odd no such interpretation is possible.
However, building on results calculating the quantum geometric tensor for the vacuum, we were able to show that the alternative {\it fidelity susceptibility/world sheet string segment area} relation universally holds for $d$ arbitrary.

For ${\rm AdS}_3$ via a detailed investigation we have also verified that the causal diamonds encode information for a unique reconstruction of the string world sheet segments in a holographic manner. 
In order to understand this reconstruction in terms  already familiar from the literature let us notice that for a particular causal diamond the bulk dual of the associated density operator $\varrho$ is the entanglement wedge.
For a region $\mathcal R$ in a fixed time slice of ${\rm AdS}_3$ the entanglement wedge ${\mathcal W}_{\mathcal R}$ is  the region bounded by the causal diamond of the region ${\bf D}$ and the relevant parts of the two light-sheets\cite{Bousso}: ${\mathcal N}_{\pm}$  i.e. the cones of Figure 1. In Figure 1. the boundary of ${\mathcal W}_{\mathcal R}$ is the region arising from the intersection of these cones and the causal diamond. The ${\mathcal N}_{\pm}$  are normal null hypersurfaces emanating from the extremal surface $X_{\mathcal R}$ of Section 3.5. For $X_{\mathcal R}$ see the red half circle of Figure 1.
For a convenient choice for ${\bf D}$
one can consider the "diamond Universe" of Section 3.8.2 with the region ${\mathcal R}$ of size $2R$.
Then one can extend the modular flow of (\ref{egyeske}) in the boundary to the bulk\cite{Wang}.

In our formalism this extension takes the following form.
Let us associate with ${\bf D}$ a causally ordered set of four consecutive events $(a_i,\bar{a}_i), i=1,2,3,4$.
We have already seen that these are in boosted inertial frames or in noninertial ones proceeding with constant acceleration exhibiting hyperbolic motion. We choose two of such events ($i=1,2$) as the ones fixing the future and past tips of ${\bf D}$ and the third and fourth ones are fixing the extra tips needed to have a pair of intersecting diamonds ${\bf D}_1$ and ${\bf D}_2$. 
According to Eqs.(\ref{ad1})-(\ref{accelerate11}) the coordinates $(a_i,\bar{a}_i)$ of these events are fixing the modular time ($\tilde{\tau}$) and acceleration parameters ($g$).
But (any three of them) also determines the normal vector $N$ of a spacelike plane where our string world sheet segment lives.
For the explicit form of $N$ see Eqs.(\ref{egyeske})-(\ref{utolsocska}).
Clearly all the vertices $V_i$ of our string world sheet segment
are on this plane.
For several segments of that kind lying in different kind of planes see Figure 5. These planes give rise to the modular slices\cite{Wang} of the entanglement wedge ${\mathcal W}_{\mathcal R}$. The simplest slice of that kind can easily be visualized using Figure 4. Here $X_{\mathcal R}$ is the half circle with largest radius featuring $V_2$ as the topmost point of the string world-sheet segment.
Now the modular slice is just the
intersection of the plane (containing the world-sheet segment) which is perpendicular to the $x-t$ plane with 
${\mathcal W}_{\mathcal R}$ (not shown in Figure 4).
Then we can conclude that the vertices $V_i$ are on a particular  modular slice, with $V_2$ on the extension of the boundary modular flow of ${\bf D}$.
The remaining vertices are coming from the lifts of the corresponding modular flows of ${\bf D}_1$, ${\bf D}_2$ and ${\bf D}_1\cap{\bf D}_2$.
Analysing these lifts gives an alternative way for deriving the expressions Eqs. (\ref{egyeske})-(\ref{utolsocska}) compatible with the equation of momentum conservation $p_1+p_2=p_3+p_4$.

The upshot of these considerations for the ${\rm AdS}_3$ case is that segmented strings seem to provide some sort tiling of the modular planes hence providing a tomography of the entanglement wedge. According to Eq. (\ref{complexlenyeg}) for $d=2$ the areas of these tiles holographically encode information on the complexity properties of the vacuum state
via the relation
\begin{equation}
\mathcal C^{(\kappa=2)}(\psi,\psi +\delta\psi)\simeq \frac{\delta A_{\diamond}}{4G^{(3)}L}     
\label{complexlenyeg2}
\end{equation}

What can we say about the $d>2$ case?
For spherical regions $\mathcal R$ one can still consider the modular flow of the corresponding causal diamond ${\bf D}_{\mathcal R}$. In the spirit of Ref.\cite{Towards} one can map the causal wedge ${\mathcal W}_{\mathcal R}$ to a domain which can be written as a slicing of the form
\begin{equation}
AdS_2\times {\mathbb H}_{d-1}
\label{kakukk123}
\end{equation}
This domain has the geometry of the exterior of a hyperbolic black hole.
Under a conformal transformation $X_{\mathcal R}$ is mapped to the horizon of the corresponding black hole.
Since the (\ref{kakukk123}) space dually corresponds to a thermal state, the  entanglement entropy can be calculated as black hole entropy via Wald's formula\cite{Towards}.
Moreover, it was shown in Ref.\cite{Han} that in this case the modular slices are 
${\rm AdS}_2$ ones taken at a fixed point in the hyperbolic space ${\mathbb H}_{d-1}$. Clearly one can interpret the pullback of these slices under the inverse conformal map as the ones where the world-sheets of segmented strings live.
Then the volumes of the corresponding balls with volume $V(B_L^{d-2})$ can show up as objects coming from the relevant part of the metric on ${\mathbb H}_{d-1}$. Then one is expecting the result
\begin{equation}
\mathcal C^{(\kappa=2)}(\psi_1,\psi_2)\simeq 
\frac{Vol({\rm AdS_2}(\diamond)\times B_L^{d-2})}{4G^{(d+1)}L}     
\label{complexlenyeg3}    
\end{equation}
where the notation ${\rm AdS_2}(\diamond)$ indicates that the string world-sheet segments should be regarded as tiles for the ${\rm AdS}_2$ modular slices.
However, one must recall that these tiles are very special ones since they can also be located on different modular slices. In any case they are patched together by the (\ref{interpol}) interpolation ansatz, due to the fact that they solve the equation of motion  coming from the Nambu-Goto action.
Notice also in this respect that according to our special result of Section
of 3.9 a variation of the discretized Nambu-Goto action leads to an equation for entanglement entropies in the boundary theory. It is in the form of a Toda equation Eq.(\ref{eq:toda}) describing the patching conditions of these tiles in different modular planes.

In this work we related  fidelity susceptibility i.e. a quantity of quantum information theoretic origin of {\it boundary} states to a geometric property of segmented strings in the bulk.
These string world-sheet segments residing in the modular planes are scanning the classical space time geometry of the {\it bulk} entanglement wedge.
Fidelity susceptibility is related to the real part of the quantum geometric tensor. However, there is the imaginary part of this quantum information theoretic quantity which also has an important physical meaning relevant in the holographic context.
Indeed, this is the Berry curvature giving rise to the famous anholonomy Berry's Phase\cite{Berry}. In holography this curvature associated with boundary physics has been holographically related to the bulk symplectic form for pure states\cite{Sarosi} associated to a full Cauchy slice. This idea has recently been generalized and reviewed for the more general setting where one can also consider  deformations of density matrices describing mixed states  corresponding to general subregions in the CFT\cite{Kirklin, Czech2}.
The result of these works is that the bulk dual of the imaginary part of the quantum geometric tensor\footnote{In the case of Ref.\cite{Kirklin} the mixed state generalization of Berry's Phase i.e. Uhlmann's Phase and the associated curvature was considered.
In this case instead of the Provost-Valee metric as the real part of the quantum geometric tensor one should consider the Bures\cite{Bures} metric.} is the same symplectic form but now it is supported on the bulk entanglement wedge.
In the light of our results on the role of segmented string world sheets (as building blocks for scanning the the entanglement wedge), 
the exploration of their relationship to the bulk symplectic form should be a natural avenue for further study.

If our simple considerations performed here (for the ${\rm CFT}_d$ vacuum and its pure ${\rm AdS}_{d+1}$ dual) can be generalized 
then these important geometric correspondences might provide valuable insight. 
We are planning to investigate these exciting possibilities in future work.

\section{Acknowledgements}
This research was supported by the Ministry of Culture and Innovation and the National Research, Development and Innovation Office within the Quantum Information National Laboratory of Hungary (Grant No. 2022-2.1.1-NL-2022-00004, and by the National Research Development and Innovation Office of Hungary within the Quantum Technology National Excellence Program (Project No. 2017-1.2.1-NKP-2017-0001). It was also supported by the \'UNKP-22-2-III-BME-3 and \'UNKP-23-3-I-BME-68 New National Excellence Program of the Ministry for Culture and Innovation from the source of the National Research, Development and Innovation Fund. One of us (P.L.) would like to express his gratitude for the warm hospitality at  the Universite de Technologie Belfort Montbeliard,
and the award of a guest professorship.

\section{Appendix A}

Here we summarize the basics of the helicity formalism used in the text, and prove Theorem 1.

When considering arbitrary two vectors $A,B\in {\mathbb R}^{2,2}$
with components $A^a,B^b$, $a,b=-1,0,1,2$ in the canonical basis we have the bilinear form $A\cdot B=-A^{-1}B^{-1}-A^{0}B^{0} +A^{1}B^{1}+A^{2}B^{2}$.
In the helicity formalism we associate to these four component vectors $A^a$ a $2\times 2$ matrix ${\mathcal A}_{m\dot{m}}$ with $m=1,2$ and $\dot{m}=\dot{1},\dot{2}$ of the form
\begin{equation}
{\mathcal A}_{m\dot{m}}=\begin{pmatrix}{\mathcal A}_{1\dot{1}}&{\mathcal A}_{1\dot{2}}\\{\mathcal A}_{2\dot{1}}&{\mathcal A}_{2\dot{2}}\end{pmatrix}=
\begin{pmatrix}A^1+A^{-1}&A^{0}+A^{2}\\A^{0}-A^{2}&A^1-A^{-1}\end{pmatrix}:=\begin{pmatrix}A^+&A\\\bar{A}&A^-\end{pmatrix}
\end{equation}
Now one can check that
\begin{equation}
A\cdot B= 
\frac{1}{2}{\rm Tr}({\mathcal A}\epsilon\otimes\epsilon {\mathcal B})=-\frac{1}{2}
{\mathcal A}_{m\dot{m}}\epsilon^{\dot{m}\dot{n}}({\mathcal B}^T)_{\dot{n}n}\epsilon^{nm}:=\frac{1}{2}{\rm Tr}({\mathcal A}\tilde{\mathcal B})
\label{spinscalar}
\end{equation}
hence the usual bilinear form\footnote{Clearly
${\rm Tr}(\tilde{\mathcal A}{\mathcal B})={
\rm Tr}({\mathcal A}\tilde{\mathcal B})$ in accord with $A\cdot B=B\cdot A$.}
in ${\mathbb R}^{2,2}$ in the helicity formalism is represented by using the metric $\frac{1}{2}\epsilon\otimes \epsilon$.
Then for example taking the quantities of Eqs.(\ref{pek}), (\ref{veujegy}) and (\ref{enegy1}) a calculation shows that
\begin{equation}
N\cdot N=\frac{1}{2}{\rm Tr}({\mathcal N }\tilde{\mathcal N})=L^2,\quad
V_1\cdot V_1=
\frac{1}{2}{\rm Tr}({\mathcal V }_1\tilde{\mathcal V}_1)=
-L^2, \quad P_i\cdot P_i=
\frac{1}{2}{\rm Tr}({\mathcal P }_i\tilde{\mathcal P}_i)
=0 
\label{bázis}
\end{equation}
hence the vectors $P_i$ with $i=1,4$ are lightlike (null) $N$ is spacelike and finally $V_1$ is timelike. 
This $AdS_3$ vector is orthogonal to all of the remaining ones
\begin{equation}
N\cdot V_1=P_1\cdot V_1=P_4\cdot V_1=0
\end{equation}
Notice also that the quantities
\begin{equation}
\begin{pmatrix}{\mathcal Q}_1\\{\mathcal Q_2}\\{\mathcal Q}_3\\{\mathcal Q}_4\end{pmatrix}=
\begin{pmatrix}{\mathcal P}_1\\{\mathcal V}_1+{\mathcal N}\\{\mathcal V}_1-{\mathcal N}\\{\mathcal P}_4 \end{pmatrix}
\label{nullbasis1}
\end{equation}
give rise to a null basis
$Q_i\cdot Q_i=0$, $i=1,2,3,4$ were $Q_1\cdot Q_4=-Q_2\cdot Q_3=2L^2$

In order to prove Theorem 1
let us give one of the points say $(a_2,\bar{a}_2)$ a special status of a fourth (running) point in ${\mathbb R}^{1,1}$ this time to be denoted by  $(a,\bar{a})$. 
Of course we still insist on our (\ref{unusual}) causality condition. 
Now we have the
\begin{equation}
\zeta:=\frac{(a_2-a_4)(a_3-a_1)}{(a_2-a_1)(a_4-a_3)}>0,\qquad
\bar{\zeta}: =\frac{(\bar{a}_2-\bar{a}_4)(\bar{a}_3-\bar{a}_1)}{(\bar{a}_2-\bar{a}_1)(\bar{a}_4-\bar{a}_3)}>0
\label{southerncross}
\end{equation}
cross ratios at our disposal
and one can also use the spinors $z=(a,L)$ and $\bar{z}=(\bar{a},L)$ regarded as row vectors. 

Let us now define

\begin{equation}
\begin{pmatrix}\psi_{11}&\psi_{12}\\\psi_{21}&\psi_{22}\end{pmatrix}=\frac{1}{\sqrt{(a_4-a_1)(a_4-a_3)(a_3-a_1)}}\begin{pmatrix}(a_4-a_3)a_1&(a_4-a_3)L\\(a_3-a_1)a_4&(a_3-a_1)L\end{pmatrix}    
\label{psii}
\end{equation}
\begin{equation}
\begin{pmatrix}\overline{\psi}_{\dot{1}\dot{1}}&\overline{\psi}_{\dot{1}\dot{2}}\\\overline{\psi}_{\dot{2}\dot{1}}&\overline{\psi}_{\dot{2}\dot{2}}\end{pmatrix}=\frac{1}{\sqrt{(\bar{a}_4-\bar{a}_1)(\bar{a}_4-\bar{a}_3)(\bar{a}_3-\bar{a}_1)}}\begin{pmatrix}(\bar{a}_4-\bar{a}_3)\bar{a}_1&(\bar{a}_4-\bar{a}_3)L\\(\bar{a}_3-\bar{a}_1)\bar{a}_4&(\bar{a}_3-\bar{a}_1)L\end{pmatrix}  \end{equation}
We denote these $2\times 2$ matrices by $\psi$ and $\bar{\psi}$.
They are $2\times 2$ matrices having the index structure
${\psi}_{\alpha m}$
and ${\bar{\psi}_{\dot{\alpha}\dot{m}}}$.
These matrices are built up from the two component spinors familiar from Eq.(\ref{vesszo12})-(\ref{vesszo22}).
Explicitly ${\psi}_{1m}$ corresponds to $\psi_1$ and ${\psi}_{2m}$ corresponds to $\psi_4$.

Then recalling the definition of Eq.(\ref{dek22}) one
can perform the following calculation
\begin{equation}
z\epsilon\psi^T=\frac{L}{\sqrt{\mathcal D}}(a_4-a_3)(a-a_1)(1,\zeta)    
\end{equation}
\begin{equation}
z\epsilon{\mathcal N}\epsilon \bar{z}^T=(z\epsilon\psi^T)\epsilon(\bar{\psi}\epsilon\bar{z}^T)=
\frac{L^2}{\sqrt{\mathcal D\bar{\mathcal D}}}(a-a_1)(\bar{a}-\bar{a}_1)
(a_4-a_3)(\bar{a}_4-\bar{a}_3)
(\bar{\zeta}-{\zeta})   
\nonumber
\end{equation}
where now in the (\ref{southerncross}) definition of the cross ratio $\zeta$ we have to use $a=a_2$.

Recall that the matrix ${\mathcal N}_{m\dot{m}}$ represents a spacelike four vector  $N\in {\mathbb R}^{2,2}$ in the helicity formalism. 
Moreover, defining 
\begin{equation}
{\mathcal Z}_{m\dot{m}}:=z_m\bar{z}_{\dot{m}}    
\end{equation}
and 
we observe that it represents a lightlike vector  $Z\in{\mathbb R}^{2,2}$. Then we have 
$z\epsilon{\mathcal N}\epsilon \bar{z}^T=-{\rm Tr}({\mathcal N}\tilde{\mathcal Z})$
hence
using (\ref{spinscalar}) one can write our result as
\begin{equation}
2N\cdot Z=\frac{L^2}{\sqrt{\mathcal D\bar{\mathcal D}}}(a-a_1)(\bar{a}-\bar{a}_1)
(a_4-a_3)(\bar{a}_4-\bar{a}_3)({\zeta}-\bar{\zeta})
\label{theorem1}
\end{equation}
Then $N\cdot Z=0$ iff ${\zeta}=\bar{\zeta}$ as claimed.

We note that by employing  the spinor matrices ${\psi}$ and $\bar{\psi}$ used in our proof of Theorem 1 the four basic null vectors of (\ref{nullbasis1}) characterizing the string world sheet segment in the helicity formalism can be arranged as\cite{Alday} 
\begin{equation}
W_{\alpha m,\dot{\alpha}\dot{m}}:=\psi_{\alpha m}\overline{\psi}_{\dot{\alpha}\dot{m}}:=\frac{1}{2}\begin{pmatrix}
{\mathcal P}_1&{\mathcal V}_1+{\mathcal N}\\{\mathcal V}_1-{\mathcal N}&{\mathcal P}_4\end{pmatrix} 
\label{combi}
\end{equation}

\section{Appendix B}

In this appendix we present some calculational details. 
First we prove the (\ref{matpluck}) identity. Using the (\ref{errjk})  definition
a calculation gives
\begin{equation}
\frac{{\mathcal Z}_i-{\mathcal Z}_j}{2r_{ij}^2}=\begin{pmatrix}
 \frac{a_i+a_j}{a_i-a_j}+\frac{\overline{a}_i+\overline{a}_j}{\overline{a}_i-\overline{a}_j}&
 \frac{2L}{(\overline{a}_i-\overline{a}_j})
 \\\frac{2L}{({a}_i-{a}_j)}&0\end{pmatrix}   
\end{equation}
Then we plug in the (\ref{hasznalni}) relation into the right hand side of Eq.(\ref{matpluck})
to obtain
\begin{equation}
    \frac{{r_{12}}r_{34}}{r_{13}r_{24}}\left[\frac{{\mathcal Z}_1-{\mathcal Z}_2}{2r_{12}^2}-\frac{{\mathcal Z}_1-{\mathcal Z}_4}{2r_{14}^2}\right]=
    \left[\frac{{\mathcal Z}_1-{\mathcal Z}_4}{2r_{14}^2}-\frac{{\mathcal Z}_1-{{\mathcal Z}_3}}{2r_{13}^2}\right]
\label{nahat}
\end{equation}
Then for the left hand side of (\ref{nahat}) we get
\begin{equation}
\frac{{r_{12}}r_{34}}{r_{13}r_{24}}
\begin{pmatrix}
a_1\frac{a_2-a_4}{(a_1-a_2)(a_1-a_4)}+\overline{a}_1\frac{\overline{a}_2-\overline{a}_4}{(\overline{a}_1-\overline{a}_2)(\overline{a}_2-\overline{a}_4)}&
L\frac{\overline{a}_2-\overline{a}_4}{(\overline{a}_1-\overline{a}_2)(\overline{a}_1-\overline{a}_4)}
\\ L\frac{{a}_2-{a}_4}{({a}_1-{a}_2)({a}_1-{a}_4)}&0
\end{pmatrix}
\end{equation}
and for the right hand side
\begin{equation}
\begin{pmatrix}
a_1\frac{a_4-a_3}{(a_1-a_4)(a_1-a_3)}+a_1\frac{\overline{a}_4-\overline{a}_3}{(\overline{a}_1-\overline{a}_4)(\overline{a}_1-\overline{a}_3)}&
L\frac{\overline{a}_4-\overline{a}_3}{(\overline{a}_1-\overline{a}_4)(\overline{a}_1-\overline{a}_3)}
\\ L\frac{{a}_4-{a}_3}{({a}_1-{a}_4)({a}_1-{a}_3)}&0
\end{pmatrix}
\end{equation}
Now one can see that by virtue of (\ref{baromijo2}) the left hand side can be
converted to the right hand side hence our (\ref{matpluck}) identity indeed holds.

Let us now prove the (\ref{important2}) identity.  
When the four points of Section 3.4. are on a hyperbola one can use the hyperbolic version of the inscribed angle theorem \cite{inscribed} to arrive at
\begin{equation}
-\frac{\Delta_{14}\bullet\Delta_{34}}{4r_{14}r_{34}}=
-\frac{\Delta_{12}\bullet\Delta_{23}}{4r_{12}r_{23}}=\cosh\theta
\label{inscribed}
\end{equation}
Indeed these expressions show that the relative rapidity $\theta$ between two observers starting from $x_1$ and later from $x_3$ will be the same as the vertex of their intended meeting points is moved from $x_4$ to $x_2$ along the hyperbola.

From Eq.(\ref{inscribed}) we obtain
\begin{equation}
\frac{r_{14}r_{34}}{r_{12}r_{23}}=\frac{\Delta_{14}\bullet\Delta_{34}}{\Delta_{12}\bullet\Delta_{32}}=\frac{r_{13}^2-r_{14}^2-r_{34}^2}{r_{13}^2-r_{12}^2-r_{23}^2}
\label{innen}
\end{equation}
where we have used the identity
\begin{equation}
    \Delta_{ij}\bullet\Delta_{kj}=-4r_{ij}^2-4r_{kj}^2+4r_{ik}^2
\end{equation}
Then from (\ref{innen}) one can see that
\begin{equation}
\frac{r_{13}}{r_{12}r_{23}}-\frac{r_{12}}{r_{13}r_{23}}-\frac{r_{23}}{r_{12}r_{13}}=\frac{r_{13}}{r_{14}r_{34}}-\frac{r_{14}}{r_{13}r_{34}}-\frac{r_{34}}{r_{13}r_{14}} 
\end{equation}
In this expression one can get rid of 
$\frac{r_{23}}{r_{12}r_{13}}$ using the second of (\ref{pl2})
and of $\frac{r_{14}}{r_{13}r_{34}}$ using the first of (\ref{pl2}).
From here one indeed arrives at (\ref{important2}).

Finally we present some details for the calculation of $N_a$ the normal vector for the string world sheet segment.
First we write (\ref{normalis})
in the following form
\begin{equation}
N_a=\frac{1}{2}\varepsilon_{abcd}{\Pi}^{cd}V_1^b=\ast{\Pi}_{ab}V_1^b
=L^2\ast{\Pi}_{ab}\left(\frac{p_3^b}{p_3\cdot p_4}\right)=\frac{r_{14}^2}{2r_{13}r_{34}r_{14}}\ast{\Pi}_{ab}Z_3^b
\label{ide}
\end{equation}
where
\begin{equation}
\Pi=\frac{p_1\wedge p_4}{p_1\cdot p_4}
\end{equation}
where we have used Eqs.(\ref{errjk}), (\ref{zeee}), 
(\ref{lambdagauge}), (\ref{pek}) and (\ref{veegy}).
Now a calculation shows that
\begin{equation}
{\Pi}^{-10}+{\Pi}^{12}=\frac{L}{a_1-a_4}+\frac{1}{L}\frac{a_1a_4}{a_1-a_4},\qquad
{\Pi}^{-10}-{\Pi}^{12}=\frac{L}{\overline{a}_1-\overline{a}_4}+\frac{1}{L}\frac{\overline{a}_1\overline{a}_4}{\overline{a}_1-\overline{a}_4}
\end{equation}

\begin{equation}
{\Pi}^{01}-{\Pi}^{-12}=\frac{L}{a_1-a_4}-\frac{1}{L}\frac{a_1a_4}{a_1-a_4},\qquad
{\Pi}^{01}+{\Pi}^{-12}=\frac{L}{\overline{a}_1-\overline{a}_4}-\frac{1}{L}\frac{\overline{a}_1\overline{a}_4}{\overline{a}_1-\overline{a}_4}
\end{equation}

\begin{equation}
{\Pi}^{-11}+{\Pi}^{02}=\frac{a_1+a_4}{a_1-a_4},\qquad
{\Pi}^{-11}-{\Pi}^{02}=\frac{\overline{a}_1+\overline{a}_4}{\overline{a}_1-\overline{a}_4}
\end{equation}
It can be shown that left and right moving parts correspond to the self-dual and anti self-dual parts of ${\Pi}_{ab}$.
Using these expressions in (\ref{ide}) the result of (\ref{egyeske})-(\ref{utolsocska}) now follows by a straightforward calculation.

\section{Appendix C}

We have seen that two $U,V\in\mathbb{R}^{2,d}$ null vectors define an $AdS_{d+1}$ minimal surface via the equations $U\cdot X=0$ and $V\cdot X=0$. In the Poincaré model these equations become:
\begin{align}
    &z^2+(x-x_u)\bullet(x-x_u)=0 \label{eq:cones_gen_app1}\\
    &z^2+(x-x_v)\bullet(x-x_v)=0 \label{eq:cones_gen_app2}
\end{align}

Due to the time and space translational invariance of the metric \eqref{eq:metric_patch} in the Poincare model the area of a minimal surface can depend only on the quantities $\Delta^\mu=x_v^\mu-x_u^\mu$ while the coordinates of the center are $\frac{1}{2}(x_u^\mu+x_v^\mu)$. Therefore we can calculate the area in the special case with $x_u^\mu=-x_v^\mu$ (hence the minimal surface lies in the origin) as a function of $\Delta^\mu=2x_v=-2x_u$ and then substitute back an arbitrary difference $\Delta^\mu=x_u^\mu-x_v^\mu$ to get the general result. In the following let us examine this special case.

Assume that $x_u^\mu=-x_v^\mu$ and denote $\Delta^\mu=2x_v=-2x_u$. The equations \eqref{masalak} can be written in the form
\begin{align}
    z^2-\left(t+\frac{1}{2}\Delta^0\right)^2+\sum_k\left(x^k+\frac{1}{2}\Delta^k\right)^2&=0\\
    z^2-\left(t-\frac{1}{2}\Delta^0\right)^2+\sum_k\left(x^k-\frac{1}{2}\Delta^k\right)^2&=0
\end{align}
Where $k=1,\dots,d-1$. In this case $x_u\bullet x_u=x_v\bullet x_v$ therefore their intersection lies in the plane
\begin{equation}\label{eq:min_subspace}
    t\Delta^0=\sum_k x^k\Delta^k
\end{equation}
Substituting this back into the equation of cones one gets the following formula:
\begin{equation}\label{eq:min_surf_zx}
    z^2-\frac{1}{(\Delta^0)^2}\left(\sum_k x^k\Delta^k\right)^2+\sum_k (x^k)^2=-\frac{1}{4}\Delta\bullet\Delta=r^2
\end{equation}
Now with a change of coordinates $x^k\rightarrow \tilde{x}^k$ by an $SO(d-1)$ transformation in the $x^k$ subspace one can make the vector $\Delta^a$ to be parallel to $\tilde{x}^1$. The sum $\sum_a (x^a)^2$ stays invariant. In this setup it can be shown that the equation simplifies to
\begin{equation}
    z^2+\frac{4r^2}{(\Delta^0)^2}(\tilde{x}^1)^2+(\tilde{x}^2)^2+\dots+(\tilde{x}^{d-1})^2=r^2
\end{equation}
The coordinate $x^0$ can be expressed by \eqref{eq:min_subspace} as
\begin{equation}
    t=\frac{\tilde{x}^1}{\Delta^0}\sqrt{\sum_k(\Delta^k)^2}
\end{equation}
The metric 
\begin{equation}
ds^2=L^2\frac{dz^2+dx\bullet dx}{z^2}
\end{equation}
Stays invariant under the change of coordinates by the $SO(d-1)$ transformation.

It is convenient to parametrize the minimal surface by $d-1$ dimensional spherical coordinates $\theta,\phi_1,\phi_{d-2}$:
\begin{equation}
\left.
    \begin{aligned}
        &z=r\cos{\theta}\\
        &t=\frac{1}{2}\sqrt{\sum_k(\Delta^k)^2}\sin{\theta}\cos{\phi_1}\\
        &\tilde{x}^1=\frac{1}{2}\Delta^0\sin{\theta}\cos{\phi_1}\\
        &\tilde{x}^2=r\sin{\theta}\sin{\phi_1}\cos{\phi_2}\\
        &\cdots\\
        &\tilde{x}^{d-2}=r\sin{\theta}\sin{\phi_1}\dots\cos{\phi_{d-2}}\\
        &\tilde{x}^{d-1}=r\sin{\theta}\sin{\phi_1}\dots\sin{\phi_{d-2}}
    \end{aligned}
    \right\}
\end{equation}
Where $\phi_1,\cdots,\phi_{d-3}\in[0,\pi]$, $\phi_{d-2}\in[0,2\pi]$ and $z>0$ implies that $\theta_{d-1}:=\theta\in[0,\pi/2]$.

The area of the minimal surface $X_R$ can be determined by calculating the following integral
\begin{equation}\label{eq:area_int}
    {\mathcal A}(X_R)=\int d^{d-1}x\sqrt{h}
\end{equation}
Where the induced metric $h_{kl}$ is given by
\begin{equation}
    h_{kl}=L^2\frac{\partial_k z\partial_l z+\partial_k \tilde{x}\bullet\partial_l \tilde{x}}{z^2}
\end{equation}

One can show that the induced metric tensor $h_{kl}$ is therefore a $(d-1)\times(d-1)$ dimensional tensor whose elements are
\begin{equation}
    \left.
    \begin{aligned}
        &h_{\theta\theta}=\frac{L^2}{\cos^2{\theta}},&\\
        &h_{kk}=\frac{L^2}{\cos^2{\theta}}\sin^2{\theta}\sin^2{\phi_1}\dots\sin{\phi_{k-2}},&\qquad\text{for }k=\phi_1,\dots,\phi_{d-2}\\
        &h_{kl}=0,&\text{for }k\neq l
    \end{aligned}
    \right\}
\end{equation}
Hence the area form on the surface is given by
\begin{equation}\label{eq:form_spherical}
    d^{d-1}\sqrt{h}=d\theta d\phi_1\dots d\phi_{d-2}L^{d-1}\frac{\sin^{d-2}{\theta}}{\cos^{d-1}{\theta}}\sin^{d-3}{\phi_1}\sin^{d-4}{\phi_2}\dots\sin{\phi_{d-3}}
\end{equation}
Now we are able to calculate the area of the minimal surface. Inserting \eqref{eq:form_spherical} into \eqref{eq:area_int} one gets the following integral
\begin{equation}
    A_{\text{min}}=L^{d-1}\int_0^{\pi/2}d\theta\frac{\sin^{d-2}{\theta}}{\cos^{d-1}{\theta}}\int_0^{\pi}d\phi_{1}\sin^{d-3}{\phi_1}\dots\int_0^{2\pi}d\phi_{d-2}
\label{moregeneral}
\end{equation}
The integration over variables $\phi_1,\dots,\phi_{d-2}$ gives the (\ref{omegawell}) surface $\Omega_{d-2}$ of a unit sphere thus
\begin{equation}
    {\mathcal A}(X_R)=L^{d-1}\Omega_{d-2}\int_0^{\pi/2}d\theta\frac{\sin^{d-2}{\theta}}{\cos^{d-1}{\theta}}
\end{equation}
Or by a change of variable $y=\cos{\theta}=z/r$ one gets
\begin{equation}
    {\mathcal A}(X_R)=L^{d-1}\Omega_{d-2}\int_0^1 dy\frac{\left(1-y^2\right)^{\frac{d-3}{2}}}{y^{d-1}}
\end{equation}

Now it is important to note that this is just a formal expression due to the divergence of the embedding metric at $z\to 0$. So to get a finite value one needs to regularize the minimal surface by limiting $z$ to $z>\delta$, where $\delta$ is small. In the $y=z/r$ coordinates this means a regularization $y>\delta/r$. Hence the regularized area of the minimal surface can be calculated by
\begin{equation}\label{eq:integral_d}
    {\mathcal A}(X_R)=L^{d-1}\Omega_{d-2}\int_{\delta/r}^1 dy\frac{\left(1-y^2\right)^{\frac{d-3}{2}}}{y^{d-1}}
\end{equation}
The integrand can be expanded into a series
\begin{equation}
    \left(1-y^2\right)^{\frac{d-3}{2}}y^{1-d}=\sum_{n=0}^\infty C_n y^{(2n+1)-d}
\end{equation}
where
\begin{equation}
    C_0=1,\qquad C_n=\frac{(-1)^n}{2^n n!}(d-3)(d-5)\dots(d-(2n+1))
\end{equation}
Therefore one need to calculate the following series
\begin{equation}
    \sum_{n=0}^\infty C_n\int_{\delta/r}^1 dy\, y^{(2n+1)-d}
\end{equation}
Now we need to calculate this series separately for even and odd dimensions. 

\subsubsection*{Case of $d$ even}
Let consider first the even $d$ case. For $n=(d-2)/2$ we need to integrate $1/y$ so we get a term proportional to $\log{y}$. Hence we need to seperate the summation into three parts
\begin{equation}
\begin{aligned}
    \sum_{n=0}^\infty C_n\int dy\, y^{(2n+1)-d}=&\sum_{n=0}^{(d-4)/2}\frac{C_n}{2n-d+2}y^{(2n+2)-d}+C_{(d-2)/2}\log{y}+\\
    &+\sum_{n=d/2}^{\infty}\frac{C_k}{2n-d+2}y^{(2n+2)-d}
\end{aligned}
\end{equation}
Therefore for the integral in the calculation for the area of the minimal surface in the case of $d$ even one gets
\begin{equation}
\begin{aligned}
    \int_{\delta/r}^1 dy\frac{\left(1-y^2\right)^{\frac{d-3}{2}}}{y^{d-1}}=&\text{const.}-\sum_{n=0}^{(d-4)/2}\frac{C_n}{2n-d+2}\left(\frac{r}{\delta}\right)^{d-(2n+2)}+\\
    &+C_{(d-2)/2}\log{\left(\frac{r}{\delta}\right)}-\sum_{n=d/2}^{\infty}\frac{C_n}{2n-d+2}\left(\frac{\delta}{r}\right)^{(2n+2)-d}
\end{aligned}
\end{equation}
Where the constant term comes from the integral evaluated at $y=1$. The logarithmic term can be written in the form
\begin{equation}
    C_{(d-2)/2}\log{\left(\frac{r}{\delta}\right)}=\frac{(-1)^{(d-2)/2}(d-3)!!}{2^{(d-2)/2}\left(\frac{d-2}{2}\right)!}\log{\left(\frac{r}{\delta}\right)}=\frac{(-1)^{(d-2)/2}}{2}\frac{(d-3)!!}{(d-2)!!}\log{\left(\frac{r^2}{\delta^2}\right)}
\end{equation}
Our main concern will be further elaborations on the meaning of this term.

\subsubsection*{Case of  $d$ odd}
For the odd dimensional case the logarithmic term disappears so the series simply becomes
\begin{equation}
    \sum_{n=0}^\infty C_n\int dy\, y^{(2n+1)-d}=\sum_{n=0}^{\infty}\frac{C_k}{2n-d+2}y^{(2n+2)-d}
\end{equation}
It is easy to show that this expression is equal to
\begin{equation}
    \sum_{n=0}^{\infty}\frac{C_n}{2n-d+2}y^{(2n+2)-d}=\frac{y^{2-d}}{d-2}\,{}_{2}F_1\left(\frac{2-d}{2},\frac{3-d}{2},\frac{4-d}{2};y^2\right)
\end{equation}
where
\begin{equation}
    {}_{2}F_1\left(a,b,c;x\right)=1+\frac{ab}{c}\frac{x}{1!}+\frac{a(a+1)b(b+1)}{c(c+1)}\frac{x^2}{2!}+\dots
\end{equation}
is the hypergeometric function. Therefore the integral in the area calculation is
\begin{equation}
\begin{aligned}
    \int_{\delta/r}^1 dy\frac{\left(1-y^2\right)^{\frac{d-3}{2}}}{y^{d-1}}=&\frac{1}{d-2}\,{}_{2}F_1\left(\frac{2-d}{2},\frac{3-d}{2},\frac{4-d}{2};1\right)-\\
    &-\frac{1}{d-2}\left(\frac{r}{\delta}\right)^{d-2} {}_{2}F_1\left(\frac{2-d}{2},\frac{3-d}{2},\frac{4-d}{2};\left(\frac{r}{\delta}\right)^2\right)
\end{aligned}
\end{equation}
The first constant term can be expressed explicitly in terms of Gamma functions:
\begin{equation}
    \frac{1}{d-2}\,{}_{2}F_1\left(\frac{2-d}{2},\frac{3-d}{2},\frac{4-d}{2};1\right)=\frac{\Gamma\left(\frac{4-d}{2}\right)\Gamma\left(\frac{d-1}{2}\right)}{(2-d)\sqrt{\pi}}=\frac{\Gamma\left(\frac{2-d}{2}\right)\Gamma\left(\frac{d-1}{2}\right)}{2\sqrt{\pi}}
\end{equation}
We are only examining again this term due to latter reasons.

Merging the even and odd case the area of the minimal surface for even and for odd $d$ is:
\begin{equation}\label{eq:area_gen}
    {\mathcal A}(X_{\mathcal R})=\dots+L^{d-1}\Omega_{d-2}\left\{
    \begin{aligned}
    &\frac{(-1)^{(d-2)/2}}{2}\frac{(d-3)!!}{(d-2)!!}\log{\left(\frac{r^2}{\delta^2}\right)},\text{ if } d \text{ is even}\\
    &\frac{\Gamma\left(\frac{2-d}{2}\right)\Gamma\left(\frac{d-1}{2}\right)}{2\sqrt{\pi}},\text{ if } d \text{ is odd}
    \end{aligned}
    \right\}+\dots
\end{equation}
One can use in this formula (\ref{omegawell}) and the identity
$\Gamma\left(\frac{1}{2}+m\right)=2^{-m}(2m-1)!!\sqrt{\pi}$ with $m=(d-2)/2$ to obtain the final form as
\begin{equation}\label{eq:area_gen2}
    {\mathcal A}(X_{\mathcal R})=\dots+L^{d-1}\left\{
    \begin{aligned}
    &\frac{(-2\pi)^{(d-2)/2}}{(d-2)!!}\log{\left(\frac{r^2}{\delta^2}\right)},\text{ if } d \text{ is even}\\
    &{\pi}^{(d-2)/2}\Gamma\left(\frac{2-d}{2}\right),\text{ if } d \text{ is odd}
    \end{aligned}
    \right\}+\dots
\end{equation}
It is important to realize that 
${\mathcal A}(X_{\mathcal R})$ can also be written in the form\cite{Perlmutter} of
\begin{equation}
    {\mathcal A}(X_{\mathcal R})=\dots+L^{d-1}
    \frac{\pi^{d/2}}{\Gamma(d/2)}
    \left\{
    \begin{aligned}
    &(-1)^{(d-2)/2}\frac{2}{\pi}\log{\left(\frac{r}{\delta}\right)},\text{ if } d \text{ is even}\\
    &(-1)^{(d-1)/2},\text{ if } d \text{ is odd}
    \end{aligned}
    \right\}+\dots
\label{unihyp1}
\end{equation}
Then we notice that the term after $L^{d-1}$ is the regularized hyperbolic volume $V({\mathbb H}^{d-1})$, i.e the volume of the space with metric
$ds^2_{{\mathbb H}^{d-1}}=du^2+\sinh^2u d\Omega^2_{d-2}$.
Then one can write for the universal contributions the formula
\begin{equation}
    {\mathcal A}(X_{\mathcal R})\simeq L^{d-1}
    V({\mathbb H}^{d-1})
\label{unihyp2}
\end{equation}

We have considered the so far a special case when the minimal surface is placed at the origin. However, due to spacial and time translational invariance \eqref{eq:area_gen} still holds for any arrangement, only $r$ need to be replaced by
\begin{equation}
    r^2=-\frac{1}{4}(x_u-x_v)\bullet(x_u-x_v)
\end{equation}

\end{document}